\newcommand{\trs}{100,000 }
\renewcommand{\baselinestretch}{1.0}

\documentclass{emulateapj}
\usepackage{graphicx,amssymb,amsfonts,amsmath,fmtcount,mathrsfs,verbatim,longtable,paralist,natbib}
\usepackage[caption=false,font=footnotesize]{subfig}

\begin{document}
\title{A novel, fully automated pipeline for period estimation  in the EROS 2 data set}

\author{Pavlos Protopapas\altaffilmark{1,2}, 
   Pablo Huijse \altaffilmark{3,4}, 
    Pablo A. Est\'evez\altaffilmark{4,3}, 
    Pablo Zegers\altaffilmark{5}, 
   Jos\'e C. Pr\'incipe\altaffilmark{6},
   Jean-Baptiste Marquette\altaffilmark{7}
}
\affil{\altaffilmark{1}Institute for Applied Computational Science, Harvard University, Cambridge, MA, USA}
\affil{\altaffilmark{2}Harvard-Smithsonian Center for Astrophysics, 60 Garden Street,  Cambridge, MA, USA}
\affil{\altaffilmark{3}Millennium Institute of Astrophysics, Chile } 
\affil{\altaffilmark{4}Department of Electrical Engineering and the Advanced Mining Technology Center, Universidad de Chile, Santiago, Chile} 
\affil{\altaffilmark{5}Universidad de los Andes, Facultad de Ingenier\'ia y Ciencias Aplicadas, Monse\~nor \'Alvaro del Portillo 12455, Las Condes, Santiago, Chile} 
%Pablo Zegers is with the Universidad de los Andes, Facultad de Ingenier\'ia y Ciencias Aplicadas, Monse\~nor \'Alvaro del Portillo 12455, Las Condes, Santiago, Chile.
\affil{\altaffilmark{6}Computational Neuroengineering Laboratory of University of Florida, FL, USA} 
\affil{\altaffilmark{7}UPMC-CNRS, UMR7095, Institut d'Astrophysique de Paris, F-75014, Paris, France}
\affil{\altaffilmark{}email: pavlos@seas.harvard.edu}

\begin{abstract}

We present a new method to discriminate periodic from non-periodic  irregularly sampled lightcurves. We introduce a periodic kernel and maximize a similarity measure derived from information theory to estimate the periods and a discriminator factor. We tested the method on a dataset containing \trs synthetic periodic and non-periodic lightcurves with various periods, amplitudes and shapes generated using a multivariate generative model.  We correctly identified periodic and non-periodic lightcurves with a completeness of $\sim 90\%$ and a precision of $\sim 95\%$, for lightcurves with a signal-to-noise ratio (SNR) larger than 0.5.  We characterize the efficiency and reliability of the model using these synthetic lightcurves and  applied the method  on the EROS-2 dataset. A crucial consideration is the speed at which the method can be executed. 
Using hierarchical search and some simplification on the parameter search we were able to analyze 32.8  million lightcurves in $\sim 18$ hours on a cluster of GPGPUs. Using the sensitivity analysis on the synthetic dataset, we infer that  0.42\% in the LMC  and 0.61\% in the SMC of the sources show periodic behavior. The training set, the  catalogs  and source code are all available in \href{http:\\timemachine.iic.harvard.edu}{http://timemachine.iic.harvard.edu}.

\end{abstract}


\section{Introduction}
Characterization of the dynamic optical sky is one of the  observational frontiers in astrophysics. Variable sources, defined as any source that its apparent magnitude changes over  time, have historically led to fundamental insights into subjects ranging from the structure of stars and the most energetic explosions in the universe  to cosmology. These changes and their characteristics, tell us a lot about the sources such as pulsating stars, supernovae, the interaction of the source with its surrounding such as AGNs or light being blocked by something between the source and the observer. However no optical telescope to date has had the capability to search for transient phenomena at faint levels over enough of the sky to fully characterize variable sources. 

A subcategory of the variable sources are the periodic variables. Those are variables that in general repeat at regular intervals. While  astronomers historically have been able to study variable and transient phenomena by examining  the behavior of individual sources, the amount of data and the large number of sources have exponentially grown in the last decade \citep{Hodapp2004, LSST2012, Catalina2003, TF2009}, making this task daunting. 

Although most stars have at least some variation in luminosity, current estimations indicate that 3\% of the stars are varying more than the sensitivity of the instruments and~$\sim$1\% are periodic \citep{Eyer1999}. EROS-2 \citep{EROS2007}, MACHO \citep{Alcock2000}, OGLE \citep{Udalski1997} were among the first generation of large scale surveys, monitoring millions of sources for many years. Pan-STARRS \citep{Hodapp2004} is currently monitoring the whole visible sky repeatedly and it will be doing it for a total of three years. In the future SDSS \citep{York2000}, LSST \citep{LSST2012} will monitor even more sources, and more frequently, generating  billions of lightcurves.  It is because of this explosion of data that there is a need for efficient and well characterized period finding techniques. 

The problem of period estimation from noisy and irregularly sampled observations has been studied before. Most approaches identify the period by some form of grid search. That is, the problem is solved by evaluating a criterion $\Phi$ at a set of trial periods and selecting the period $p$ that yields the best value for $\Phi(p)$. Commonly used techniques vary in the form and parametrization of $\Phi$, the evaluation of the fit quality between model and data, the set of trial periods searched, and the complexity of the resulting procedures. Two methods that are popular are the LS periodogram \citep{Scargle1982,Reimann1994} and the phase dispersion minimization (PDM) \citep{Stellingwerf1978}, both known for their success in empirical studies. The LS method is relatively fast and is equivalent to maximum likelihood estimation under the assumption that the function has a sinusoidal shape. It therefore makes a strong assumption on the shape of the underlying function. On the other hand, PDM makes no such assumptions and is more generally applicable, but it is slower and is less often used in practice.  

In this paper we adopted the correntropy kernelized periodogram (CKP), an information theoretical criterion introduced in \citet{Huijse2012} to assess periodicity in lightcurves. The CKP combines the generalized autocorrelation function \citep{Principe2010} with a periodic kernel yielding a generalized periodogram. The CKP measures similarity over time using statistical information contained in the probability density function (pdf) of the samples. This gives the CKP an advantage over methods that rely on second-order statistical descriptors\footnote{To fully characterize non-gaussian random processes the higher order moments are needed.}. By adjusting the kernel parameters of the CKP one can adapt the metric to different noise regimes and periodicities. The selection of these parameters for the case of lightcurves is thoroughly discussed in the present work.

To fully qualify the method we generated a large set of  synthetic lightcurves (110K) using  parameter distributions motivated from the data. To do so, we used a model free multivariate generative model and sampled the parameters. We also use a smaller but manageable subset from the real data in order to compare our results with reality. These subsets were used to optimize the free parameters of the pipeline and to characterize the efficiency and completeness of the process. 

Astronomy and many experimental sciences are now collecting more data that can be possibly analyzed by human experts in reasonable time. We are not really interested in the data per se, but in the information it contains about the natural phenomena. Machine learning and signal processing are becoming an integral part of the process of extracting information from data, because they are quantitative methods based on statistics and function analysis methods. This synergism is in its early stages, and this paper shows an effective methodology to speed up the discovery of periodic stars in large data bases as the EROS2.

Section \ref{sec:framework}  describes the theoretical framework that this work is based on, Section \ref{sec:method} describes the pipeline and methodology, Section \ref{sec:synthetic} describes the synthetic data set, Section \ref{sec:data} describes the data, Section \ref{sec:results} contains the results obtained from our runs and finally conclusions are in Section \ref{sec:conclusion}.

%--------------------------- Theoretical framework ----------------------
\section{Theoretical framework} \label{sec:framework}

The structure of a time series can be quantified by measuring the signal similarity over time. The first measure that comes to mind is the autocorrelation function of the time series \citep{Jenkins1968}. Let us define the time series as a realization of a stochastic process $\{x_n,n=0,1,\ldots,N\}$, where $x$ is a random variable in $\mathbb{R}$. The autocorrelation function for stationary processes is defined as 
\begin{equation} \label{correlation}
R[m] = {\mathbb E}[\langle {x_{n},x_{n-m}}\rangle],
\end{equation}
where ${\mathbb E}[\cdot]$ indicates the expectation value. The autocorrelation coefficient\footnote{Covariance normalized by the variance} normally is 
estimated for stationary and ergodic time series as a simple sum of lagged products over a window of data.
\begin{equation} \label{correlation-estimator}
\hat{R}[m] =\frac{1}{N+1-m} \, \, \frac{1}{ \sigma^2} \sum_{n=m}^{N} (x_n-\mu)(x_{n-m} - \mu),
\end{equation}
%[PP: WHY +1 .... I AM COUNTING AND IT DOES NOT WORK]
where $N+1$ is the number of measurements in the time series and the true mean $\mu$ and true variance $\sigma^2$ are time-independent.

Looking more closely at the autocorrelation definition one finds out that only second order information of the random variable $x$ is utilized in the definition, and as it is well known, only a few distributions such as the Gaussian are fully described by their (first and) second order moments. Therefore, one compromises the simplicity of the autocorrelation definition with a loss of a more in depth description of the signal similarity. This paper will use more powerful definitions of similarity for a better quantification of time series structure, which is pivotal to achieve the reported results. The ideas are founded in the mathematical theory of information and a descriptor of entropy that exploits the full statistical information from samples \citep{Principe2010}, which is utilized to define similarity metrics. 

Let us consider a stationary stochastic process $\{x_n\}$, and define the generalized autocorrelation as 
\begin{equation}\label{gcf}
	V[m]=  {\mathbb E}[\kappa(x_{n},x_{n-m})],
\end{equation}
where $\kappa(x,y)$ is a positive definite function of two arguments called a kernel \citep{Scholkopf2002,Cristianini2004}. If we define $\kappa(x,y) = \langle {x,y} \rangle $, i.e. the first order polynomial kernel one obtains the autocorrelation function of Eq. \eqref{correlation}, \eqref{correlation-estimator}. Instead let us select $\kappa(x,y)$ as a translation invariant kernel \citep{Scholkopf2002}, i.e. $\kappa(x,y) = \kappa(x-y,0)$. For simplicity we will use $\kappa(x-y)$ for translation-invariant kernel functions. The Gaussian kernel defined as

\begin{equation} \label{Gausskernel}
	G_{\sigma}(x-z)= \frac{1}{\sqrt{2\pi}\sigma}  \exp \left(- \frac{ \|x - z\|^2 }{2\sigma^2} \right),
\end{equation}
is a popular kernel that fits the conditions, where $\sigma$ is the covariance, and will be called in this context as {\em the kernel size}. In \cite{Principe2010} this class of functions is called autocorrentropy, or more simply correntropy, and here we will always assume the use of the Gaussian kernel. One of the advantages of correntropy is that it is still very easy to estimate directly from data assuming the random process is ergodic. Using the sample mean we can estimate Eq. \eqref{gcf} as 
\begin{equation}\label{autocorr}
	\widehat{V}_{\sigma}[m] = \frac{1}{N+1-m} \sum_{n=m}^{N}{ G_{\sigma}(x_{n}-x_{n-m}) }.
\end{equation}

\begin{figure*}
	\centering
	\subfloat[]{\label{fig-normV}\includegraphics[scale=0.45]{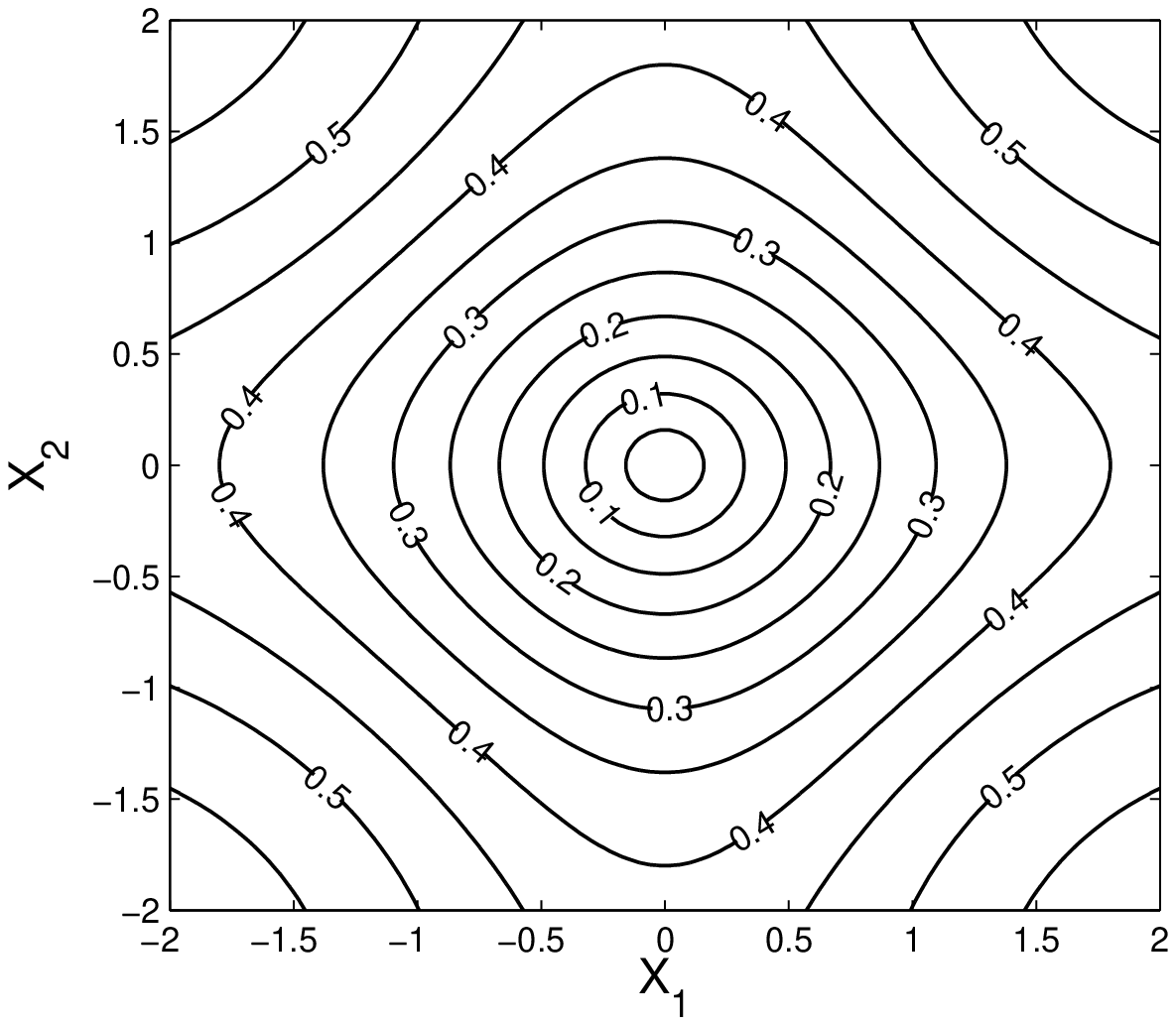} }
	\subfloat[]{\label{fig-normL1}\includegraphics[scale=0.45]{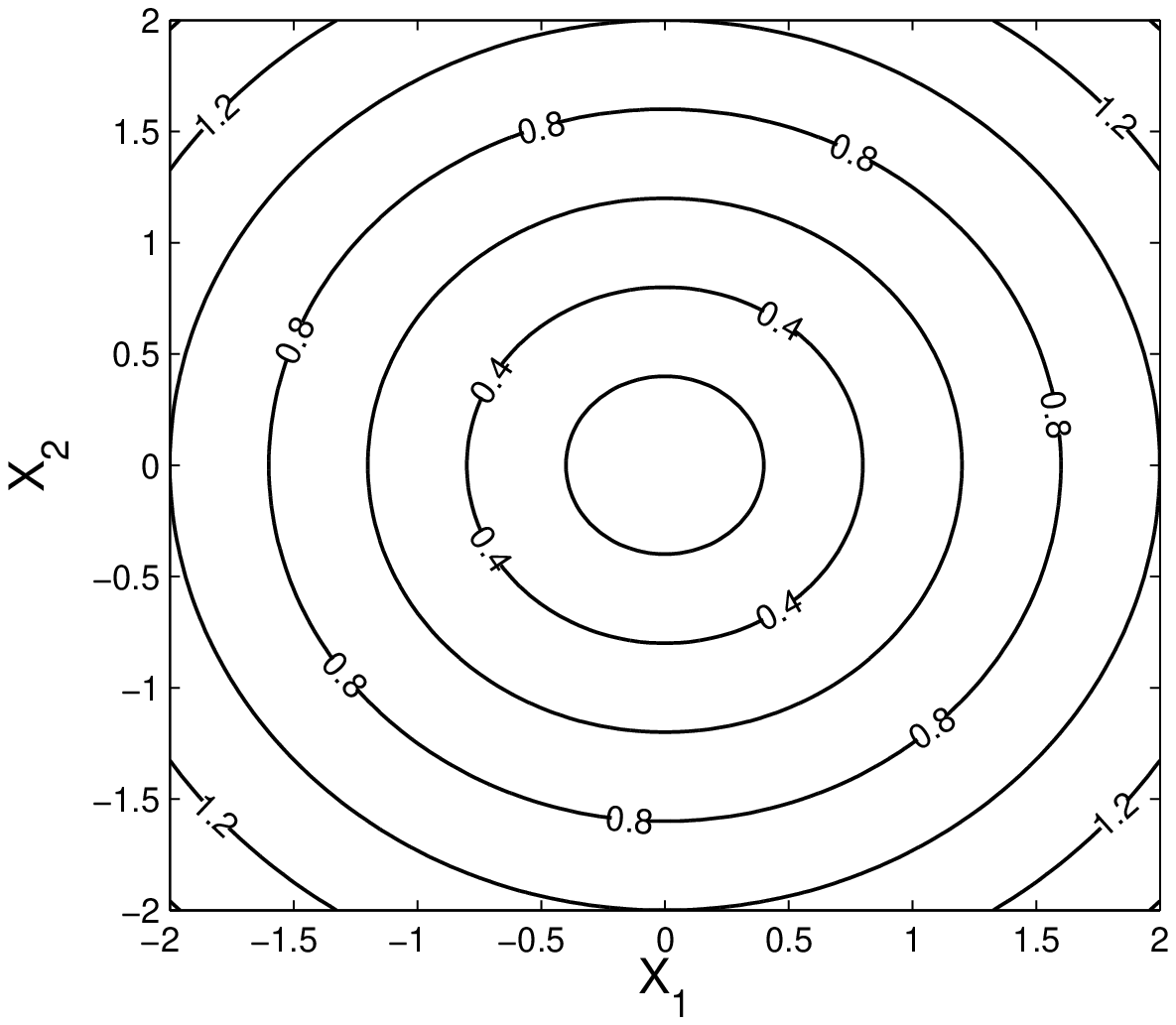} } 
	\subfloat[]{\label{fig-normL2}\includegraphics[scale=0.45]{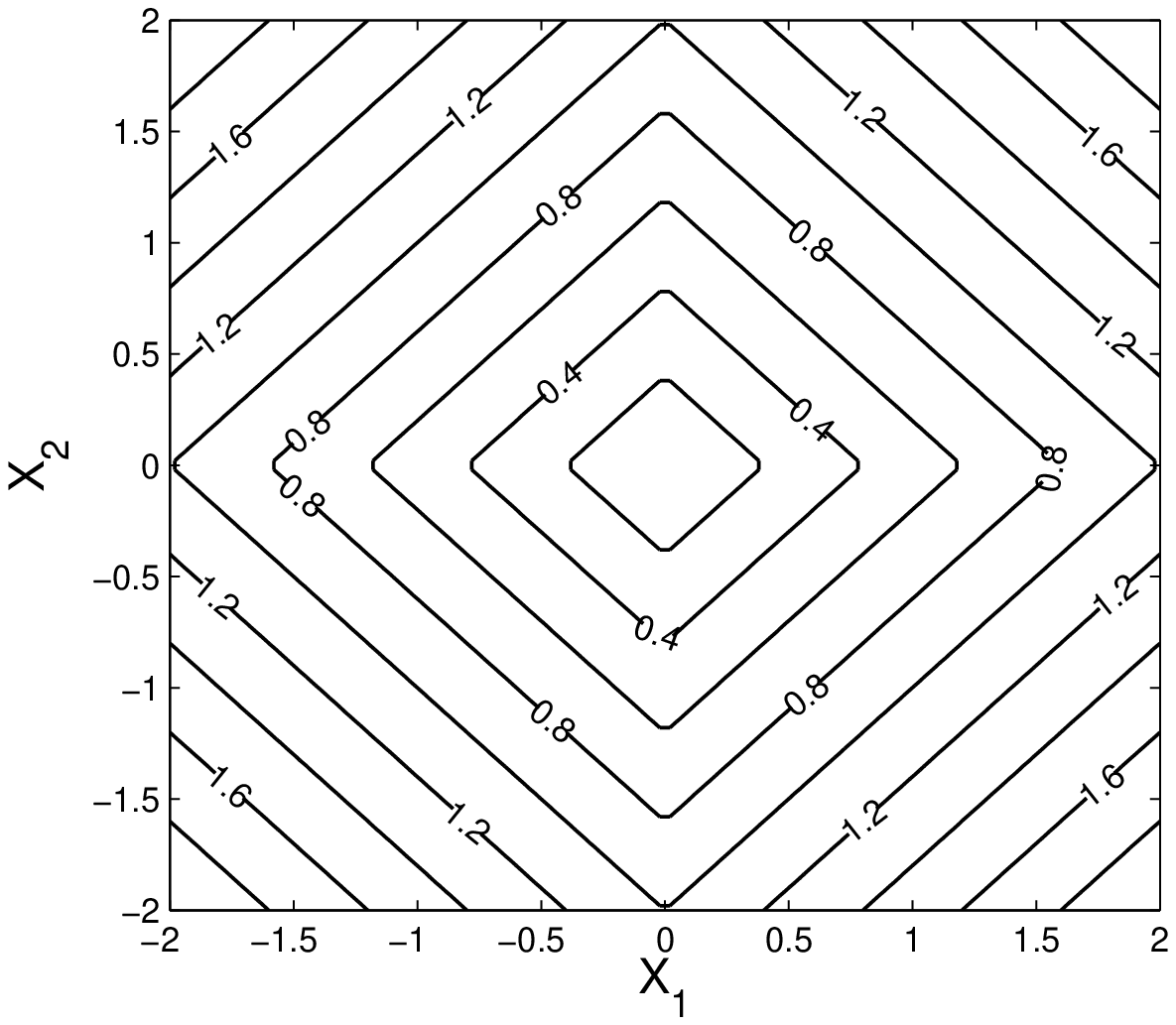} }    
	\caption{\label{fig:norm} Distances to the origin (contours) in a bidimensional sample space using the CIM(X,0) (a), $L_2$ norm (b) and $L_1$ norm (c). For the CIM (Eq. \ref{CIM}) a Gaussian kernel function with $\sigma=1$ is considered. Note how the CIM incorporates the $L_1$, $L_2$ and $L_0$ norms at different scales. }
\end{figure*}

%[PP: WHY N-m+1 ?] 
%PH: Because the index goes from 0 to N, ie N+1 samples
The difference between autocorrelation and autocorrentropy seems pretty minor, but it is very significant, as fully discussed in \citet{Principe2010}. For this work, the important correntropy properties are the following: 

\begin{enumerate}

\item Correntropy with the Gaussian kernel includes a weighted sum of all the even moments of the random variable, including %In fact if one expands the Gaussian kernel in correntropy using a Taylor series, it is easy to show that 
% arrive at
%\begin{equation} \label{CorrTaylor}
%	V_{\sigma}[m]= \frac{1}{\sqrt{2\pi} \sigma} \sum_{j=0}^{\infty}{ \frac{(-1)^j}{2^j \sigma^{2j} j!} E \left[ \|x_n-x_{n-m}\|^{2j} \right]},
%\end{equation}
%note that 
the second order moment (the autocorrelation) of $\|x_n-x_{n-m}\|$.% is included in correntropy. % ($j=1$).
\item Correntropy is a positive definite function %and with minor modifications is square integrable, therefore it 
can replace the autocorrelation function in the definition of the Power spectrum, yielding %what has been called  
the correntropy spectral density (CSD) \citep{Principe2010}, as
\begin{equation}\label{CSD}
	P_{\sigma}[f] = \sum_{m=-\infty}^{\infty}{ {U}_{\sigma}[m] \cdot \exp \left(- i \,2\pi f \frac{m}{F_s} \right) },
\end{equation}
where $F_s$ corresponds to the sampling frequency. The function $U_{\sigma}[m]$ corresponds to $V_{\sigma}[m]-IP$, where $IP$ corresponds to the mean value of the autocorrentropy function over the lags\footnote{This also the argument of Renyi's quadratic entropy \citep{Principe2010}.}. %Subtracting the IP is required to make correntropy square integrable.

\item Correntropy has a free parameter that can be interpreted as a scale parameter,
%in wavelet decompositions, 
therefore needs to be defined according to the time series data. 
%In supervised learning when correntropy is used as a cost function \citep{Principe2010}, the kernel size can be adapted online to achieve the smallest error. But for the correntropy function the same procedure cannot be applied, and 
%This work presents an experimental methodology to select the kernel size from time series data. 

\item Correntropy quantifies similarity using the correntropy induced metric (CIM) defined as 
\begin{equation}\label{CIM} 
CIM(x,y) = \left( \kappa(0,0) - {\mathbb E}[\kappa(x,y)] \right)^{1/2}.
\end{equation} 
The CIM is a  metric very different from the $L_p$ norms that define the Minskowski spaces where the distances are always weighted the same (Fig \ref{fig:norm})
%In the context of time series $L_p$ norms are $L_1=|x-y|, L_2=\sqrt{(x-y)^2}$ etc 
%[PH]: "In the context of time series is a bit confusing", it could mean distances between time series, in the examples we show distances to the origin |x-0| in R2.  The thing is that for x in R (unidimensional) L1 and L2 are the same!
\footnote{For $\vec x \in \mathbb{R}^n$, the $L_p$ norms are defined as $L_p=\|\vec x \|_p = \left( \sum_{i=1}^N x_i^p \right)^\frac{1}{p}$, $p \in (0,\infty)$. In the limit $p \rightarrow 0$, the $L_0$ norm is defined as the number of non-zero components in the vector (counting norm).}.
This means that distances between the arguments of the CIM are weighted nonuniformly, i.e. if the distance between the arguments is small then the CIM approximates the $L_2$ norm, but if the difference is larger then it will approximate the $L_1$ norm, and for very large difference between the arguments, the CIM tends to the $L_0$ norm. The transitions between the norms are smooth, and the assessment of `small' and `large', the scale in this space is controlled by the kernel size, which impacts drastically the assessment of similarity. 
\end{enumerate}

\begin{figure}[t]
	\centering
	\subfloat[]{\label{fig-esin}\includegraphics[scale=0.55]{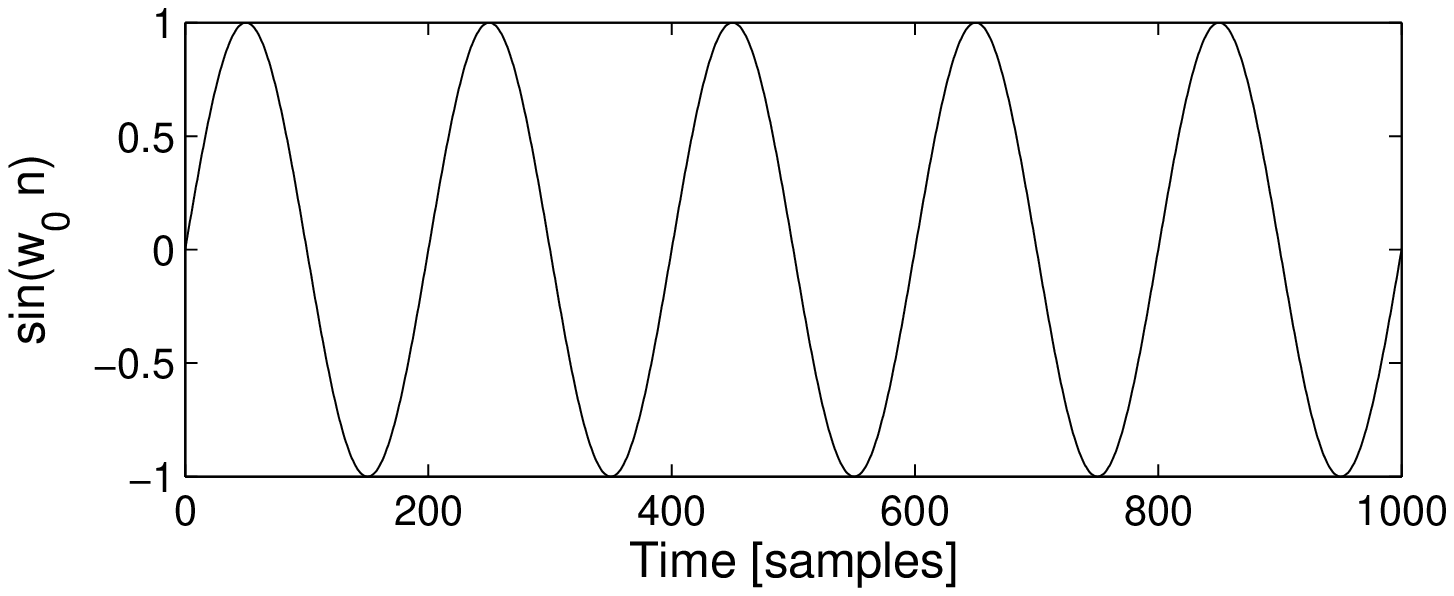} } \\ \vspace{-5pt}
	\subfloat[]{\label{fig-eR}\includegraphics[scale=0.55]{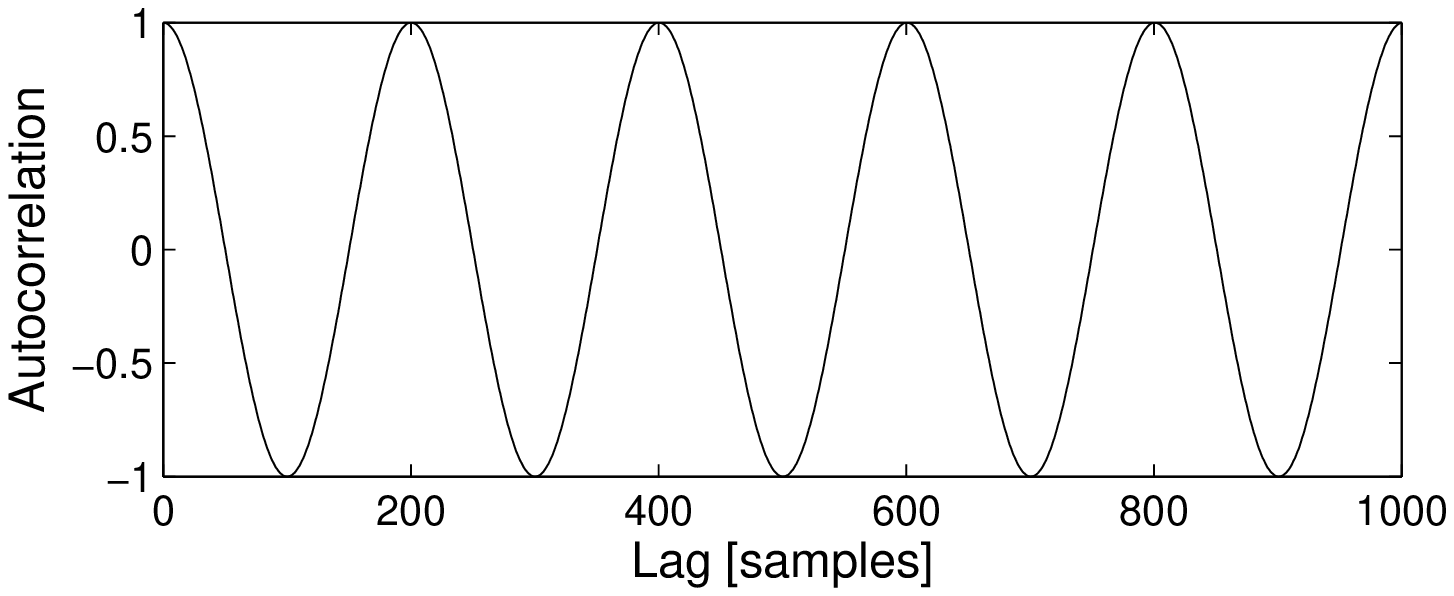} } \\ \vspace{-5pt}
	\subfloat[]{\label{fig-eV}\includegraphics[scale=0.55]{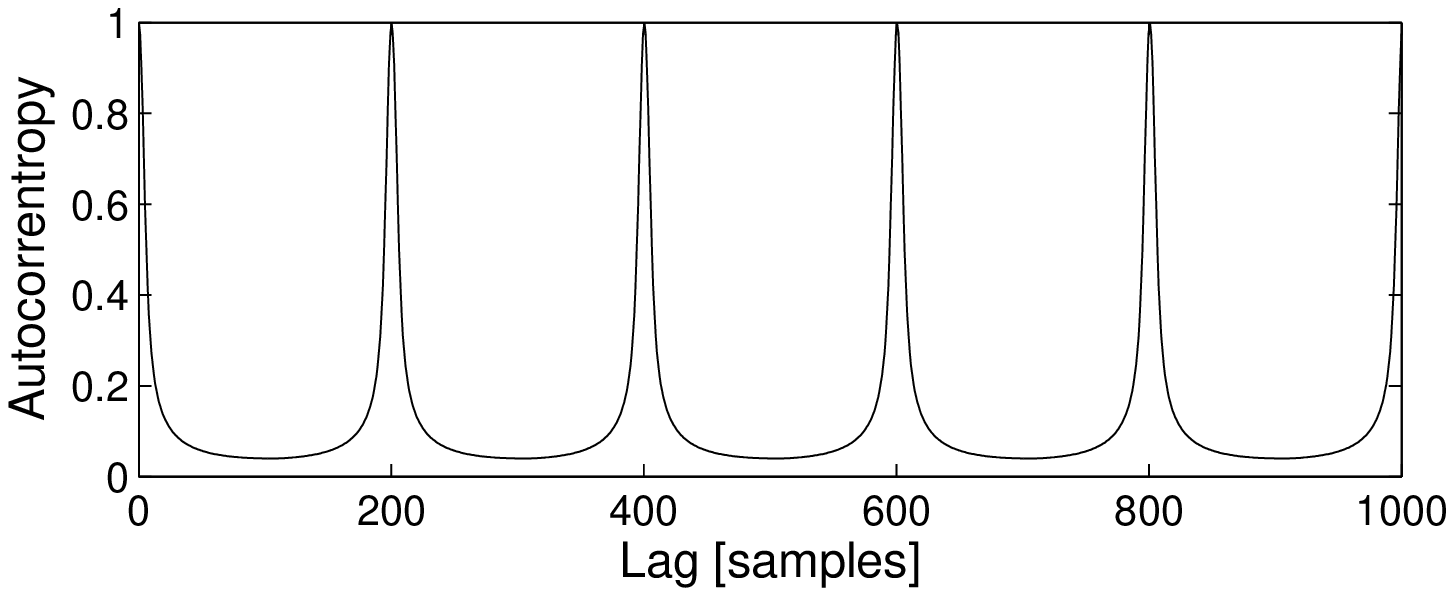} }    \vspace{-1pt}
	\caption{\label{fig:sinewave}  (a) Plot of $x_n= A \sin (w_0 n + \varphi)$ with unit amplitude, $w_0=2\pi /200$ and where $\varphi$ is a random variable uniformly distributed in $[-\pi,\pi]$. (b) Autocorrelation of $x_n$, note that the autocorrelation function of a sinewave is a sinewave. (c) Autocorrentropy of $x_n$, note that the autocorrentropy of a sinewave is a train pulse in which the periodicity is represented by the peaks The sharpness of the peaks can be controlled using $\sigma$.}
\end{figure}

It is appropriate to present a synthetic example to illustrate the difference between autocorrelation and autocorrentropy in assessing similarity over time, and also to elucidate the role of the kernel size. Let us take the case of the stochastic process with uniform random amplitude in $[-A, A]$ and a random phase in $[-\pi,\pi]$ defined as $x_n= A \sin (w_0 n + \varphi)$. As it is well known, the autocorrelation function of sinewaves is a sinewave with the same period. But should it be a sinewave if we are interested in assessing the degree of similarity of the signal time structure? Since the sinewave is periodic, the similarity is maximum when the delay is exactly one period, but for intermediate shifts, the two functions are very dissimilar, and autocorrelation does not show this very clearly (and the similarity is not normalized nor always positive, hence the use of the correlation coefficient). Therefore, if we are seeking a discriminative measure of similarity, the autocorrelation function is not exploiting optimally the information available in the statistics of the data.
% (if one recalls the probability density function of the sinewave which is far from Gaussian, this becomes obvious). 
It turns out that correntropy is more discriminative, as shown in Fig \ref{fig:sinewave}. The autocorrentropy of a sinewave (or any other periodic function) is a periodic pulse train defined by the data period, where the pulses can be made arbitrarily sharp by decreasing the kernel size to zero. This can be easily explained by observing Eq. \eqref{autocorr}. When $x_n$ and $x_{n-m}$ are similar the argument is close to zero and the Gaussian yields a value close to the argument square; when the difference increases, the Gaussian function produces exponentially smaller results proportional to the difference in arguments %around the value $(x_n-x_{n-m})^2/\sigma^2 = 1$ 
; and for larger differences, the Gaussian gives back very small values close to zero %effectively exponentially decreasing the differences 
(see Fig \ref{fig-normV}). Of course if white noise is added to the sinewave, one immediately sees that the kernel size can not be made arbitrarily small, otherwise the correntropy becomes always very small, not capturing the periodic nature of the noisy signal. But if the kernel size needs to be made very large to accommodate large noises, then the autocorrentropy approaches the autocorrelation function.

\subsection{Periodic kernel} \label{sec:pkernel}

With this introduction in mind, we move on  specifying the kernel that best encapsulates the information in the data for periodic signals.
Periodic kernel functions are known to be appropriate for  nonparametric estimation, modelling and regression of periodic time series \citep{Michalak2010}. A kernel function is periodic with period $P$ if it repeats itself for inputs  separated by $P$.  Periodic kernel functions have also been proposed in the Gaussian processes literature \citep{Rasmussen2006,Mackay1998,Wang2012}.

A periodic kernel function can be obtained by applying a nonlinear mapping (or warping) $u(t)$ to the input vector $t$. In \citet{Mackay1998} a periodic kernel function was constructed by mapping a unidimensional input variable $t$ using a periodic two-dimensional warping function defined as
\begin{equation} \label{warp} \nonumber
u_f \left(t \right) = \left (\cos \left( 2\pi f t\right) , \sin \left( 2\pi f t\right) \right).
\end{equation}
The periodic kernel function $G_{\sigma}^{P}(f,t_z-t_y)$ with period $1/f$, is obtained by applying theis warping function to the inputs of the Gaussian kernel function (Eq. \ref{Gausskernel}). The periodic kernel function is defined as:

\begin{eqnarray}  \label{periodickernel}
G_{\sigma}^{P}(f,t_z-t_y) &=& G_{\sigma}(u_f(t_z)-u_f(t_y)) \\ \nonumber
	&=& \frac{1}{\sqrt{2\pi}\sigma} \exp \left ( - \frac{2\sin^2 \left( \pi f  (t_z - t_y) \right)}{\sigma^2}\right ),
\end{eqnarray}
where the following expression is  used
\[
\left \|u_f \left( z \right) - u_f \left( y\right)   \right \|^2 = 4 \sin^2 \left( \pi f (z - y) \right).
\]
Note that the periodic kernel is a function $\delta t=(t_z-t_y)$ and frequency, the inverse of the period. The Taylor series expansion at $\delta t=0$ of Eq. \eqref{periodickernel} is defined as 

\begin{eqnarray} \label{taylor}
&&G_{\sigma}^{P}(f,\delta t) = \lim_{N \rightarrow \infty} \\ \nonumber 
  & &\sum_{k=0}^{N}{ \frac{(-1)^k}{k! ~\sigma_t^{2k} ~2^{k-1}} \left[ \sum_{m=0}^{k}{\binom{2k}{k-m} (-1)^m g_m \cos(2\pi m f \delta t) }\right] },
\end{eqnarray}
where
\[ 
	g_m=
	\begin{cases}
	1/2, & \text{if $m=0$}. \\
	1, & \text{otherwise}.
	\end{cases}
\]
Note that for large values of $\sigma$, only the first terms contribute to the sum and thus the periodic kernel  tends to a constant plus $cos(2 \pi f \delta t)$, which corresponds to the real part of the Fourier basis.

\section{Method} \label{sec:method}

We base our methodology on the work described in \citet{Huijse2012}. In this section we %firstly 
summarize the key points from that work,% which are based on the overall description given in the previous sections and 
then introduce the new concepts, particularly an intuitive interpretation of the parameters of the CKP, simple rules to select these parameters and a normalization term that is needed to perform ensemble comparisons. % THIS IS NEW

The correntropy kernelized periodogram (CKP) used in \citet{Huijse2012} is a %metric for 
period detection function developed for unevenly sampled time series. 
The CKP is computed from the available samples following a direct quadratic estimator approach as proposed in \citet{Marquardt1984}\footnote{The basic idea is that for uneven samples, one can calculate the periodogram without having to regularize the data.}.
%The correntropy kernelized periodogram (CKP) used in  \citep{Huijse2012} is a metric for period detection developed for unevenly sampled time series. The CKP is a function of the trial period and is computed directly from the available samples and does not require  any resampling or folding scheme. 
%[PP: THIS IS A PRODUCT OF KERNELS BUT SOMEHOW RELATED TO THE FT OF THE CORRENTROPY]
For a discrete unidimensional random process $\{x_n,n=1,\ldots,N\}$ with kernel sizes $\sigma_t$ and $\sigma_y$, and a period $1/f$, the CKP is computed as:
\begin{eqnarray}   \label{HP}
&& \text{CKP}_{\{\sigma_t,\sigma_y\}}(f) = \\ 
	&& \frac{1}{N^2}  \sum_{i=1}^N \sum_{j=1}^N  \left( G_{\sigma_y}(\Delta y_{ij}) -IP_{\sigma_y} \right)  G_{\sigma_t}^P(f,\Delta t_{ij}) , \nonumber
\end{eqnarray}
where $\Delta y_{ij} = y_i - y_j$, $\Delta t_{ij} = t_i - t_j$,  $G_{\sigma_y}(\cdot)$ is the Gaussian kernel function (Eq. \ref{Gausskernel}), $G_{\sigma_t}^P(\cdot,\cdot)$ is the periodic kernel function (Eq. \ref{periodickernel}), and $IP_{\sigma_y}$ is the information potential
\begin{equation} \label{IP}
	IP_{\sigma_y}= \frac{1}{N^2} \sum_{i=1}^{N}{\sum_{j=1}^{N}{ G_{\sigma_y}(\Delta y_{ij})  } }.
\end{equation}

Note that Eq. \eqref{HP} is similar to the CSD (Eq. \ref{CSD}) with two main differences: a) the CKP is estimated in a direct approach and b) 
the basis functions, $\exp \left(- i \,2\pi f m/F_s \right)$ have been replaced by the periodic kernel (Eq. \ref{periodickernel}). In this sense the CKP can be interpreted as the result of transforming the autocorrentropy function through a basis defined by the periodic kernel. 

By comparing magnitude values through the autocorrentropy function, the CKP is effectively using a CIM (Eq. \ref{CIM}) metric to measure magnitude distances. The kernel size $\sigma_y$ has influence in the assessment of magnitude similarities as explained in the previous section. The CKP compares time differences with the trial period through the periodic kernel. % This is similar as how the Fourier basis compares the time lags with a given frequency. 
The periodic kernel size $\sigma_t$ allows the user to choose how this comparison is made. 

%[PH:] Can I add the proof that shows that the periodic kernel tends to the real part of the fourier transform when ot is large ?
%[PP: ADD IT AND I WILL SUMMARIZE IT]
% PH: I put the proof at the end of the periodic kernel section, I made a conection to the proof in the discussion about sigma_t in the following page

By summing in the time and magnitude index, a function of the trial period is obtained, thus the CKP can be considered a generalized periodogram. %If all the magnitude values separated by a given trial period $1/f$ are not very different, it will be reflected as a peak in the CKP located at $f$. 
%The two components of the CKP address two different aspects of the periodic pattern of a lightcurve. The Gaussian kernel assess similarities between two magnitudes, and the periodic kernel  to select the time instants separated by a given trial period. In other words two points of a lightcurve will contribute heavily if the two points are separated by the trial period and their $y$'s values are not very different. 
%The Gaussian kernel makes the CKP robust to outliers. 
%The summation of all those values from all possible pairs is the measure of how periodic a lightcurve is. This estimation is related to the generalized entropy as shown in \cite{} and therefore is also known as the information potential. 
%Consequently, in order to detect periods in lightcurves the CKP is maximized over the  parameters of the model, namely the two kernel bandwidths ($\sigma_y, \sigma_t$) and the period. 
%PH: Reviewer 2 is right, it is not maximized over the parameters
Consequently, in order to detect periods in lightcurves the CKP is maximized over the frequency (inverse of the period) for a given combination of parameters, namely the two kernel bandwidths ($\sigma_y, \sigma_t$). 
 %In \citep{Huijse2012} the CKP was tested in set of lightcurves drawn from the MACHO survey, whose period were known a priori. 

One of the major advantages of the CKP over conventional methods is its adaptability given by the kernel parameters. In what follows, we describe heuristic approaches that use the available information on the lightcurve to set the kernel sizes. %These approximations are not included in the previous work but they are necessary when dealing with large datasets. 
Without them the maximization of the CKP would have been a very expensive procedure. 

\begin{figure*}
	\centering	
	\subfloat[]{\label{fig-exampleKSa}\includegraphics[scale=0.7]{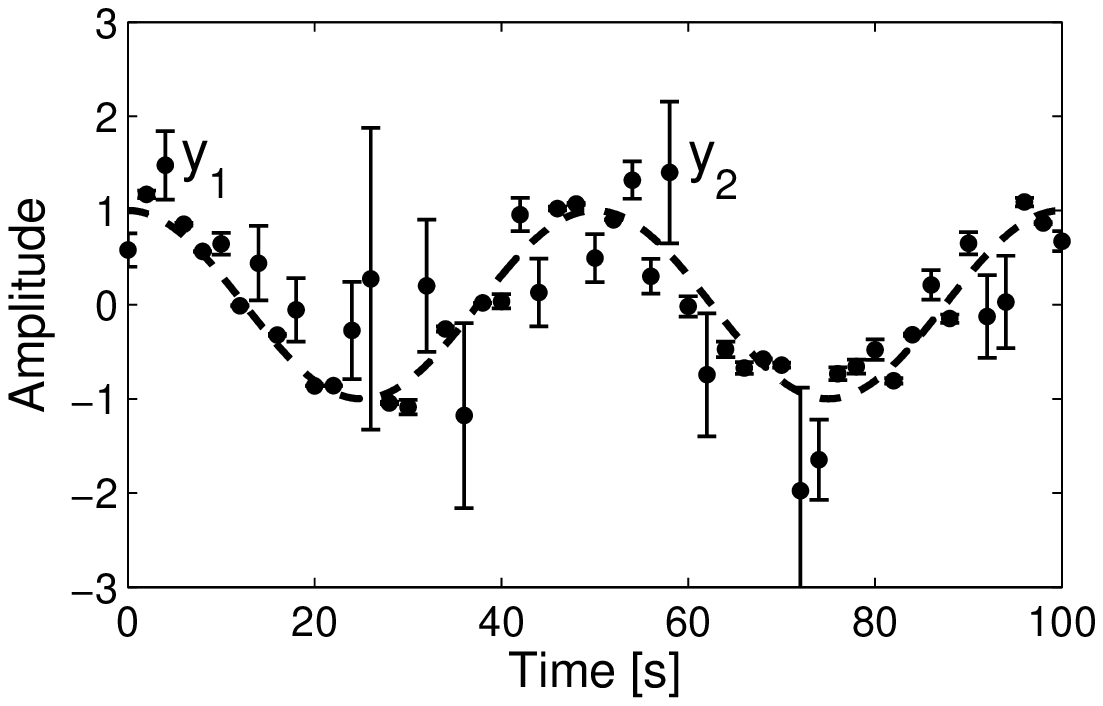} } 
	\subfloat[]{\label{fig-exampleKSb}\includegraphics[scale=0.7]{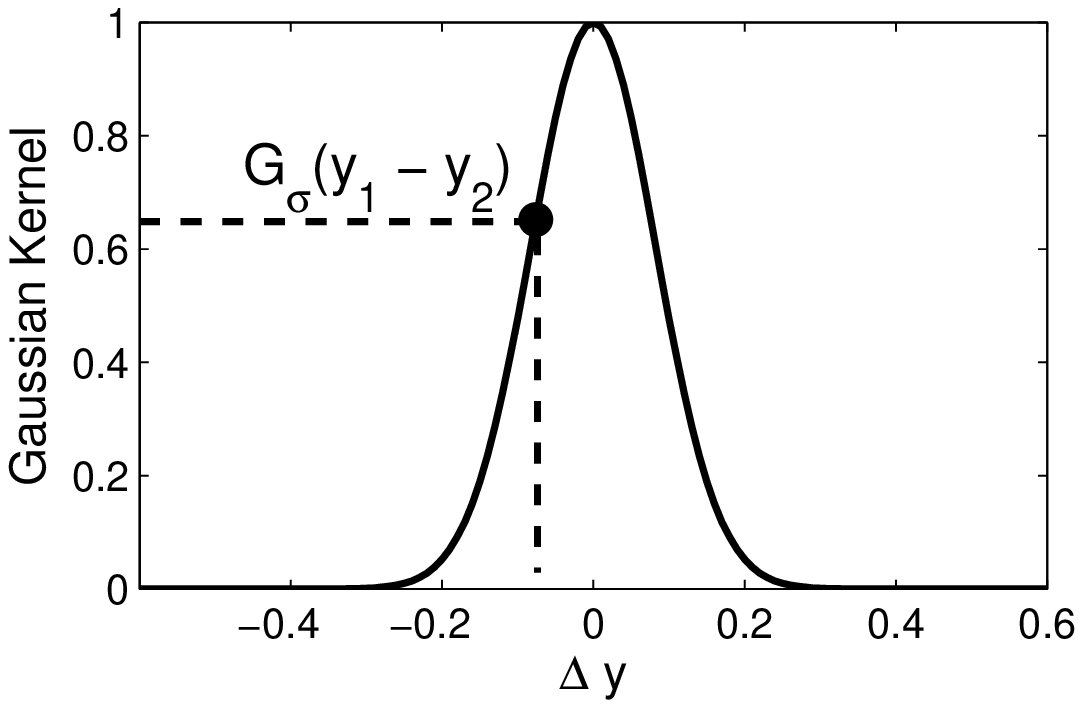} }  
	\caption{ (a) Periodic synthetic time series $\cos(2\pi f t) + N(0,0.5)$, the dotted line corresponds corresponds to the underlying signal. In this example the median of the errorbars is 0.08. Samples $y_1$ and $y_2$ are compared using the Gaussian kernel (b). The kernel size is set to 0.08. }
\end{figure*}

\noindent$\star$ The kernel bandwidth, $\sigma_y$, controls the observation window that is used to compare the magnitude values of the lightcurve. This parameter needs to be set small enough so that outliers are filtered, but large enough to compensate for the observational and other measurements errors. Conveniently those errors are usually available for most measurements in lightcurves (these are the magnitude errors). 
 For a given lightcurve the Gaussian kernel bandwidth is selected as
\begin{equation} \label{pavlosrule}
	\sigma_y = \text{med}(\{e\}),
\end{equation}
where $\text{med}$ is the median, and $\{e\}$ are the error bars of the measurements in lightcurve. Fig. \ref{fig-exampleKSa} shows a synthetic periodic lightcurve with random error bars. Samples $y_1$ and $y_2$ are compared using the Gaussian kernel, where the median of the error bars is $0.08$ and the $\sigma_y$ is set to be 0.08. Fig.~\ref{fig-exampleKSb} shows the equivalent Gaussian kernel value for this pair. 
In reality the observational errors are not constant and therefore eq. \ref{pavlosrule} should not be the same for all pairs and  should be a combination of the two observational errors added in quadrature. Practically%, though the variations of the error bars in real dataset are not that significant and therefore 
the difference of this approximation and the correct approach is  insignificant. 

\noindent$\star$ The kernel bandwidth, $\sigma_t$, controls the observation window that is used to compare the time differences of the lightcurve with the trial period. When $\sigma_t\to 0$  only the samples whose time differences are equal to the trial period will be picked by the periodic kernel. The smaller the $\sigma_t$ is, the more precise the estimation will be, although in practice fewer samples will be available. When $\sigma_t$ grows large,  the exponential in Eq. \eqref{periodickernel} takes less relevance and the periodic kernel tends to a sinusoidal function\footnote{As shown in Section \ref{sec:pkernel} through the Taylor expansion of Eq. \eqref{periodickernel}.}. Intuitively, this parameter has influence on the periodicity's shape. A smaller $\sigma_t$ is beneficial to pick up shapes that have many features or abrupt changes, such as the narrow eclipses of an Algol-type eclipsing binary. On the contrary a large $\sigma_t$ is used for smoother shapes, \emph{i.e.} wiggles and high derivatives are ignored. In summary the $\sigma_t$ needs to be set small enough so that the features of the periodicity will not be missed, but large enough so that there will be enough samples representing the period and to avoid picking up  structures due to the noise. 

\begin{figure}
	\centering
	\subfloat[]{\label{fig-exsigmat1}\includegraphics[scale=0.5]{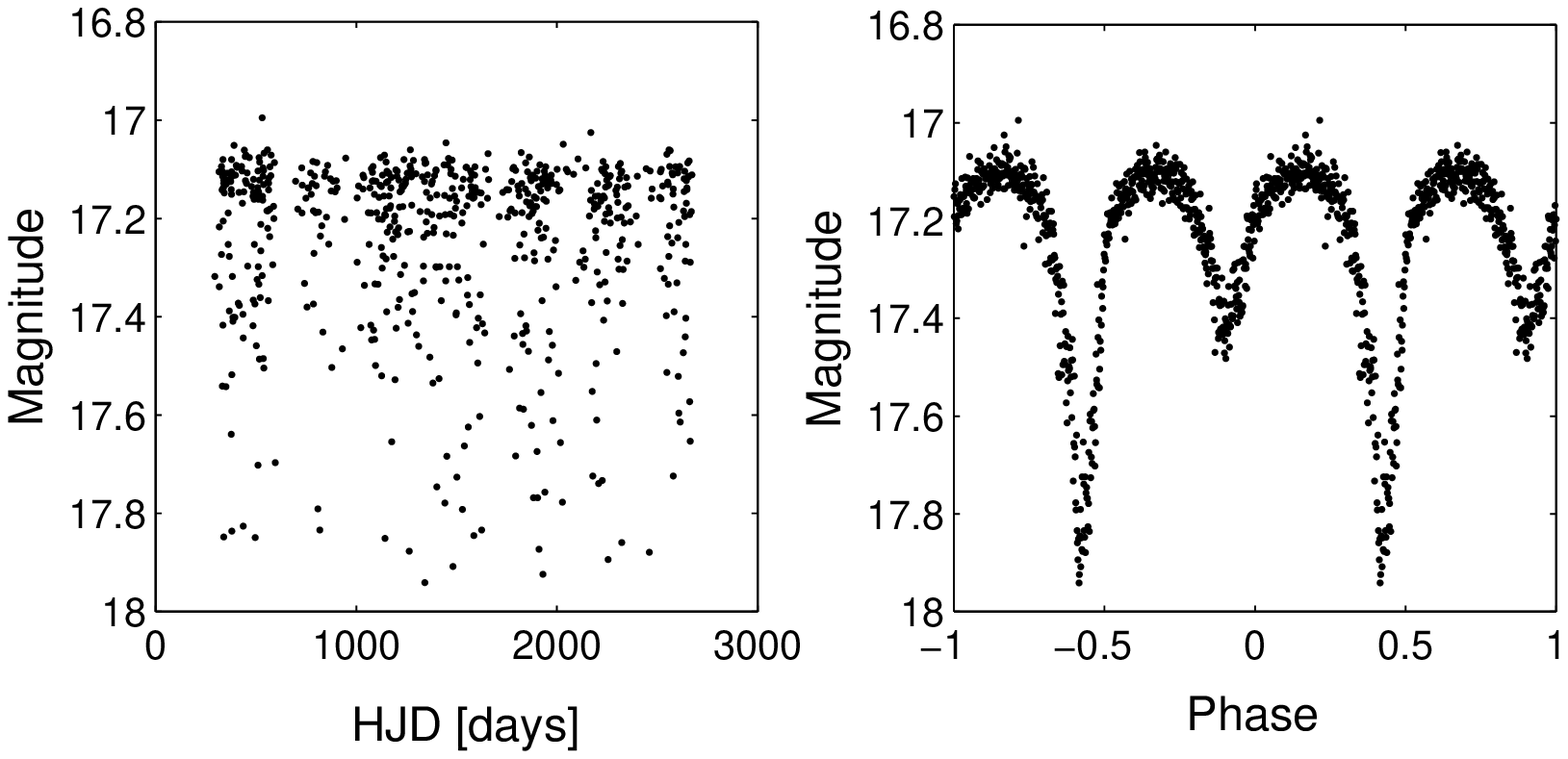} } \\ \vspace{-5mm}
	\subfloat[]{\label{fig-exsigmat2}\includegraphics[scale=0.5]{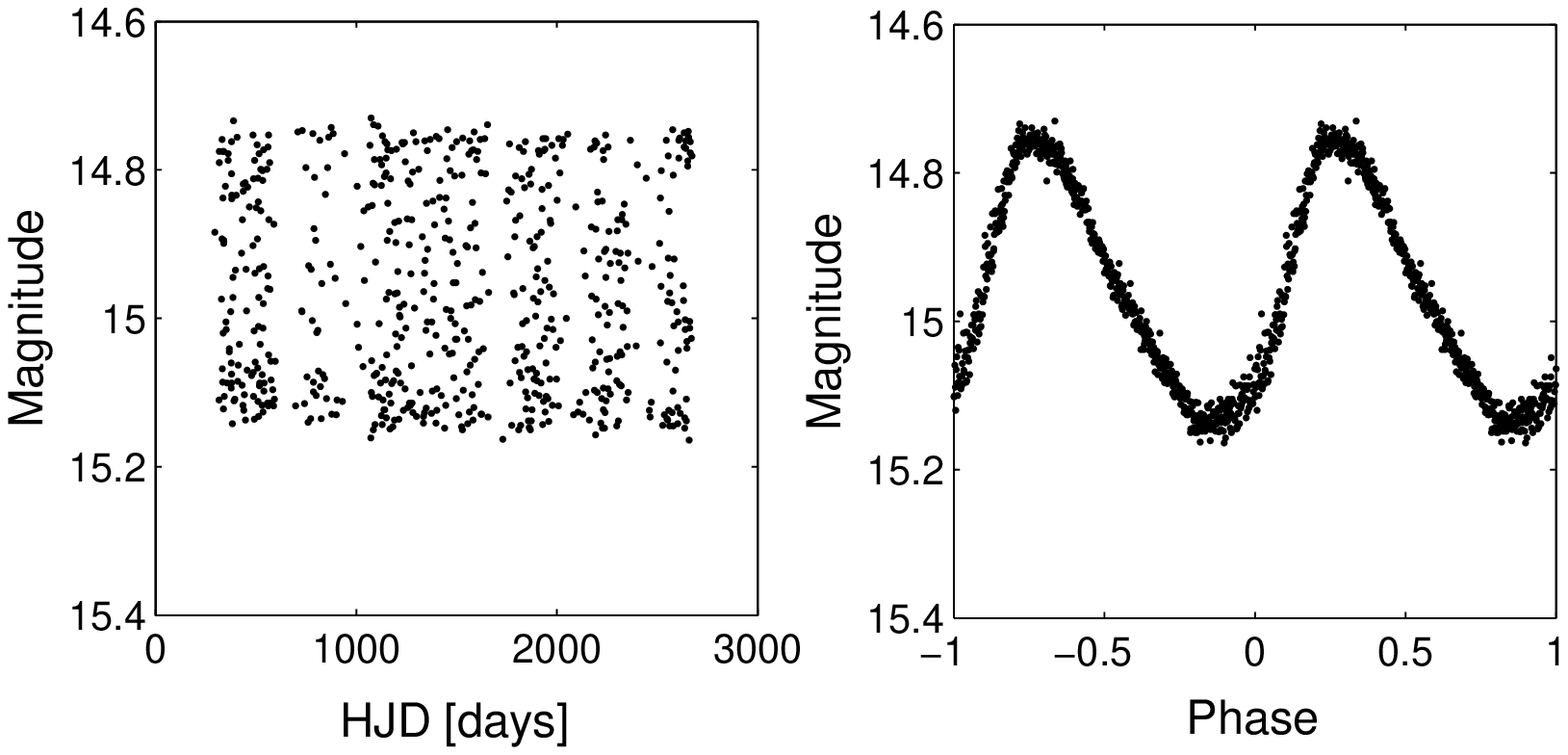} }  
	\caption{ (a) lightcurve lm0090l7821 folded with a period of 1.4255 days. This lightcurve has a highly positive skewed distribution. A time kernel bandwidth of $\sigma_t=0.115$ is selected for this lightcurve. (b) lightcurve lm0090n9337 folded with a period of 4.3949 days. This lightcurve has a symmetric distribution. In this case a time kernel bandwidth of 0.475 is selected.}
\end{figure}

\vspace{.5cm}
Since $\sigma_t$ describes the smoothness of the 
shape of the lightcurve, a way to estimate $\sigma_t$ is to find the variation of $\delta t$'s in a given 
y-band. %A method to do this was examined, but it was not robust against  low sampled lightcurves.
 Empirically, we observed that for almost all periodic lightcurves, the CKP is maximized 
for  $\sigma_t \sim 0.1 - 0.6$ and that the value of $\sigma_t$ is strongly correlated with
   the third moment or the skewness of the distribution of the magnitudes of the lightcurves. 
%Consequently we propose an alternative heuristic rule to set $\sigma_t$ that uses information of
%Because lightcurves with
%high skewness tend to be less smooth we use that information in the following way:
%For a given lightcurve the periodic kernel bandwidth is selected as
%\begin{equation} \label{skewnessrule}
%	\sigma_t = \lambda+ 0.5  \exp(-\mu ~ S(\{x\})^2), 
%\end{equation}
%where $\lambda,\mu$ to be determined by maximize the overall performance of the system.\footnote{For the EROS-2 dataset we found $\lambda = 0.1$ and $\mu =12$. [PP 
%darn how we explain this]}
%
%\noindent The magnitudes of the lightcurve are $\{x\}$, and $S(\{x\})$ is the quartile estimator of the skewness \citep{Bowley1920}
%\begin{equation} \label{skewness}
%	S(\{x\}) = \frac{Q_3(\{x\}) + Q_1(\{x\}) - 2 Q_2(\{x\}) }{ Q_3(\{x\}) - Q_1(\{x\}) },
%\end{equation}
%where $Q_1(\{x\})$, $Q_2(\{x\})$ and $Q_3(\{x\})$ are the first, second and third quartiles of $\{x\}$, respectively.\footnote{The second quartile is equivalent to the median. The first and third quartiles are the medians of the first half and second half of the rank-ordered magnitudes, respectively.} 
%Note that $S(\{x\}) \in [-1,1]$ and for a symmetric pdf $S(\{x\})=0$
%and therefore $\sigma \sim 0.6$ and when a lightcurve is very asymmetric $\sigma_t \sim 0.1$. 
Lightcurves with skewed distributions, such as those corresponding to eclipsing binaries (Fig. \ref{fig-exsigmat1}), get a small $\sigma_t$ value.
% when Eq. \eqref{skewnessrule} is used. 
On the other hand, lightcurves with very symmetric distributions (Fig. \ref{fig-exsigmat2}) will get a larger $\sigma_t$. 
Finally, we will address ensemble comparisons for period discrimination.
%In what follows we use the CKP for period estimation and periodicity discrimination for lightcurves for the EROS-2 dataset. To accomplish the task for periodicity discrimination, lightcurves are compared through their CKP values. 
The kernel sizes are selected for each lightcurve differently as described above % using Eq. \eqref{pavlosrule} and Eq. \eqref{skewnessrule}
and in order to compare different lightcurves, the CKP is required to be invariant under  $\sigma_y$, $\sigma_t$ and the sample size. 

\begin{figure}
	\centering
	\subfloat[]{\label{fig-exhistN1}\includegraphics[scale=0.55]{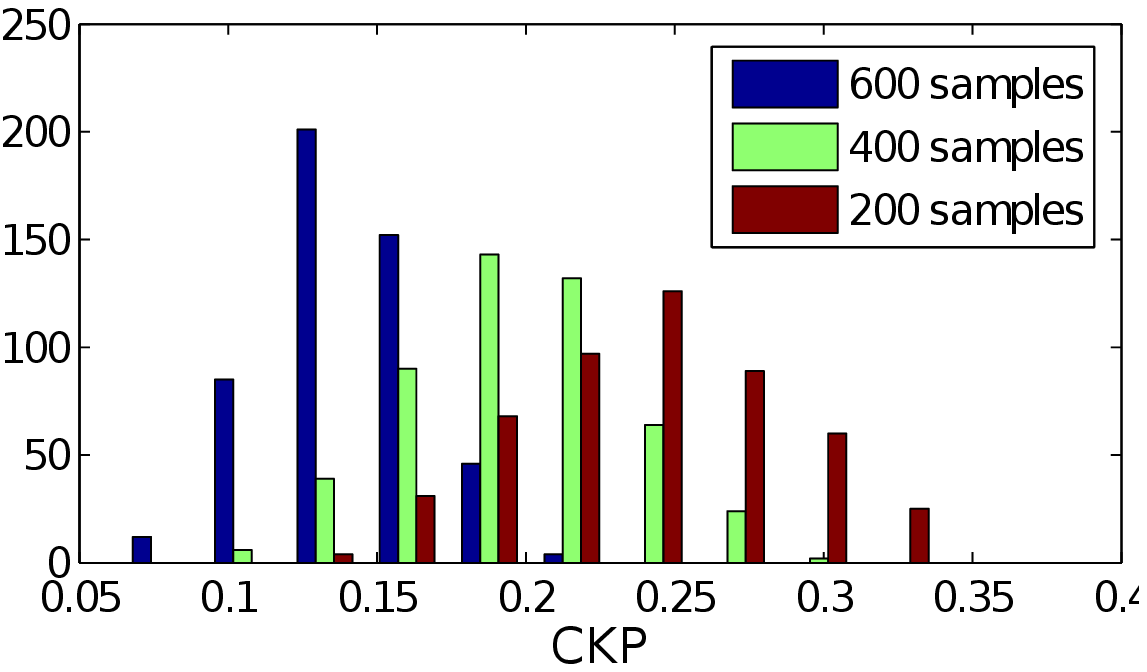} }  \hspace{2mm}
	\subfloat[]{\label{fig-exhistN2}\includegraphics[scale=0.55]{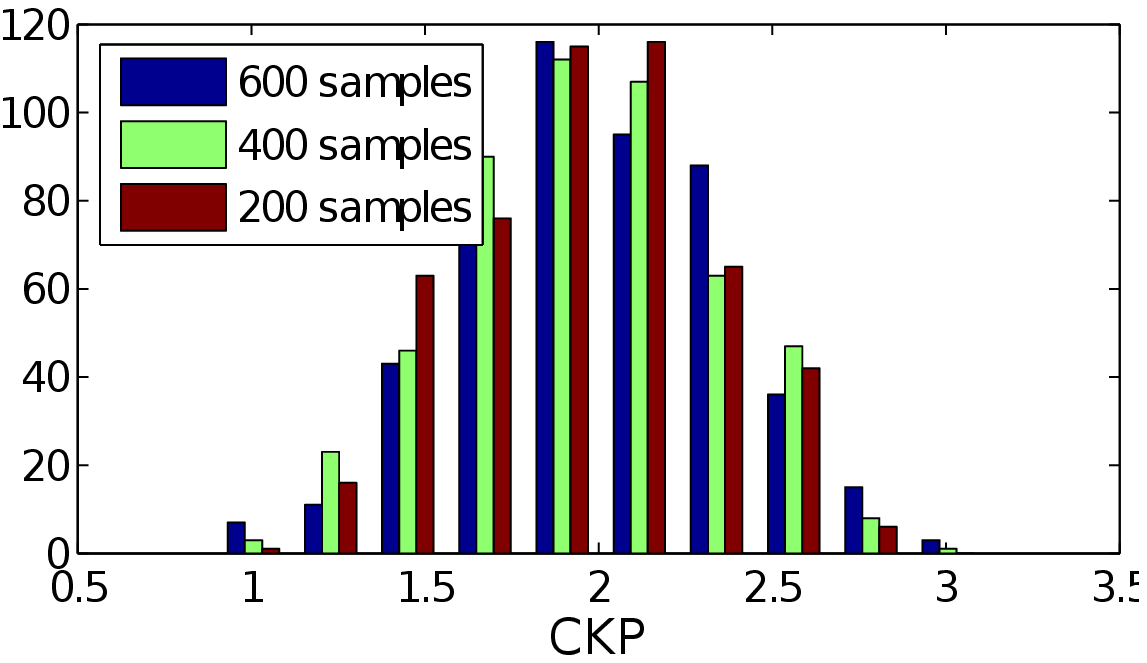} }  
	\caption{ Distribution of the maximum CKP values on a set of 1500 synthetic light curves. The light curves are generated with the same period and SNR but using different number of samples (N). Three sets of 500 light curves are generated using 200, 400 and 600 number of samples, respectively. Fig (a) shows the distribution of the unnormalized CKP. It is clear the CKP is not invariant to N. Light curves with higher N have higher CKP values. Fig (b) shows the distribution of the normalized CKP. }
\end{figure}

For that we propose a properly normalized CKP metric as:
\begin{eqnarray}   \label{CKP}
&& \text{nCKP}_{\{\sigma_t,\sigma_y\}}(f) =  \\ 
&& \frac{\sqrt{N \sigma_t}}{IP_{\sigma_y}} \frac{1}{N^2}  \sum_{i=1}^N \sum_{j=1}^N  \left( G_{\sigma_y}(\Delta y_{ij}) -IP_{\sigma_y} \right)  G_{\sigma_t}^P(f,\Delta t_{ij}), \nonumber
\end{eqnarray}
where $1/IP_{\sigma_y}$ normalizes against $\sigma_y$, $\sqrt{\sigma_t}$ normalizes against $\sigma_t$ and $\sqrt{N}$ normalizes against the number of samples. The normalization factors were confirmed empirically by comparing the distribution of the CKP across different sets of surrogate lightcurves, generated with the procedures described in Section \ref{sec:synthetic}. Fig. \ref{fig-exhistN1} shows a histogram of $ \max \text{CKP}_{\{\sigma_t,\sigma_y\}}(f)$ for three sets of surrogates generated with different $N$ values. In this figure the unnormalized CKP is used (Eq. \ref{HP}). For the histogram shown in Fig. \ref{fig-exhistN2} the normalized CKP (Eq. \ref{CKP}) is used, in this case the distribution of the CKP is equivalent, thus it is invariant to the different $N$ of the surrogates. 

\subsection{Trial period extraction, the bands method} \label{sec:bands} 

The parameter to be estimated by maximizing the CKP is the period. Unfortunately the dependence of CKP on period is not uniform and difficult to model \citep{Huijse2012}, therefore any clever optimization technique fails to converge faster than the brute force approach. 

To alleviate this problem, a fast search algorithm is adopted. The basic idea is that two points  in an ideal lightcurve having the same magnitude, have to be apart in time by an integer multiple of the period. For the ideal lightcurve case, finding the period is as simple as finding the greatest common divisor of the times of two points with the same magnitude\footnote{This is the famous Euclid algorithm (oldest known).}. 
However, the ideal case is not applicable to astronomical data because: a) lightcurves comprise of a nominal part and a signal part as in the case of planetary transits and eclipsing binaries, b) the observations are not performed continuously and c)  measurements are not perfect but suffer from observational errors. 

What follows, is an approximation %to the ideal case 
tailored for real lightcurves. %The idea is that 
Instead of looking at pairs of points with the same 
magnitude, %we select 
subsets of points with similar magnitudes are selected. These subsets, called bands, should contain points that have time differences that are multiples of the period, and therefore, in Fourier space these periods are enhanced. To avoid bands that the lightcurve is in its nominal state we select bands where the derivatives are higher. %A set of trial periods is extracted from the lightcurve. These trial periods are evaluated by the CKP metric in the next step of the pipeline. We refer to the trial period extraction method as the bands method. In what follows, the steps of the bands method are described.

\noindent The details of the method are as:\\
For an unidimensional time series $\{t_i,x_i\}$ with $i=1,\ldots,N$ 
\begin{compactitem}
\item Compute the first derivatives $d_i = \frac{x_{i+1} - x_{i}}{t_{i+1} - t_i}$.
\item Divide the ordinate axis in $10$ uneven-width bands, such that each band has a 10\% of the lightcurve samples.
\item Compute the sum of the first derivatives that belong to band-$j$ ($B_j$), $D_j = \sum_{i \in B_j} |d_i|$, with $j=1,\ldots,10$.
\item Sort the bands in descending order of $D_j$ and keep the first $N_b$ bands.
\item For each band compute the spectral window function \citep{Jenkins1968} on a linearly spaced frequency grid from 0.00125 1/days to 3 1/days (periods between 0.3 days and 800 days),
\begin{equation} \label{bands}
S_j(f) = \left| \sum_{i \in B_j}  \exp \left(\jmath 2 \pi f t_i \right) \right|^2
\end{equation}
\item Save the frequencies associated with the $N_{t}$ highest local maxima of $S_j(f)$. Periods that comply with $\|P-1\|< 1e-4$ are omitted \footnote{The one day pseudo sampling period is strongly represented in all the bands.}. This gives a total of $N_b \, N_{t}$ trial frequencies. 
%[PP: WHY P-1 ? ]
%PH: one day period, appears in all the bands with great strength due to the pseudo one day sampling
\end{compactitem}

The number of analyzed bands, $N_b$, and the amount of trial periods extracted per band, $N_{t}$, are user defined parameters, that represent a trade-off between efficiency and computational time. We expect to find  the correct period in the first sorted bands, however the true period may be captured by different bands although with different amplitudes, \emph{i.e} the rank of true period may vary across bands. For example the true period may be ranked $100^{th}$ in the first band and $10^{th}$ in the third band. Synthetic lightcurves (see Section \ref{sec:synthetic}) are analyzed with the period detection pipeline using different combinations of $N_t$ and $N_b$. 
%Computational time per lightcurve (average over 10,000 ligvurves) and hit rates are measured as a function of $N_t$ and $N_b$. Hit rate is the percentage of lightcurves in which the underlying period is chosen as the best trial period at the end of the pipeline.

\begin{figure}[h]
	\centering
	\includegraphics[scale=0.45]{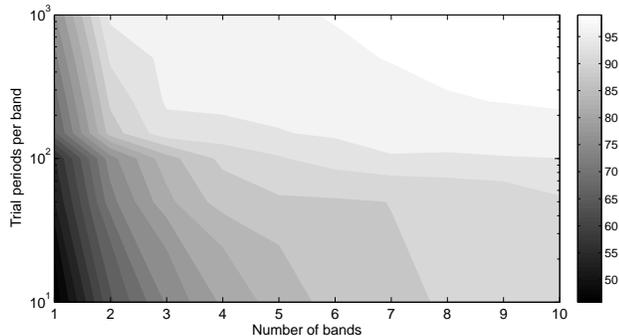} 
	\caption{Hit rate as a function of the parameters of the bands methods. These parameters are the number of bands $N_b$ and the number of trial periods extracted per band $N_t$.}
	\label{res:bands1}
\end{figure}

\begin{figure}[h]
	\centering
	\includegraphics[scale=0.45]{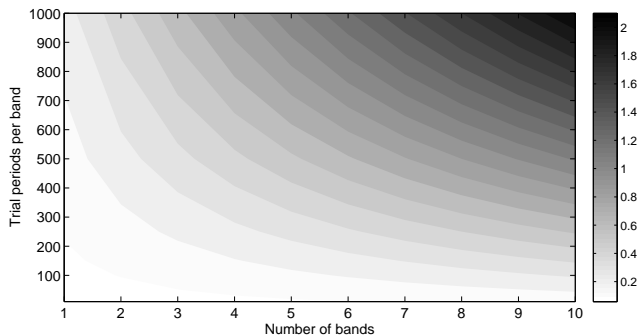} 
	\caption{Computational time in seconds required to process one lightcurve (600 samples) as a function of the parameters of the bands methods. These parameters are the number of bands $N_b$ and the number of trial periods extracted per band $N_t$.   }
	\label{res:bands2}
\end{figure}

Fig. \ref{res:bands1} shows a contour plot of the hit rate as a function of $N_b$ and $N_t$. As expected, hit rates increase with $N_b$ and $N_t$. For every $N_t$, the hit rate {\bf gain} obtained by adding additional bands decreases with $N_b$, which indicates that the bands are correctly sorted. Fig. \ref{res:bands2} shows a contour plot of the computational time required to analyze one lightcurve as a function of $N_b$ and $N_t$. For two points with equal $N_b N_t$ the point with lower $N_b$ requires less computational time. In terms of computational time, adding bands is less desirable than increasing $N_t$. 
The maximum hit rate achieved is 98.1\%. We find the best operation point to be $N_b = 3$ and $N_t = 150$, which yields a hit rate of 95.1\% with a computational time of 0.162s per lightcurve. This point represent the best compromise between efficiency and computational time and is found by maximizing $HR + 1/c_t$, where $c_t$ is the computational time.

\begin{figure}[h]
	\centering
	\subfloat[]{\label{fig-bands1a}\includegraphics[scale=0.6]{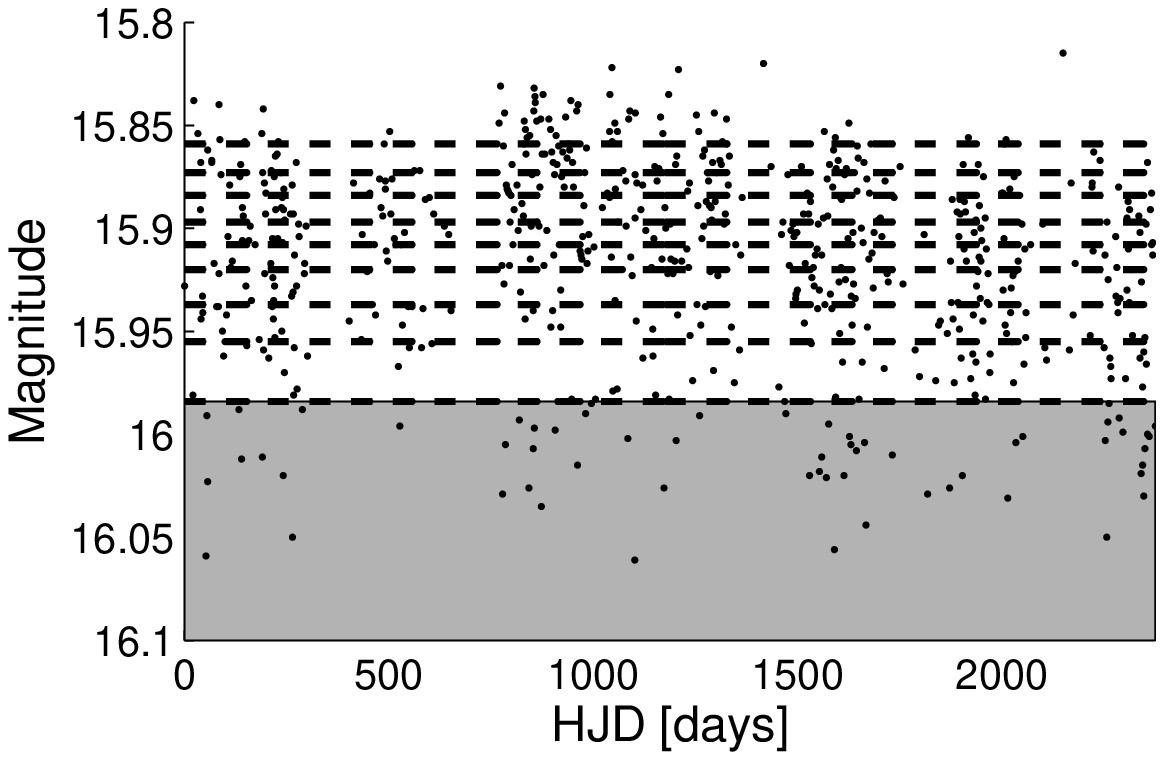} }  \qquad
	\subfloat[]{\label{fig-bands1b}\includegraphics[scale=0.6]{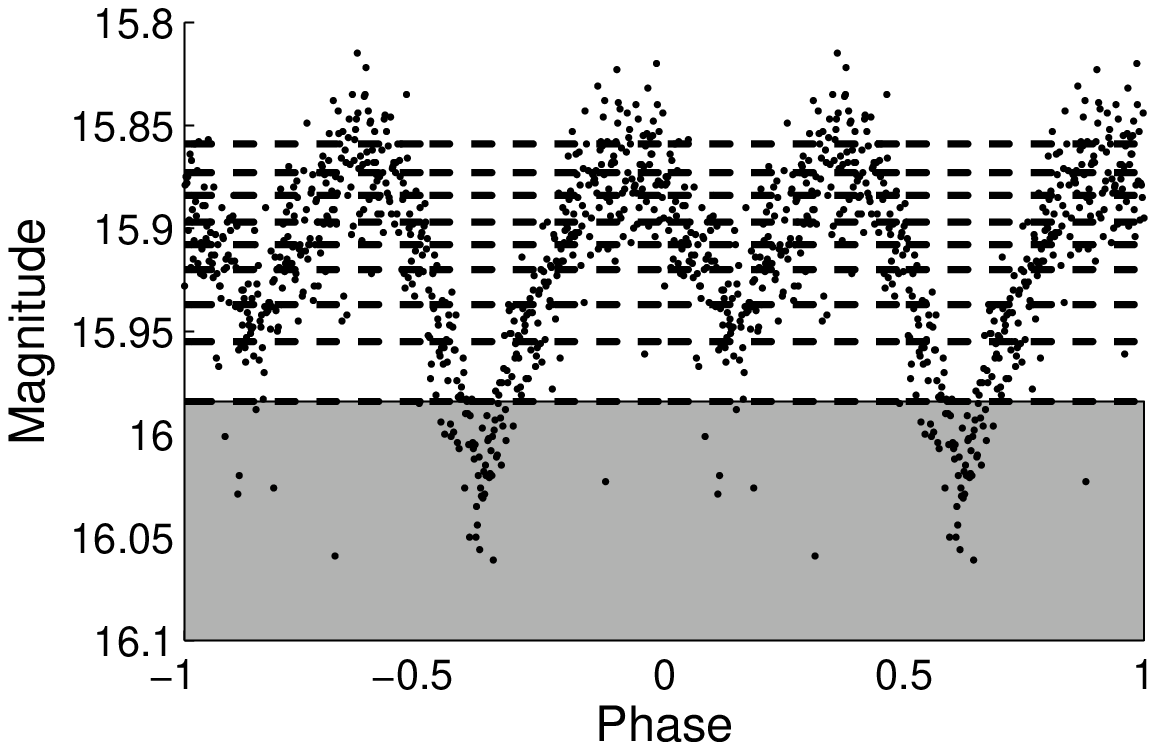} }  
	\caption{ (a) EROS-2 lightcurve lm0090m4818. The dotted lines show the band divisions. The shaded region shows the best band in terms of the first derivatives criterion. (b) Same lightcurve folded with a period of $1.54192$ days.} 
\end{figure}

\begin{figure}[h]
	\centering
	\includegraphics[scale=0.7]{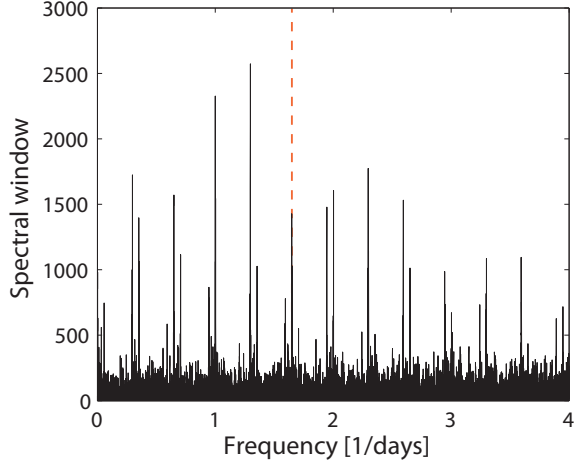} 
	\caption{ Spectral window of the tenth band from lightcurve lm0090m4818. The red dotted line shows the location of the underlying period ($1/P = 0.6485$).  The underlying period is associated to the  eighth highest local maximum of the spectrum. }
	\label{fig-bands2}
\end{figure}

Fig. \ref{fig-bands1a} shows a plot of an EROS-2 lightcurve, lm0090m4818. Fig. \ref{fig-bands1b} shows the same lightcurve folded with a period of 1.54192 days. The black dotted lines mark the band divisions on the magnitude axis. The shaded region shows the best band in terms of the first derivatives criterion. Fig. \ref{fig-bands2} shows a plot of the spectral window function of the time instants extracted from the best band of lm0090m4818. The true period of the lightcurve is associated with the eighth highest local maximum of the spectral window. In this case, if $N_t>8$ then the underlying period will be within the trial period set that is to be evaluated by the CKP in the next step of the pipeline.

\subsection{Performance criteria}
\label{sec:training}
The task of discriminating periodic lightcurves can be viewed as a binary classification problem where the classes are periodic (true) and non-periodic (false) lightcurves. In this case: true positives (TP) are the periodic lightcurves classified as periodic, false positive (FP) are the non-periodic lightcurves classified as periodic, true negative (TN) are the non-periodic lightcurves classified as non-periodic and false negative (FN) are the periodic lightcurves classified as non-periodic.

\noindent To evaluate the performance of our method we use the definitions of recall, $r$,  precision $p$ 

\begin{equation} \label{recall}
	r = \frac{\text{TP}}{\text{TP}+\text{FN}}, \hspace{1cm} p = \frac{\text{TP}}{\text{TP}+\text{FP}}
\end{equation}

\noindent and  F-score
\begin{equation} \label{fbeta}
	F_\beta = \frac{(1+\beta^2) \,p \, r}{\beta p + r}.
\end{equation}

The denominator of $r$ in  Eq. \eqref{recall} corresponds to the number of periodic lightcurves in the dataset. Recall,  is the ratio of recovered periodic lightcurves over the total number of periodic lightcurves in the dataset. The denominator of $p$ in Eq. \eqref{recall} corresponds to the number of lightcurves that are classified as periodic. Precision or completeness,  is the ratio of recovered periodic lightcurves over the total amount of lightcurves that are classified as periodic. The F-score (Eq. \ref{fbeta}) is a weighted average of recall and precision. The parameter $\beta$ controls the importance of recall over precision on the weighted average. In what follows we use the $F_1$ score ($\beta=1$).

We also define hit rate as:

\begin{equation} \label{hr}
	HR = \frac{\text{TP}^*}{\text{TP}^*+\text{FN}},
\end{equation}

\noindent where $\text{TP}^*$ are the periodic lightcurves classified as periodic and at the same time the true period is recovered\footnote{Note that a light curve can be classified as periodic even if the true period is not recovered, such as when a multiple of the true period is found. }.

\section{Synthetic lightcurves} \label{sec:synthetic}
In order to evaluate the actual efficiency of the system and determine the true number of periodics in our dataset, we build a synthetic set containing both non-periodic and periodic lightcurves. 

\vspace{0.5cm}
\noindent \underline{Periodic set:} 
The periodic synthetic lightcurves are generated using a multivariate Gaussian generative model with a covariance matrix similar to the periodic kernel in eq. \ref{periodickernel}. To generate a periodic synthetic lightcurve, with period $P$, signal-to-noise ratio $S$, and smoothness $\sigma$ we
follow the procedure below. 

\begin{enumerate}
\item Randomly select a lightcurve from the database and extract its time instants $\{t_i\}$ and error bars $\{e_i\}$. This defines the number of samples, $N$, of the generated lightcurve.
\item Use the time instants $\{t_i\}$, period $P$, smoothness $\sigma$ and generate an $N\times N$ covariance matrix as,
\[
	\Sigma_1(i,j) = \frac{1}{\sqrt{2\pi} \sigma } \exp \left(- \frac{2\sin^2 \left( \pi (t_i - t_j)/P\right)}{\sigma^2} \right).
\]
\item Generate a random periodic vector, $Y_s$, of length $N$ using a multivariate normal random generator with $N\times1$ zero mean vector and $\Sigma_1$  covariance matrix. 
\item Use the error bars to generate a $N \times N$ diagonal covariance matrix with diagonal elements, 
\[
	\Sigma_2(i,i) = e_i^2
\]
\item Generate a random noise vector $Y_n$ of length $N$ using a multivariate normal random generator with a $N \times 1$ zero mean vector and $\Sigma_2$  covariance matrix. 

\item The synthetic lightcurve $Y$ is obtained by summing the noise vector and the signal vector as follows
\begin{equation}
	Y =  S\frac{\text{med}({e_i})}{0.7413 ~\text{iqr}(Y_s)} Y_s + Y_n, 
\end{equation}
where $S$ is the desired signal-to-noise ratio, med is the median function and iqr is the interquartile range. 
Note that the resulting lightcurve has signal-to-noise ratio $S$ by construction.
\end{enumerate}

%The time instants and error bars for the synthetic lightcurves are obtained from randomly selected EROS-2 lightcurves.

For our purpose we generated a set of 10,000 synthetic periodic lightcurves, using 
the following parameter ranges,
\begin{compactitem}
\item Ten linearly spaced values for $\sigma$ in the range $[0.1,0.6]$.
\item Twenty logarithmically spaced values for $P$ in the range $[0.4,1000]$ days.
\item Ten values for $S$ extracted from the distribution of the signal-to-noise ratio of EROS-2 lightcurves. 
\end{compactitem}
Five synthetic lightcurves are generated for each combination of $S$, $P$ and $\sigma$. %The time instants and the length of the synthetic lightcurves are obtained directly from randomly selected EROS-2 lightcurves.

\begin{figure}[h]
	\centering
	\subfloat[]{\label{fig:synthp1}\includegraphics[scale=0.47]{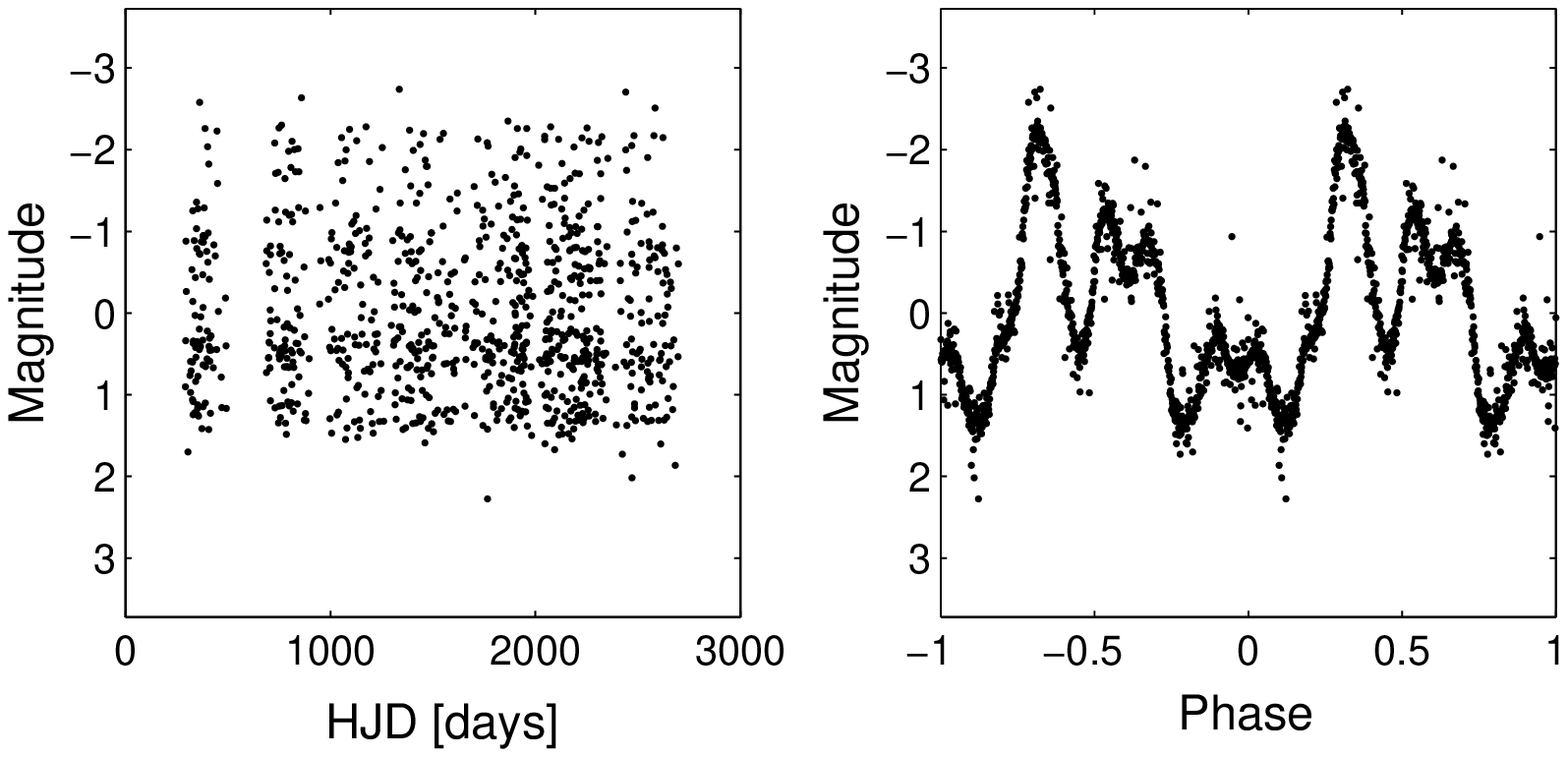} } \\
	\subfloat[]{\label{fig:synthp2}\includegraphics[scale=0.47]{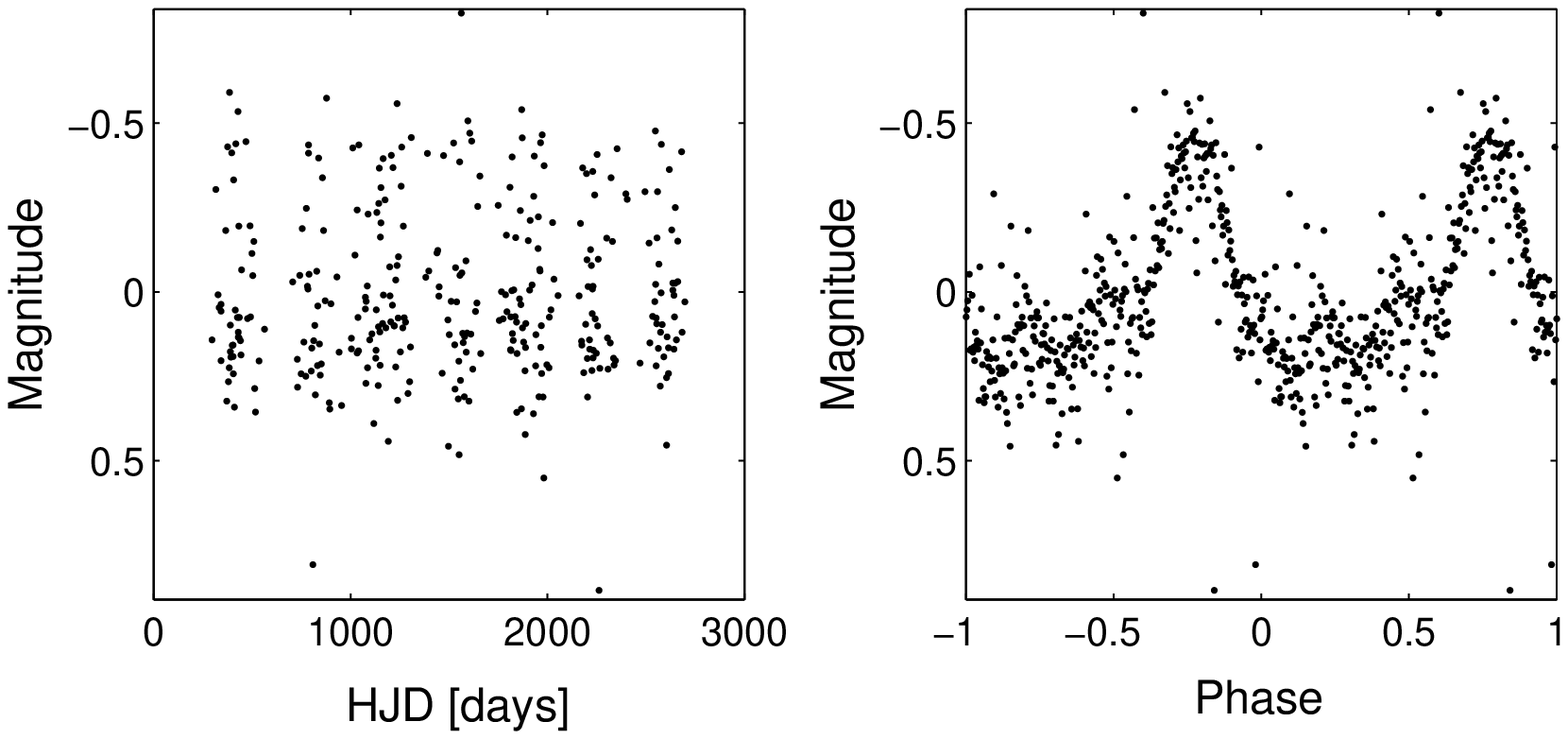} } \\
	\subfloat[]{\label{fig:synthp3}\includegraphics[scale=0.47]{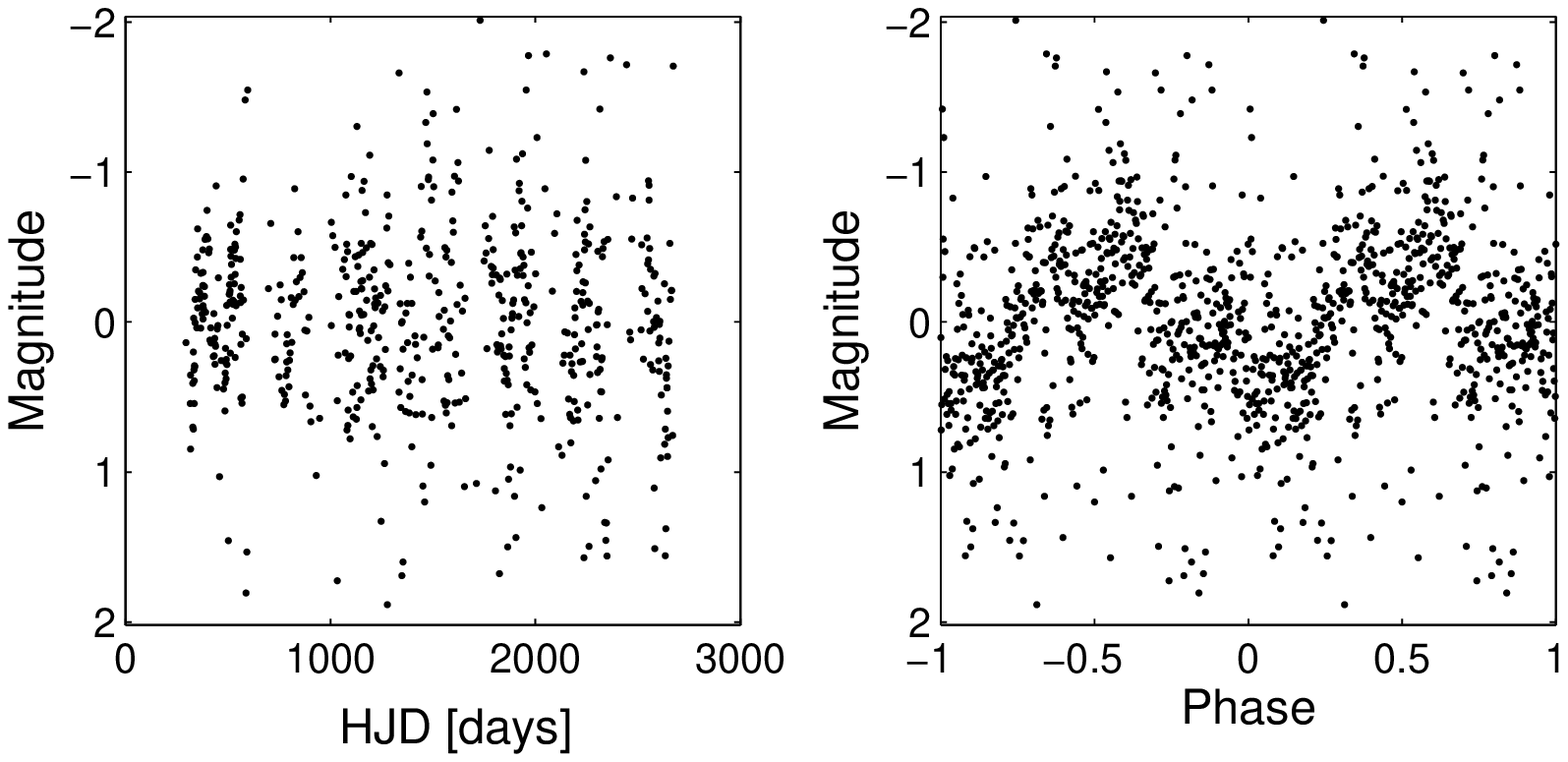}}  
	\caption{ 	\label{fig:synthp} Example of a synthetic periodic lightcurves. (a) shows a lightcurve created using P= 2.432d, $\sigma_t$ = 0.2, SNR =10 and N= 642. (b) shows a lightcurve created using P= 10.24d, $\sigma_t$ =0.5, SNR =4 and  N= 342. (c) shows a lightcurve created using P= 154d, $\sigma_t$ =0.4, SNR =2 and  N= 932.} 
\end{figure}

We present examples of the synthetic lightcurves generated using this procedure in Fig \ref{fig:synthp}.  Fig \ref{fig:synthp1} shows a synthetic lightcurve with a period of $2.432$ days, a smoothness value of $0.2$ and a SNR of $10$. Using a low smoothness value yields a shape with many features. Due to the high SNR the periodicity is very clear. Fig \ref{fig:synthp2} shows a synthetic lightcurve with a period of $10.42$ days, smoothness of $0.5$ and SNR of $4$. In this case, a higher $\sigma$ value yields a smoother shape as seen in the folded lightcurve. Fig \ref{fig:synthp3} shows a synthetic lightcurve with a period of $154$ days, smoothness of $0.4$ and SNR of $2$. 

\vspace{0.2cm}
\noindent \underline{Non-periodic set:} 
The non-periodic synthetic lightcurves are generated using block-bootstrap surrogates \citep{Schmitz1999,Schreiber1999,Buhlmann99}.
The procedure to generate a non-periodic synthetic lightcurve  is as follows

\begin{enumerate}
\item Randomly select a lightcurve and extract its time instants $\{t_i\}$ and error bars $\{e_i\}$. This defines the number of samples $N$ of the generated lightcurve.
\item Compute slotted autocorrelation function (ACF) \citep{Edelson1988} of the lightcurve. 
\item Find the time lag associated to the ACF value of $\exp(-1)$, this time lag is used as the block length (BL) for the block bootstrap method below.
\item Until at least N magnitude values have been created, do
\begin{enumerate}
\item Randomly select the block starting point $i_s$, such that $i_s \in [1,N-N')$. Find $N'$ as the last lightcurve sample that complies with
\[
t(N) - t(N') > BL
\]

\item Find the end point of the block $i_e$ as the first time instant that complies with
\[ 
t(i_e+1) -t(i_s) > BL
\]
\item Grab the time instants, magnitudes, and error bars of the original lightcurve segment in $[i_s,i_e+1]$.
\item Subtract the initial time $t_{i_s}$ to the selected time instants. After this the block starts at zero days.
\item Add the time from the previous block $t_{PB}$ to the selected time instants ($t_{PB} =0$ for the first block). After this the block starts where the last block ended.
\item Update $t_{PB} = t(i_e+1)$. Delete the time instant, magnitude and error bar of sample $i_e+1$ from the block. 
\item Add the newly constructed block to the surrogate. 
\end{enumerate}
\end{enumerate}

\begin{figure}
	\centering	
	\subfloat[]{\label{fig:surrogate1}\includegraphics[scale=0.53]{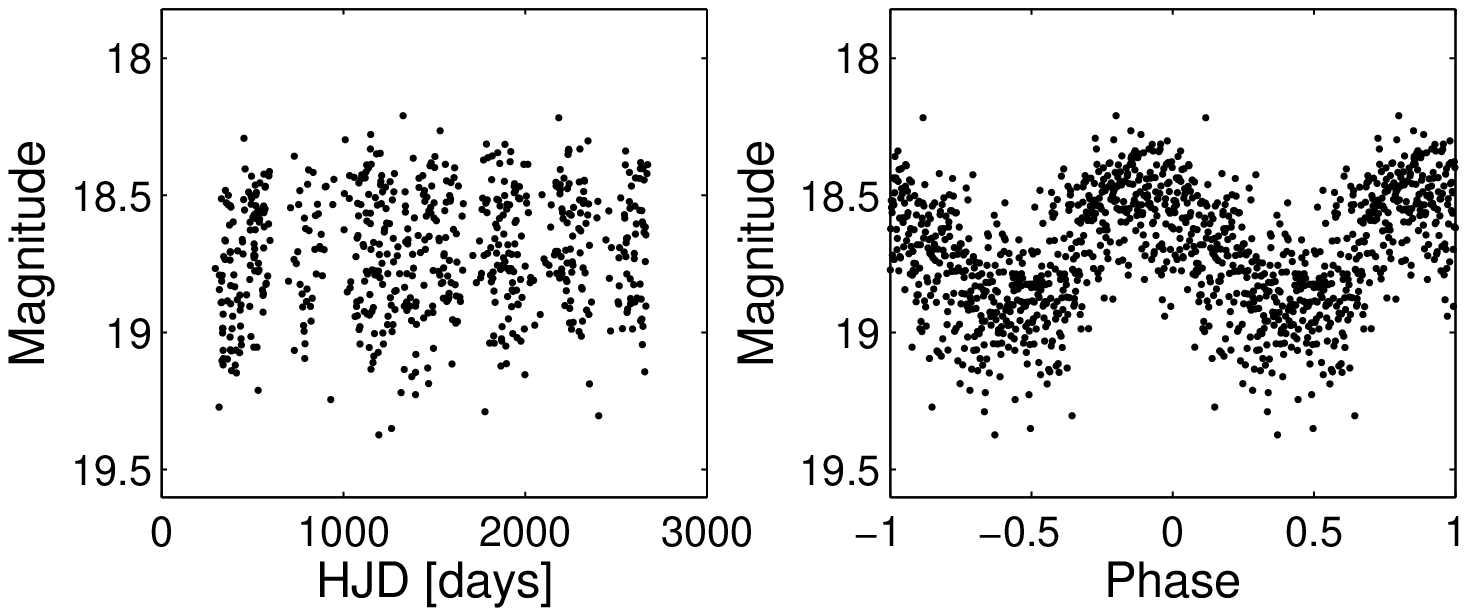} } \\
	\subfloat[]{\label{fig:surrogate2}\includegraphics[scale=0.47]{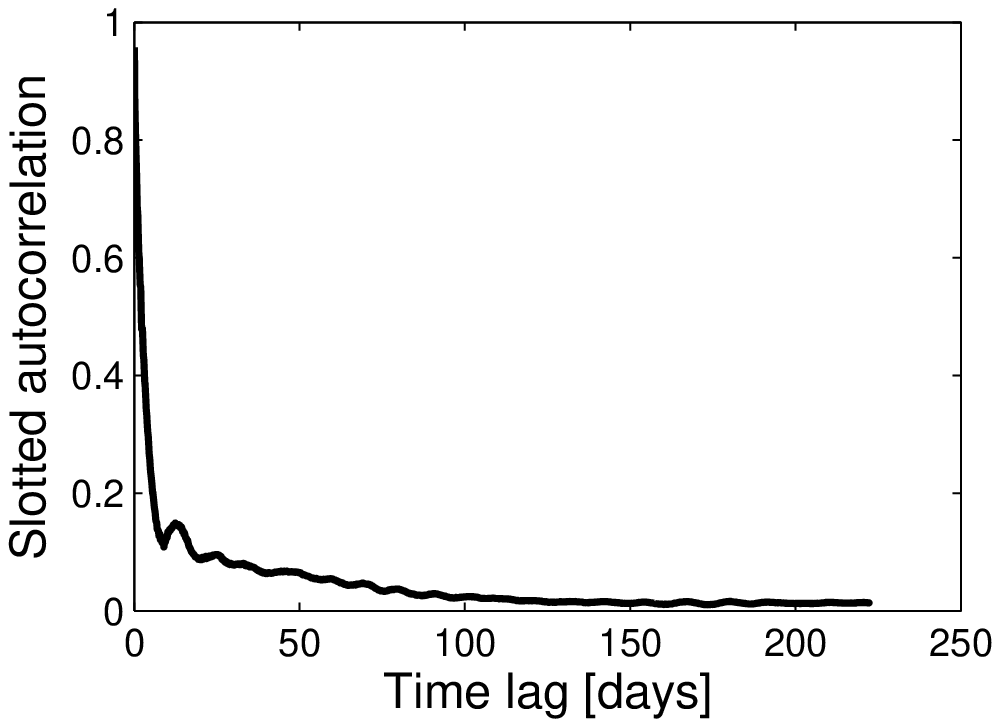}  } \\
	\subfloat[]{\label{fig:surrogate3}\includegraphics[scale=0.53]{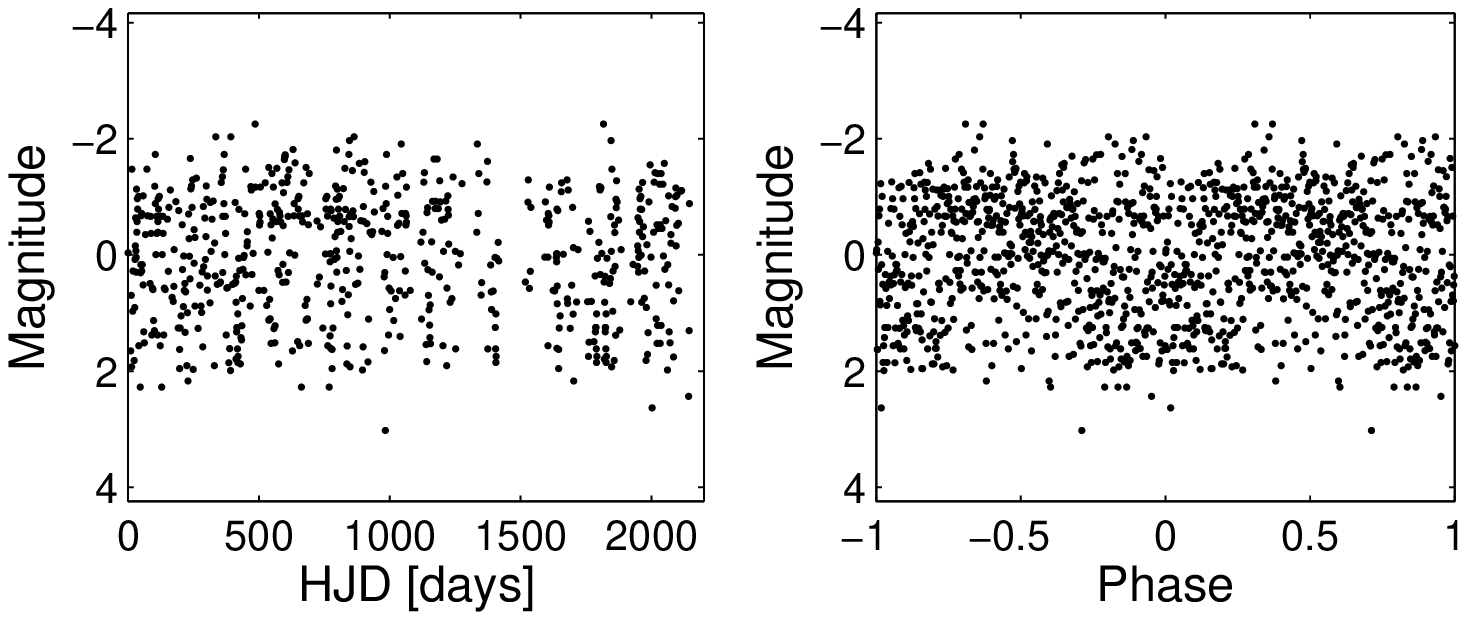} }  
	\caption{ (a) Periodic lightcurve EROS-2 lm0090l27524 folded with the period of 0.337443 days, this period has a CKP value of 2.7424. (b) Slotted autocorrelation function of lightcurve lm0090l27524. Using the slotted ACF, a window length of 3.67 days is selected to create the surrogates.  (c) A surrogate created from lm0090l27524. The CKP value of the surrogate is 0.4532, which is below the corresponding periodicity threshold. }
\end{figure}

For each EROS-2 lightcurve selected, ten surrogates were created. Ten thousand EROS-2 lightcurves were used to create a training set of 100,000 non-periodic synthetic lightcurves. To demonstrate that the resulting surrogates are not periodic and retain the same spectra characteristics as the originals lightcurves, we perform the procedure described above with a lightcurve of a periodic star. Fig~\ref{fig:surrogate1} shows EROS-2 lightcurve lm0090l27524 folded with a period of 0.337443 days. The associated CKP value is 2.7424. The block bootstrap method was used to create a non-periodic synthetic lightcurve. Fig. \ref{fig:surrogate2} shows the slotted ACF and the block length selected for this lightcurve is 3.67 days. Ten surrogates are generated using the procedure described above. Fig. \ref{fig:surrogate3} shows one of the  surrogates. The surrogate is folded with its best period and clearly the periodicity of the original lightcurve is not retained by the surrogate. %This example shows that even a periodic lightcurve can be used to generate a non-periodic surrogate.

\subsection{Obtaining the periodicity discrimination thresholds} \label{pth}

A lightcurve is labelled as periodic if the CKP value associated to its best trial period is above a given periodicity discrimination threshold. We determine the threshold by optimizing the $F_1$ score (Eq. \ref{fbeta}) with a training set created as described above and  following the guidelines in Section \ref{sec:training}.
The periodicity threshold is a function of the SNR and therefore we obtain a periodicity threshold per SNR. To do so, the SNR values are discretized in eight bins: $S=\{[0,1.5]$, $[0,1.5]$, $[1.5,2]$, $[2,2.5]$, $[2.5,3.5]$, $[3.5,5]$, $[5,10]$, $[10,20]$, $[20,\infty]\}$ and compute the periodicity threshold according to the following procedure:

\begin{compactitem}
\item Evaluate the CKP values for each lightcurve in the training set whose SNR fall in bin $S$.
\item Construct a threshold array of 5000 points in $[\min(\text{CKP}),\max(\text{CKP})]$.
\item Compute the $F_1$ score (Eq. \ref{fbeta}) at each threshold value.
\item Select the threshold $th(S)$ as the CKP value that maximizes the $F_1$ score. 
%Save the precision $p(S)$ (Eq. \ref{prec}) and recall $r(S)$ (Eq. \ref{recall}) values associated to the threshold.
\end{compactitem} 

Once the thresholds have been computed, a lightcurve whose SNR falls in bin $S$ is labelled as periodic if:
\[
CKP(P_{best}) > th(S),
\]
where $P_{best}$ is the detected period that maximizes the CKP for the given lightcurve.

\subsection{Estimating the true number of periodic lightcurves} \label{sec:true-number-periodics}

In this section we elaborate on how to estimate the number of periodic lightcurves in a dataset. This is not to be confused with the number of lightcurves labeled as periodic by the proposed method. The true number of periodic lightcurves in a dataset, $N_p$, is the number of true positives plus the false negatives, which is equivalent to the denominator of  $r$ in Eq. \eqref{recall}. The number of lightcurves classified as periodics, $\tilde{N}_p$, is the number of true positives plus false positives, which is equivalent to the denominator of $p$ in Eq. \eqref{recall}. 

Using Eq. \eqref{recall} we can estimate the actual number of periodics in a given SNR bin $S$ as
\begin{equation} \label{Np}
N_p(S)  =  \tilde{N}_p (S) ~ \frac{p(S)}{r(S)},
\end{equation}
where $p(S)$ and $r(S)$  are the precision and recall values for bin S, respectively, which we assume we 
can determine from the training set. The precision and recall values are computed following the procedure given in Section \ref{pth}. Given an $\tilde{N}_p$, we can estimate the true number of periodic lightcurves in a dataset as:
\begin{equation}
\tilde{N}_p = \sum_S  \tilde{N}_p (S) ~ \frac{p(S)}{r(S)},
\end{equation}

\noindent Table \ref{tab:th} shows the thresholds $th(S)$ and associated F-score, recall and precision values obtained for each SNR bin $S$.  The overall precision and recall (across the SNR bins) are 95.3\% and 92.7\%, respectively.

\begin{table}[t]
	\begin{center}
	\caption{Periodicity thresholds and associated precision and recall values for each SNR bin.}  	
	\begin{tabular}{l c c c c}
	\hline
	S	& th(S)	&max F-score	&p(S) [\%]	&r(S) [\%]	\\ \hline
	$[0,1.5]$	&0.4584	&0.92	&94.26	&89.15	\\
	$[1.5,2]$	&0.4565 &0.94	&95.14	&92.15	\\
	$[2,2.5]$	&0.4537	&0.95	&96.42	&92.98	\\
	$[2.5,3.5]$	&0.4581	&0.96	&96.82	&94.26	\\
	$[3.5,5]$	&0.5875	&0.97	&97.52	&96.12	\\
	$[5,10]$	&1.1153	&0.98	&98.12	&97.51	\\
	$[10,20]$	&1.6464	&0.98	&98.22	&97.81	\\
	$[20,\infty]$	&2.4112	&0.97	&98.54	&96.15	\\
	\hline
	\end{tabular}
	\label{tab:th}
	\end{center} 
\end{table}

\subsection{Efficiency as a function of parameters}

In the following tests we assess the efficiency of the proposed method as a function of the parameters of the synthetic lightcurves. Hit rate (Eq. \ref{hr}) is measured as a function of the total time span divided by the period, number of samples, smoothness, and SNR for the 10,000 synthetic periodic lightcurves. Hit rates are computed as a function of one of the parameters while summing for the other three. The CKP is compared with the LS periodogram on each test.

\begin{figure}[h]
	\centering \vspace{-0pt}
	\subfloat[]{\label{fig-efP}\includegraphics[scale=0.56]{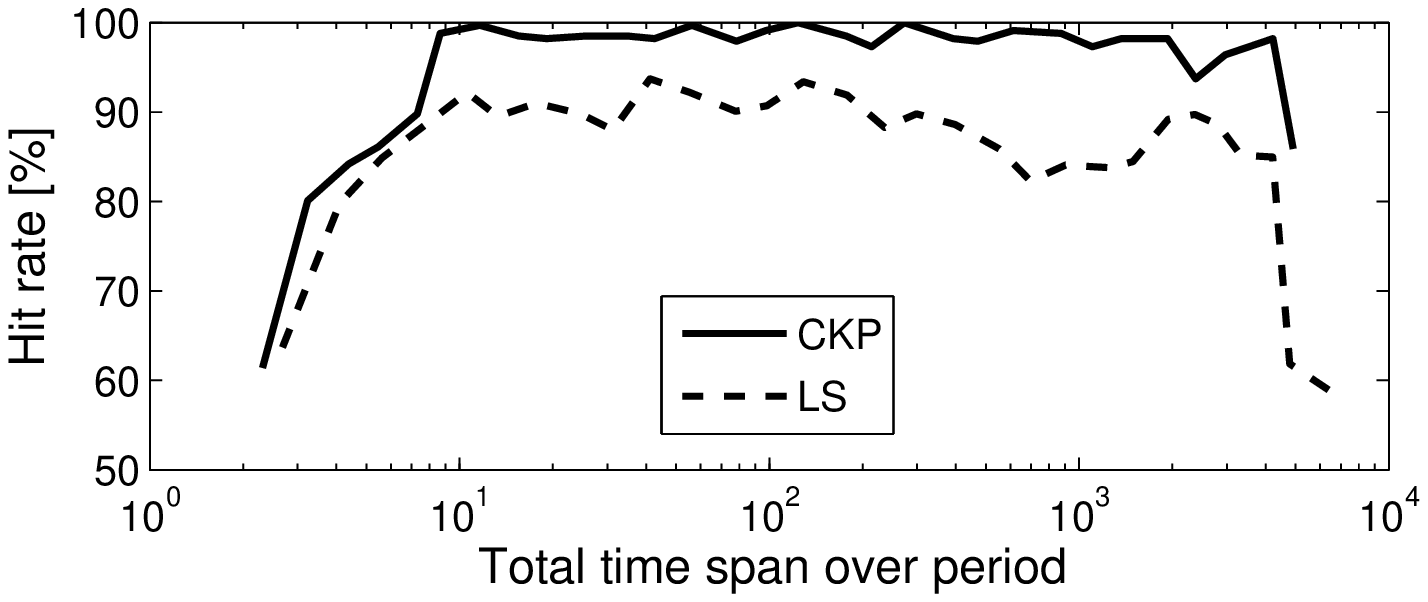}}  \\ \vspace{-10pt}
	\subfloat[]{\label{fig-efN}\includegraphics[scale=0.56]{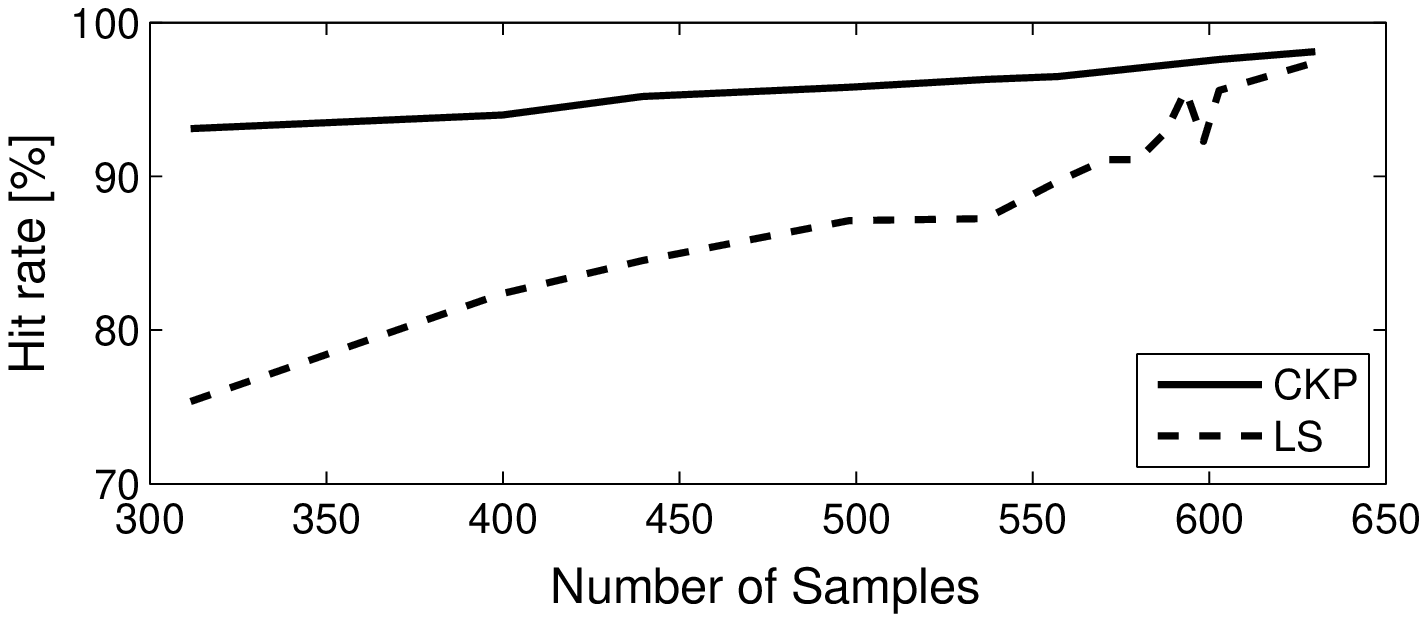}} \\ \vspace{-10pt}
	\subfloat[]{\label{fig-efs}\includegraphics[scale=0.56]{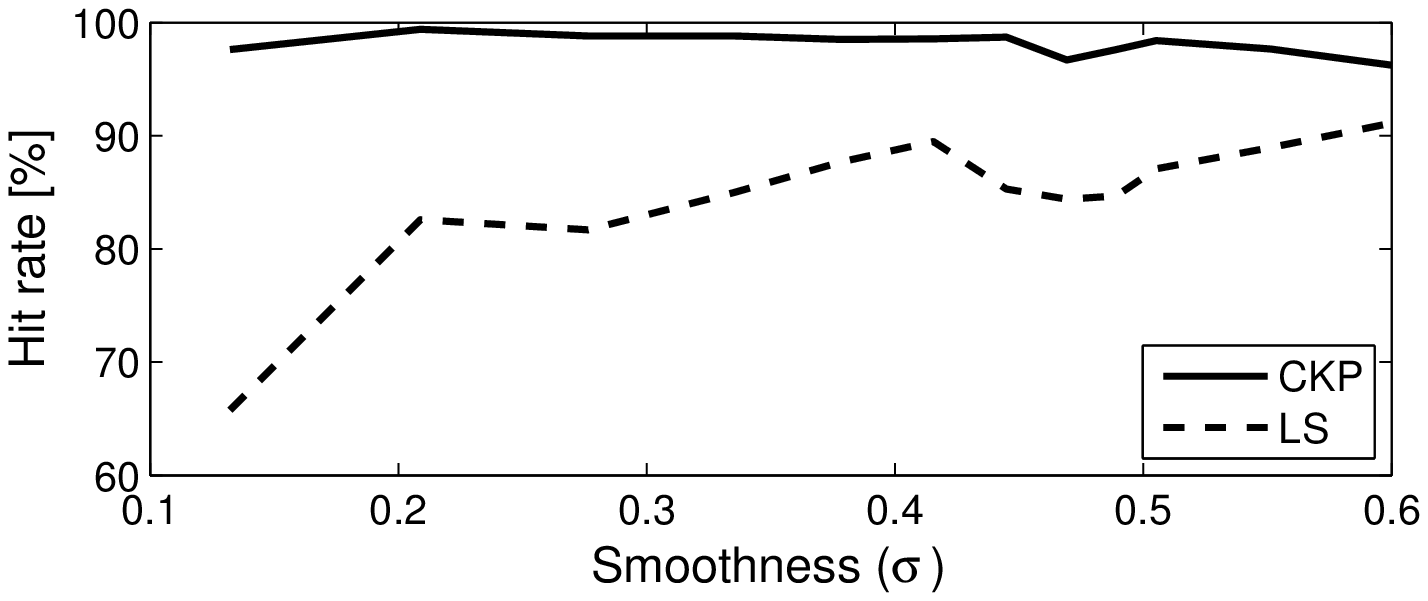}} \\ \vspace{-10pt}	
	\subfloat[]{\label{fig-efS}\includegraphics[scale=0.56]{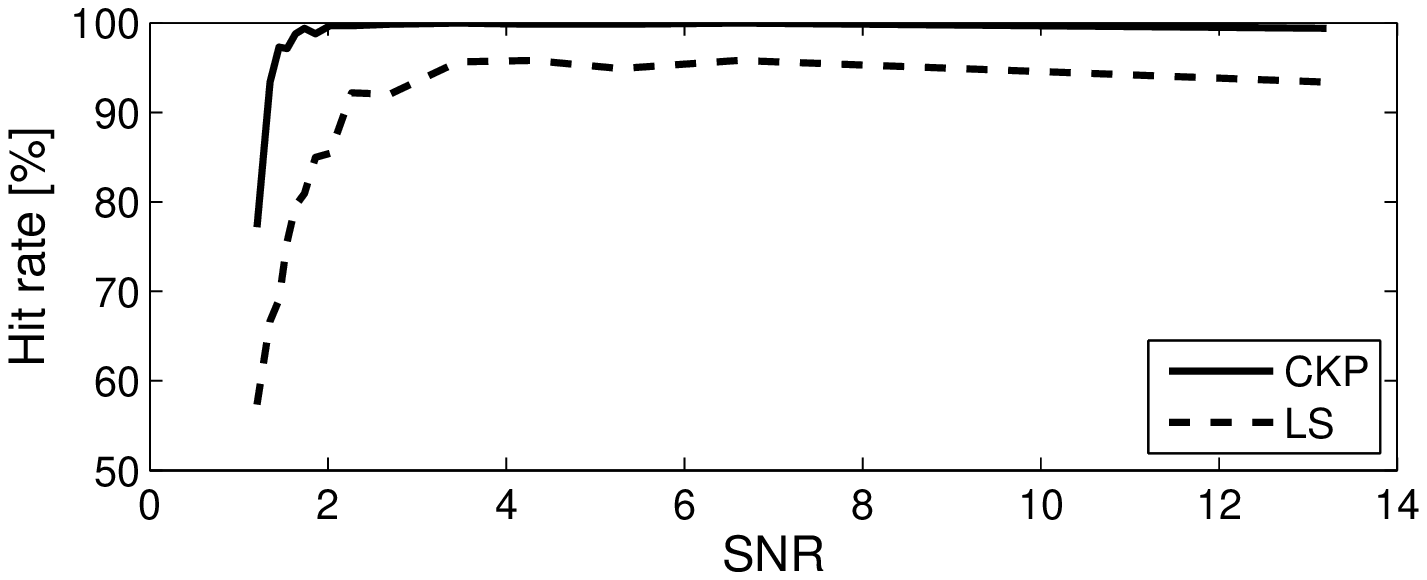}}   \vspace{-0pt}
	\caption{ Hit rate in the synthetic periodic lightcurves as a function of the value of the parameters used to generate the set. The parameters are the number of samples (a), the smoothness (b), period over total time span (c), and SNR (d). The proposed method is compared with the LS periodogram.  }
\end{figure}

Fig. \ref{fig-efP} shows a plot of the HR as a function of the ratio between the total time span of the lightcurve and its period (T/P). The total time span of the lightcurves in EROS-2 survey is approximately 2500 days, and the sampling rate is approximately 1.2 samples per day. The ratio T/P can be viewed as the number of times the underlying signal repeats itself. The period range in the training set goes from 0.4 days to 1000 days. HR is stable across the given range except for T/P below 10 and above 2300. Intuitively, the fewer times a signal is repeated across T the more difficult it is to assess its periodicity. This can be seen in the plot for periods above 280 days. 
There is also a limit in the resolution due by the sampling rate, which is reflected as a hit rate drop for periods below 0.5 days. 
%The lower SNR lightcurves are the most affected by the resolution limit. 
The same hit rate drop can be observed for the LS periodogram.

Fig. \ref{fig-efN} shows a plot of HR as a function of the number of samples of the synthetic lightcurve. HR increases with the number of samples. The hit rate rises by 5\% when the number of samples  increases from 300 to 600. In comparison with the LS periodogram, the CKP is less affected by $N$. Intuitively, the less information available on the process the harder it is to assess its periodicity.

Fig. \ref{fig-efs} shows a plot of the hit rate as a function of the smoothness ($\sigma$) of the synthetic lightcurves. The hit rate is stable across the given range, decreasing slowly for the very large and very small values of $\sigma$. Overall, the smoothness does not have great influence on the CKP hit rate. The LS-periodogram hit rate increases with $\sigma$. This is expected, as smaller values of $\sigma$ produce lightcurves with highly non-sinusoidal shapes, as shown in  Fig. \ref{fig:synthp1}\ref{fig:synthp2}, \ref{fig:synthp3}. 

Finally, Fig. \ref{fig-efS} shows a plot of the hit rate as a function of the SNR (Eq. \ref{pSNR}) of the synthetic lightcurves. HR is stable for the given SNR range, dropping abruptly for SNR below $1.8$. For SNR of $1.2$ hit rate has decreased by a almost 25\%. A similar behaviour can be seen for the LS-periodogram. 

\section{Data} \label{sec:data}
\subsection{Description of the data}

The EROS-2 project \citep{EROS2007,EROS2009} was designed to search for gravitational microlensing events caused by massive compact halo objects (MACHOs) in the halo of the Milky Way. To do this, 32.8 million stars in the Magellanic clouds were surveyed over 6.7 years. The objective of the EROS-2 survey was to test the hypothesis that MACHOs were a major component of the dark matter present in the Halo of our galaxy.

The EROS-2 project surveyed 28.8 million stars in the Large Magellanic Cloud (LMC) and 4 million stars in the Small Magellanic Cloud (SMC), distributed in 88 and 10 observational fields, respectively. Each field is divided in 32 chips (8 CCDs and 4 quadrants per CCD). Each lightcurve file has 5 columns: time instant, red channel magnitude, red channel error bars, blue channel magnitude and blue channel error bars. In what follows, only the blue channel is used. The average number of samples per lightcurve is 430 and 780 in the LMC and SMC, respectively.

%\subsection{A pipeline for automated periodic lightcurve discrimination} \label{pipeline}

\subsection{Preprocessing and intricacies of the data}

\underline{Fixing the error bars:}
As described above the kernel size was estimated using the errorbars of the magnitudes 
or the estimate of the observational errors. 
If these observational errors were underestimated or overestimated (as is often the case) 
the kernel size will be also wrongly-estimated. For example if the error bars
are for some reason underestimated then the kernel bandwidth will be also underestimated
and will not account of the true scatter of the lightcurve resulting into low CKP values. 
%\begin{itemize}
%\item EXPLAIN THE MEANING OF HAVING ERRORBARS BIGGER THAN STANDARD DEVIATION, WHY THEY WERE OVERESTIMATED IN THE FIRST PLACE?
%\item WHY THE OVERESTIMATION SEEMS TO HAPPEN IN THE FIELDS WITH MORE PERIOCICS/ MORE CROWDED (BAR)? IS OVERESTIMATION RELATED TO HOW CROWDED THE FIELD IS?
%\item OVERESTIMATED ERRORBARS AFFECT THE SELECTION OF THE KERNEL SIZE
%\item CORRECTION: FUDGE FACTOR, WHAT IS THE OFFICIAL NAME FOR THIS? ERRORBAR CORRECTION FACTOR?
%\item ONLY THE LOCUS? IS USED TO CALCULATE THE ERROR BAR CORRECTION FACTOR(MAGNITUDES BETWEEN 17 AND 21)
%\end{itemize}

\begin{figure}
	\centering	
	\subfloat[]{\label{fig-fudge1a}\includegraphics[scale=0.6]{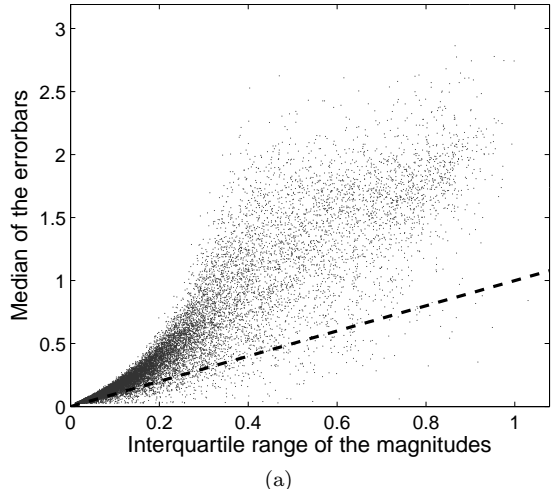} }  \qquad
	\subfloat[]{\label{fig-fudge1b}\includegraphics[scale=0.6]{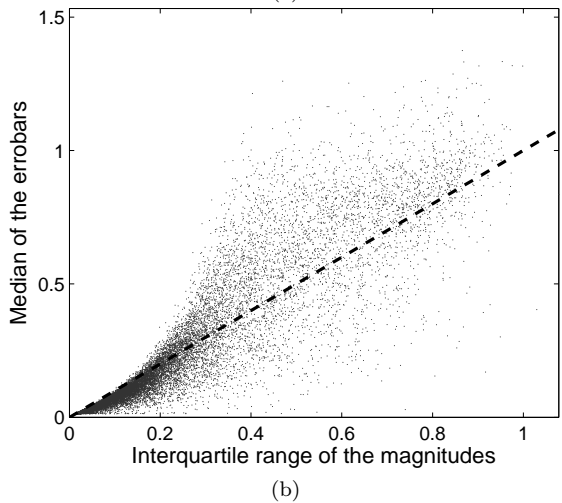} }  
	\caption{ (a) Median of the magnitude's error bars as a function of the interquartile range of the magnitudes for chip lm0090k. The dotted line has a slope of one. The error bar correction factor for lm0090k is $0.42$. (b) The same plot after correcting the error bars.  }
\end{figure}

For a lightcurve that is not variable the sample variance and the error bars should have very 
similar values. Another way of expressing this is that for a given non-variable lightcurve the median of the error bars
should be equal to the inter-quartile range. Since we know most sources are not variable a plot
of those two quantities should be distributed around the bisector\footnote{Line with slope of one.}.
Fig. \ref{fig-fudge1a} shows a plot of the median of the error bars as a function of the interquartile range of the magnitudes for a randomly selected chip, lm0090k. Each dot corresponds to a lightcurve. The locus of the points (lightcurves with magnitudes between 17 and 21) is over the bisector, \emph{i.e.} the error bars are larger than the dispersion of the lightcurve. This is an example of a field with overestimated error bars.

For a given field with $N_{lc}$ lightcurves, the error bar {\em correction factor} is defined as the constant that minimizes
\begin{equation} \label{ff1}
\alpha_{cf} = \arg\min_{\alpha} \sum_{k=1}^{N_{lc}} \left( \text{iqr}(\{y\}_k) - \alpha ~\text{med}(\{e\}_k) \right )^2,
\end{equation}
where $\{y\}_k$ and $\{e\}_k$ are the magnitudes and error bars of lightcurve $k$, respectively, $\text{iqr}$ is the interquartile range and $\text{med}$ is the median. 
%The error bar correction factor can obtained by differentiating Eq. \eqref{ff1} wrt to $\alpha$ and setting it, 
%\begin {equation} \label{ff2}
%\alpha = \frac{\sum_{k=1}^{N_{lc}}{ \text{iqr}(\{x\}_k) ~ \text{med}(\{e\}_k)  }  }{ \sum_{k=1}^{N_{lc}}{ \text{med}(\{e\}_k)^2}}.
%\end{equation}

\begin{figure}[t]
	\centering
	\includegraphics[scale=0.6]{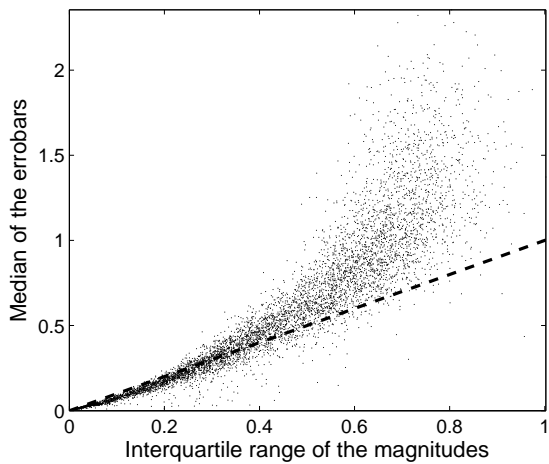} 
	\caption{ Median of the magnitude's error bars as a function of the interquartile range of the magnitudes for chip lm0140k. The error bar correction factor for lm0140k is $\sim 1$. }
	\label{fig-fudge2}
\end{figure}

%Fig. \ref{fig-fudge1a} shows a plot of the median of the error bars as a function of the interquartile range of the magnitudes for chip lm0090k. This chip is on the LMC bar and is crowded wrt other fields. Each dot corresponds to a lightcurve. The locus of the chip (lightcurves with magnitudes between 17 and 21) is over the slope one line, \emph{i.e.} the error bars are larger than the dispersion of the lightcurve. 
For the field shown in Fig. \ref{fig-fudge1a} an error bar correction factor of 0.42 is obtained for this field. Fig. \ref{fig-fudge1b} shows the plot of the same field after correcting the error bars. Fig. \ref{fig-fudge2} shows the same plot for chip lm0140k. 
%This chip is on the LMC bar and is crowded. 
This chip is on the periphery of the LMC. 
%[PP: IS IT ON THE PERIPHERY OR ON THE CORE]
%PH: Periphery, look at the LMC fields figure
The error bar correction factor for this field is $\sim 1$, \emph{i.e.} there is no need for correction.

Using the error bar correction factor, we define the pseudo signal-to-noise-ratio (pSNR) of a given lightcurve as
\begin{equation} \label{pSNR}
pSNR =  \frac{0.7413 ~\text{iqr}(\{y\})}{\alpha ~\text{med}(\{e\})},
\end{equation}
where $y$ and $e$ are the magnitudes and error bars, respectively, and $\alpha$ is computed per field using Eq. \ref{ff1}. 
%This metric is used in the following steps of the periodic lightcurve discrimination pipeline.

\underline{Removing outliers and bad points:}
%The lightcurve's blue channel is imported. 
The mean $\bar{e}$ and the standard deviation $\sigma_e$ of the error bars are computed per lightcurve and samples that do not comply with 
\[
e_i < \bar{e} + 3\sigma_e,
\]
where $e_i$ is the error bar of a sample $i$, are removed from the lightcurve. At this point, lightcurves with less than fifty samples are discarded from the analysis. 

\begin{figure*}[ht]
	\centering	
	\subfloat[]{\label{fig-detrend1}\includegraphics[scale=0.7]{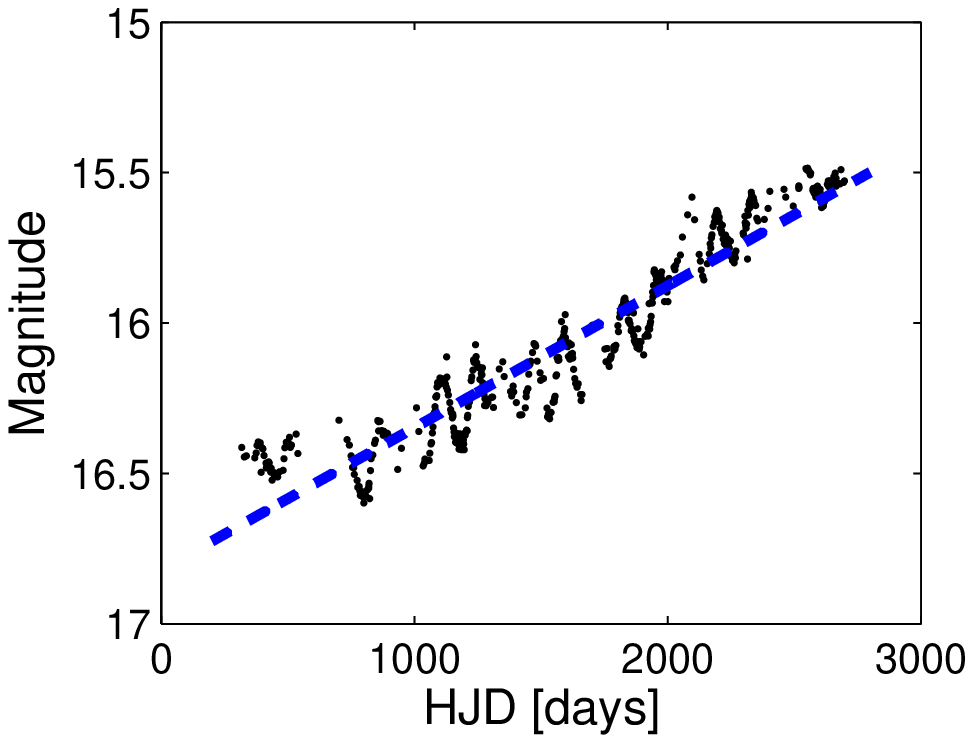} } 
	\subfloat[]{\label{fig-detrend2}\includegraphics[scale=0.7]{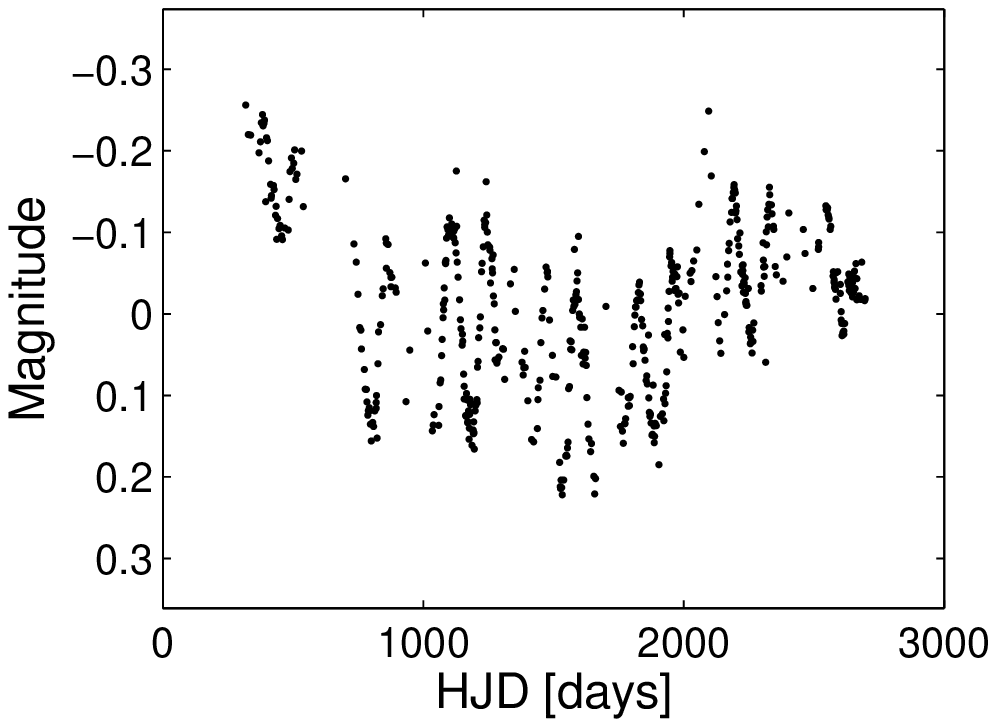} } 
	%\\ \subfloat[]{\label{fig-detrend3}\includegraphics[scale=0.7]{lineardetrendFolded.eps} } 	
	\caption{ (a) Lightcurve lm0324k13673 from the EROS-2 survey. A linear $\chi^2$ fit is computed for this lightcurve (blue dotted line). The correlation coefficient for the linear fit is 0.9493. (b) lightcurve lm0324k13673 after the linear trend subtraction.}
\end{figure*}

\underline{Simple detrending:}
After that, the coefficients of a least square linear $\chi^2$ regression on the magnitudes are computed

\begin{equation} \label{chisquare}
\chi^2 = \sum_{i=1}^N\frac{(a_0 + a_1 t_i - x_i)^2}{e_i^2},
\end{equation}

\noindent where $a_0$ is the intercept and $a_1$ is the slope. The coefficients of the linear fit are obtained by differentiating Eq. \eqref{chisquare} wrt $a_1$ and $a_0$. The linear $\chi^2$ fit is subtracted from the lightcurve only if the correlation coefficient between the lightcurve and its linear fit is above 0.5 (goodness of fit). Fig. \ref{fig-detrend1} shows EROS-2 lightcurve lm0324k13673. The signal is mounted on a monotonically increasing linear trend. The dotted line in Fig. \ref{fig-detrend1} shows the $\chi^2$ linear fit. Fig. \ref{fig-detrend2} shows the lightcurve after the linear fit subtraction, further evaluation shows that the lightcurves is periodic with a period of 120.38 days.

\begin{comment}
The values of the intercept and slope are obtained by differentiating Eq. \eqref{chisquare} with respect to $a_1$ and $a_2$

\[
	a_1 = \frac{\left(\sum_{i=1}^N x_i/e_i^2 \right) \left(\sum_{i=1}^N t_i^2/e_i^2\right) - \left(\sum_{i=1}^N (x_i t_i)/e_i^2\right) \left(\sum_{i=1}^N t_i/e_i^2\right)} { \left(\sum_{i=1}^N 1/e_i^2\right) \left(\sum_{i=1}^N t_i^2/e_i^2\right) - \left(\sum_{i=1}^N t_i/e_i^2\right)^2 },
\]

\[
	a_2 = \frac{\left(\sum_{i=1}^N 1/e_i^2\right) \left(\sum_{i=1}^N (x_i t_i)/e_i^2\right) - \left(\sum_{i=1}^N t_i/e_i^2\right) \left(\sum_{i=1}^N x_i/e_i^2\right)} { \left(\sum_{i=1}^N 1/e_i^2\right) \left(\sum_{i=1}^N t_i^2/e_i^2\right) - \left(\sum_{i=1}^N t_i/e_i^2\right)^2 },
\]

and the correlation coefficient between the lightcurve and the linear fit

\[
	r = \frac{ N \sum_{i=1}^N x_i (a_1 + a_2 t_i)  - \left(\sum_{i=1}^N x_i\right)  \left(\sum_{i=1}^N a_1 + a_2 t_i\right)  } { \sqrt{N \sum_{i=1}^N x_i^2 -\left(\sum_{i=1}^N x_i\right)^2 } \sqrt{N \sum_{i=1}^N \left(a_1 + a_2 t_i\right)^2 -\left(\sum_{i=1}^N a_1 + a_2 t_i\right)^2 }}.
\]
\end{comment}

%--------------------------- Results ---------------------- 
\section{Results} \label{sec:results}

\subsection{Filtering of spurious periods} 

\begin{figure}
	\centering
	\subfloat[]{\label{fig-histspa}\includegraphics[scale=0.6]{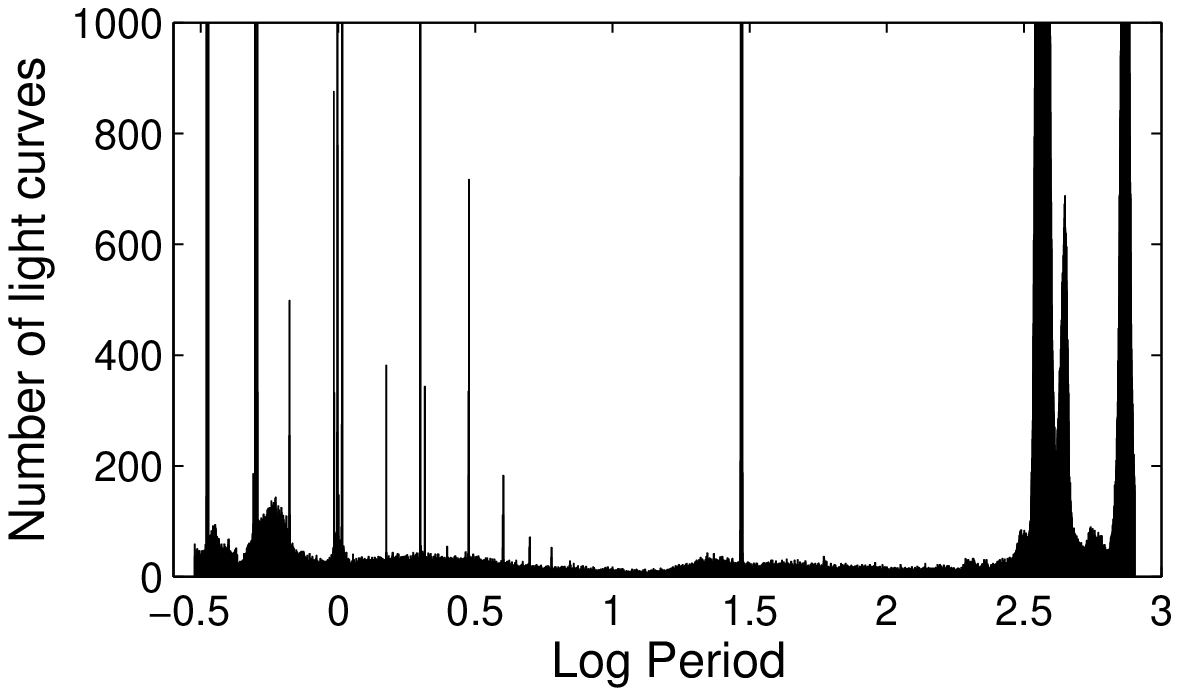} }  \qquad
	\subfloat[]{\label{fig-histspb}\includegraphics[scale=0.6]{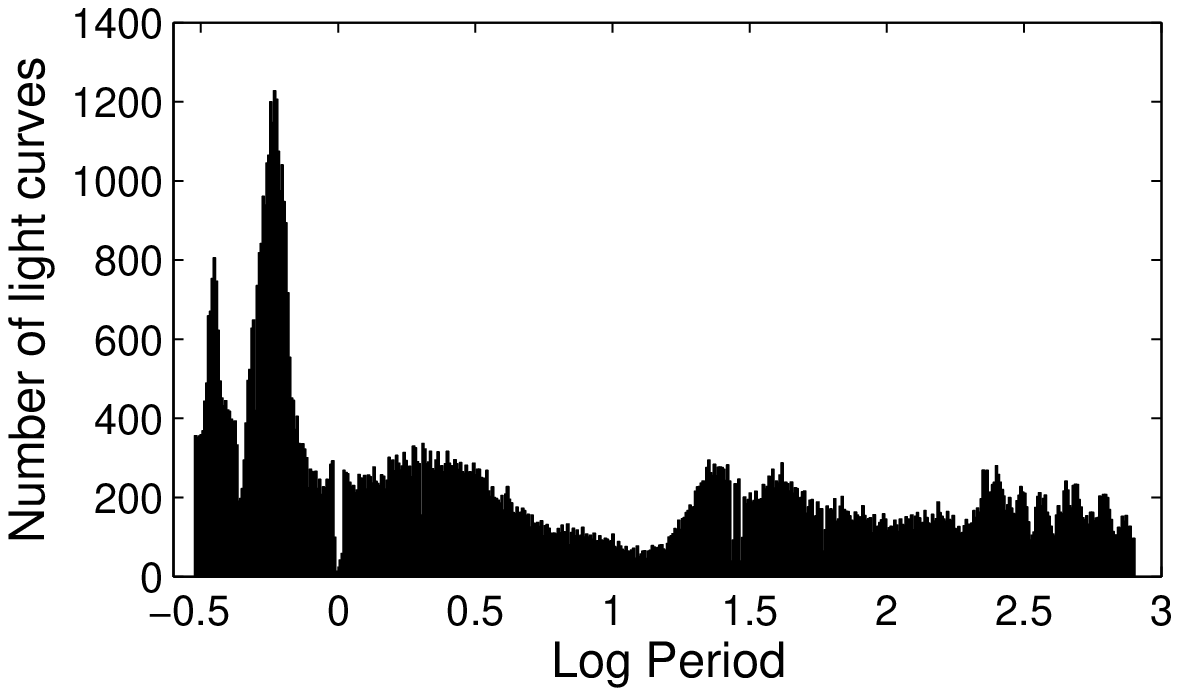} }  \\
	\caption{ (a) Histogram of the periodic lightcurves detected with the proposed method on the LMC. The spurious periods have not been filtered in these results. The vertical columns corresponds to the spurious periods, their multiples and aliases. (b) Histogram of the periodic lightcurves detected in the LMC after carrying out the spurious period removal scheme.  } 
\end{figure}

The trial periods extracted with the bands method are evaluated using the CKP (Eq. \ref{CKP}) contain spurious periods related to the solar day, the moon phase, the year, and their multiples are filtered. Additional spurious periods were found by analyzing the histogram of the periodic lightcurves detected by the proposed method (Fig. \ref{fig-histspa}). These additional spurious periods, which are given in Table \ref{tab:spuriousperiods}, correspond to aliases of the known spurious periods. 

A Gaussian mask centered around the spurious period is created for each of the spurious periods. Periods whose CKP fall inside the masks are filtered as spurious periods. The standard deviation and the amplitude of the masks are set so that the associated spurious peak in the period histogram is flattened \footnote{The parameters of the filters can be found alongside the catalogs at http:\\timemachine.iic.harvard.edu}. The trial period that maximizes the CKP and does not fall in any of the spurious period masks is selected as the best trial period for the lightcurve.

\begin{table}[t]
	\begin{center}
	\caption{Description of the spurious periods}  	
	\begin{tabular}{c l}
	\hline
	Period [days]	& Description\\ \hline
	$1$		&	Solar day ($P_d$)\\
	$29.5305$	&	Moon phase or Synodic month ($P_m$)\\
	$365.24$	&	Tropical year ($P_y$)\\
	$2,335$		&	Average time span of EROS-2 lightcurves ($T$)\\
	$0.4917$	&	$( (P_d/2)^{-1} + P_m^{-1}   )^{-1}$\\
	$0.5086$	&	$( (P_d/2)^{-1} - P_m^{-1}   )^{-1}$\\
	$0.9672$	&	$( P_d^{-1} + P_m^{-1} )^{-1}$\\
	$1.0351$	&	Lunar day, $( P_d^{-1} - P_m^{-1} )^{-1}$\\
	$0.9973$	&	Sidereal day, $( P_d^{-1} + P_y^{-1} )^{-1}$\\
	$1.0027$	&	$( P_d^{-1} - P_y^{-1} )^{-1}$\\	
	$27.31$		& 	Sidereal month, $( P_m^{-1} + P_y^{-1} )^{-1}$\\
	$32.13$		& 	$( P_m^{-1} - P_y^{-1} )^{-1}$\\
	$315.65$	& 	$( P_y^{-1} + T^{-1} )^{-1}$\\
	$432.63$	& 	$( P_y^{-1} - T^{-1} )^{-1}$\\
	
	\hline
	\end{tabular}
	\label{tab:spuriousperiods}
	\end{center} 
\end{table}

\subsection{Results for selected fields}

In this experiment the proposed method is evaluated on three fields from the EROS-2 survey. The objectives are to measure the accuracy of the method and to compare the number of periodic lightcurves in the fields with the expected number of periodic lightcurves computed from the synthetic results by performing visual inspection to a large but manageable number of lightcurves. The first six chips from fields lm009, lm012 and sm001 are used in this experiment. Table \ref{tab:summaryFields} shows the number of lightcurves, the average number of samples and the average SNR from the selected fields. 

Table \ref{tab:resFields} shows the results obtained for the selected fields. Column two ($\tilde{N}_p$) corresponds to the number of lightcurves labelled as periodic by our method. These lightcurves are folded with the detected period and visually checked in order to find the number of false positives (column three). Column four is the precision in the detected periodic lightcurves set. Column five gives an estimate of the false negatives (FN) in the field. The FNs are estimated by visually inspecting the folded lightcurves of the objects that are below the periodicity thresholds. Because it is impracticable to check all the non-periodic objects, the search for FNs is stopped if 50 consecutive non-periodic lightcurves are found for each SNR bin. Column six is the recall calculated using the observed number of true positives ($\tilde{N}_p$ -FP) and the FN. Column seven corresponds to the observed number of periodic lightcurves ($\tilde{N}_p$ -FP+FN). Column eight shows an estimation of the true number of periodic variables ($N_p$) using the synthetic precision and recall values given in Section \ref{sec:true-number-periodics}. Column seven is also an estimation of $N_p$ because the true amount of FNs is not known.

\begin{figure*}
	\centering
	\subfloat[]{\label{fig-mapLMC}\includegraphics[scale=0.7]{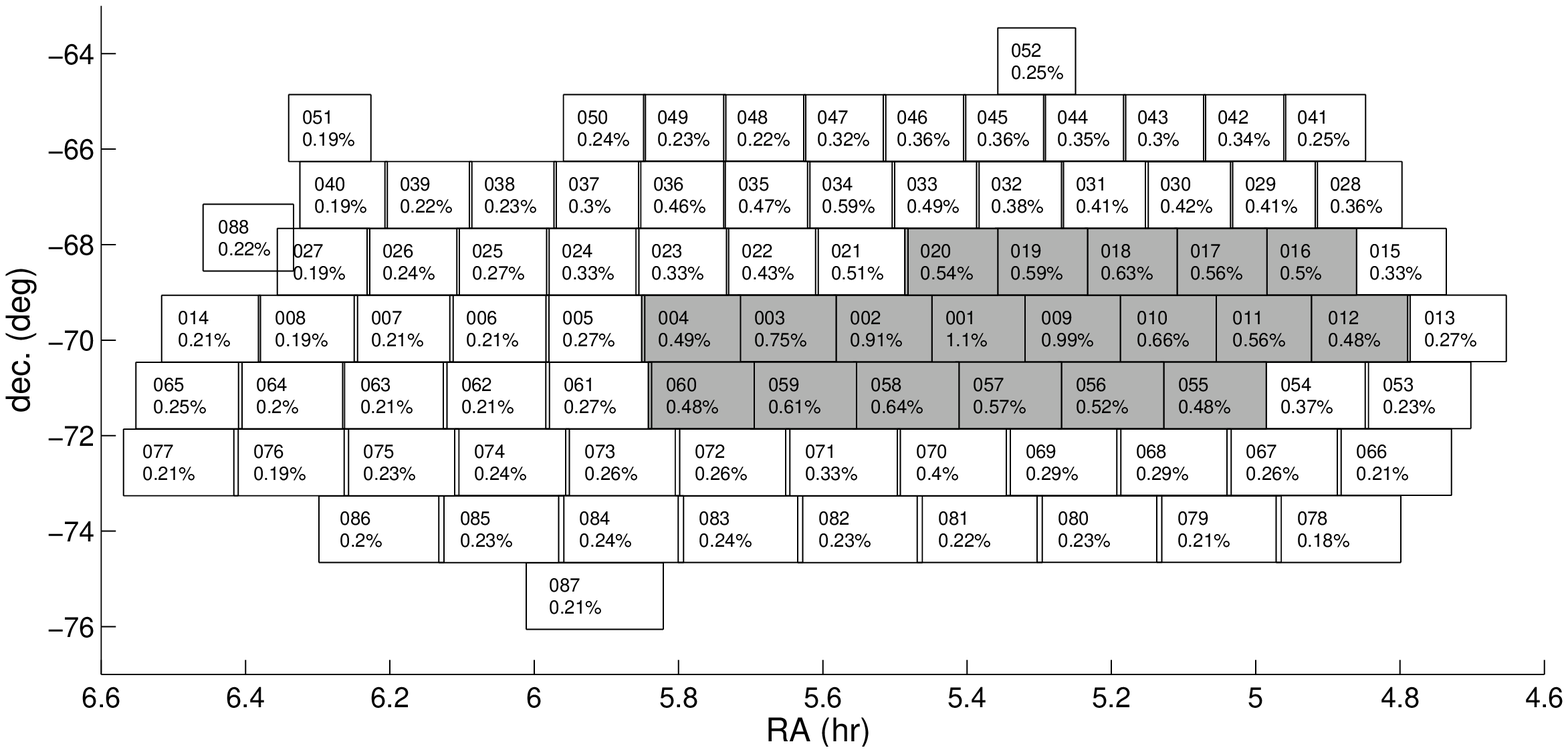} } \\
	\subfloat[]{\label{fig-mapSMC}\includegraphics[scale=0.7]{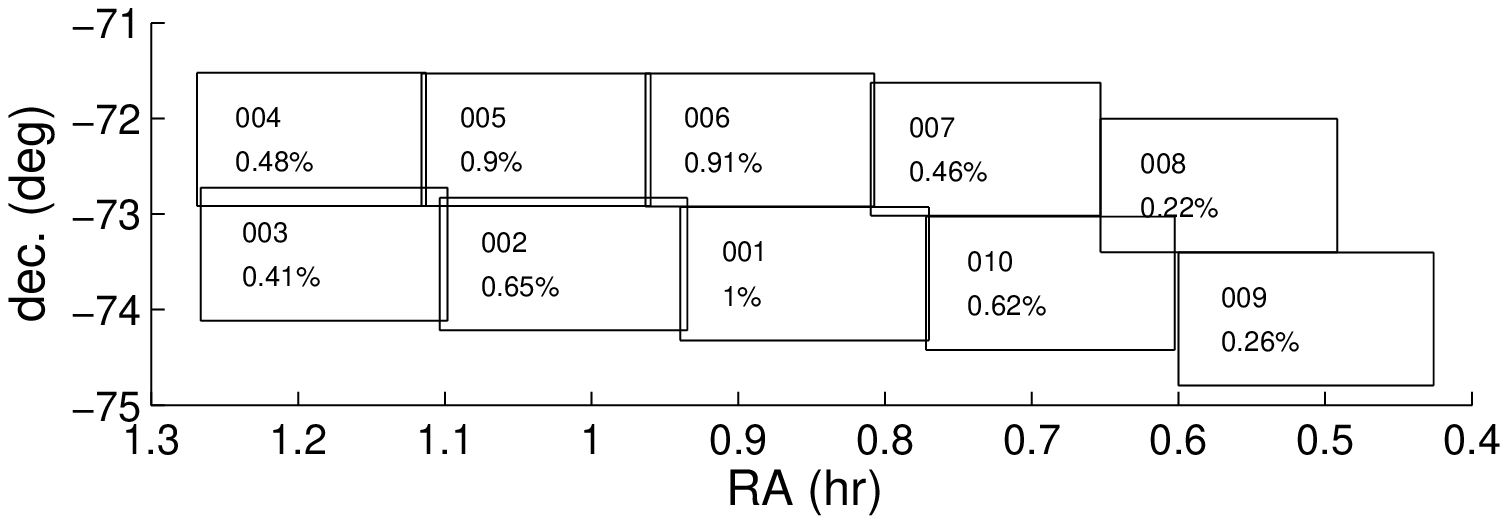} } 	
	\caption{ Maps of the EROS-2 LMC (a) and SMC (b) fields, respectively. The percentage of periodic lightcurves is shown below the name of the field.   }
\end{figure*}

A grand total of 1160 periodic lightcurves is recovered from field lm009, which corresponds to a 1.06\% of the field. The percentage of periodics lightcurves in lm012 and sm001 is 0.75\% and 1.69\%, respectively\footnote{These chips have a higher number of periodics than the average found in the LMC and SMC as it can be seen in Fig. \ref{fig-mapLMC}. This issue is discussed in the next section. }. The overall precision and recall in all the fields is within 2\% of the overall precision and recall found in the synthetic dataset. For comparison we ran the Lomb-Scargle periodogram\footnote{The vartools software with the -LS option is used.} on the lm009 field. The spurious periods are filtered as described in previous Sections. The filtered periods found with the LS periodogram are sorted according to their normalized LS statistic. By imposing a threshold on this statistic the periodic light curves obtained the CKP plus 298 falses positives and 14 additional true positives are obtained. This corresponds to a drop of 16.5\% in precision and a negligible increase in recall (1\%) with respect to the CKP.
%In all the selected fields, the estimated $N_p$ given by the synthetic results is higher than the observed $N_p$, especially in field sm001 which is the field with the higher number of periodics. The true $N_p$ is expected to be bigger than the observed $N_p$ (FN estimation)...
%[PP1: THIS IS HIGHER THAN THE OVERALL NUMBER WE REPORT AT THE END. ]
%[PH1: Yes, note that these fields are near bar of their respective galaxies, and we have seen that the \% periodics is higher in the bar. Also, don't know why, but in all the cases, the first 6 chips have higher \% of periodics than the whole field (32 chips).]
%[PP2: ALSO 2\% SEEMS IMPRESSIVE AND WE SHOULD EMPHASIZE]

\begin{figure}
	\centering	
	\subfloat[]{\label{fig-exfp1}\includegraphics[scale=0.49]{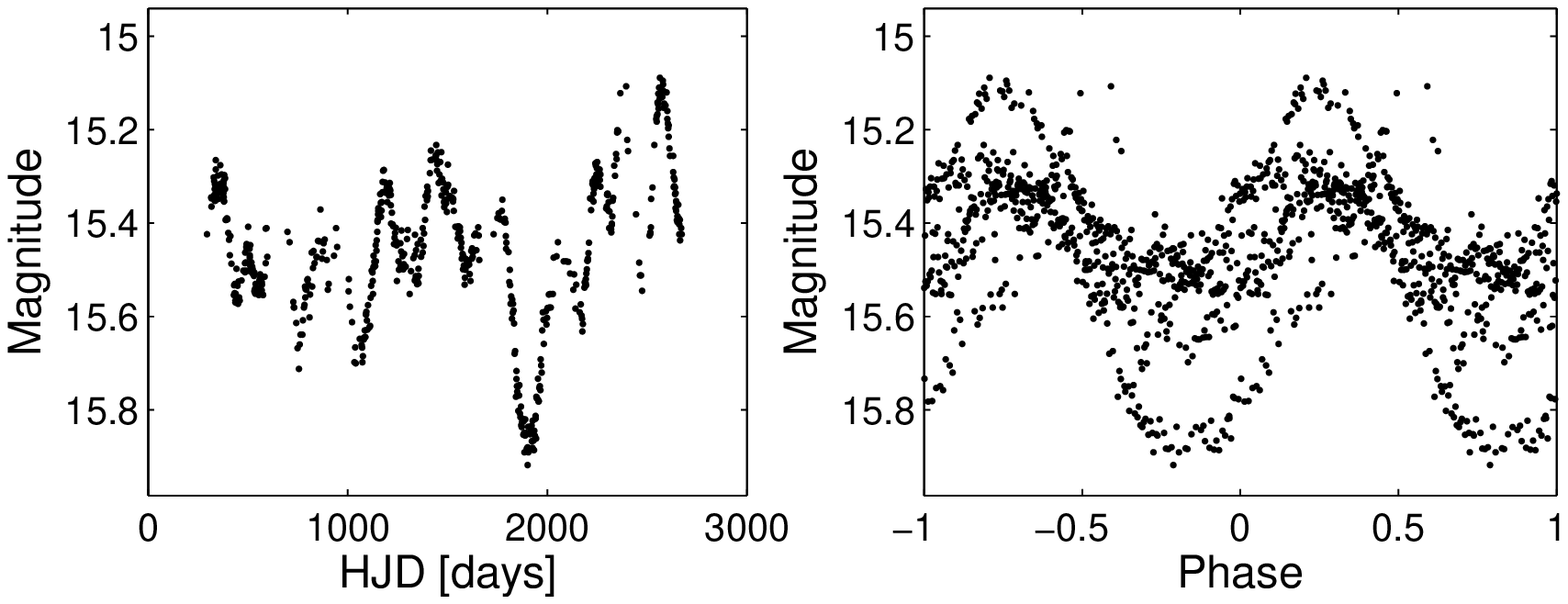}} \\
	\subfloat[]{\label{fig-exfp2}\includegraphics[scale=0.49]{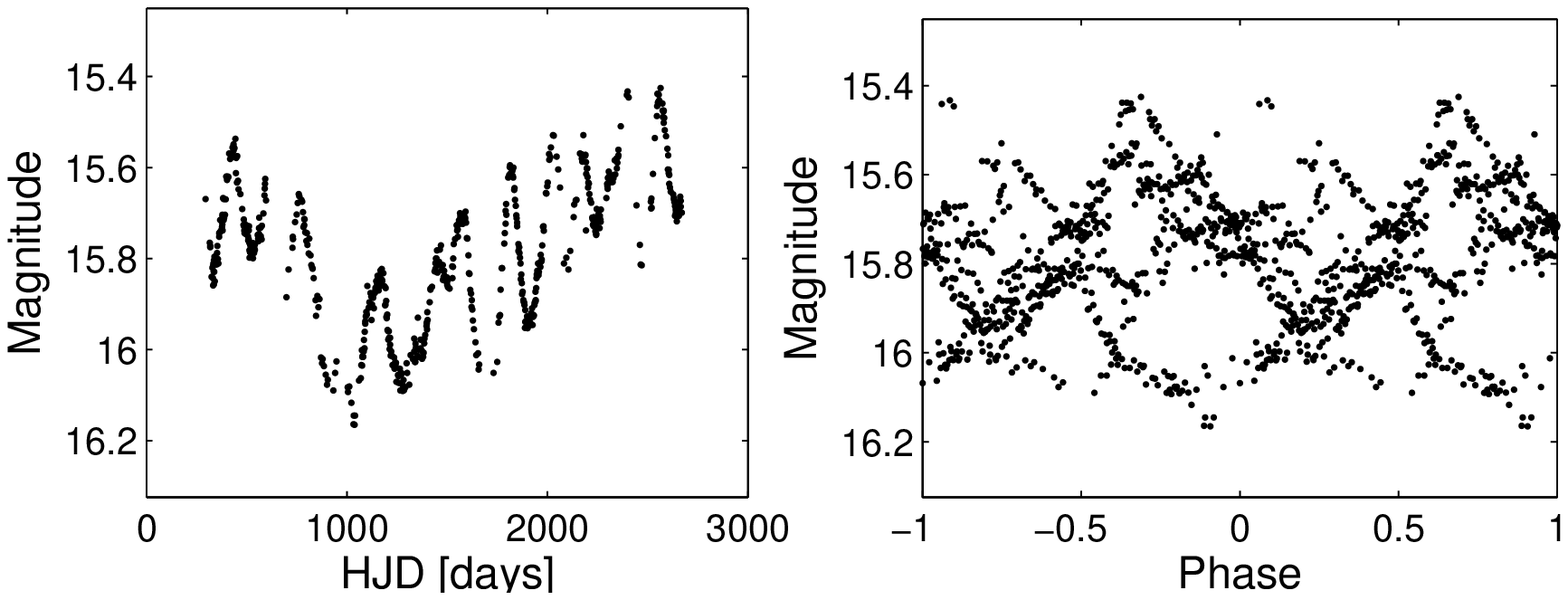}} \\
	\subfloat[]{\label{fig-exfp3}\includegraphics[scale=0.49]{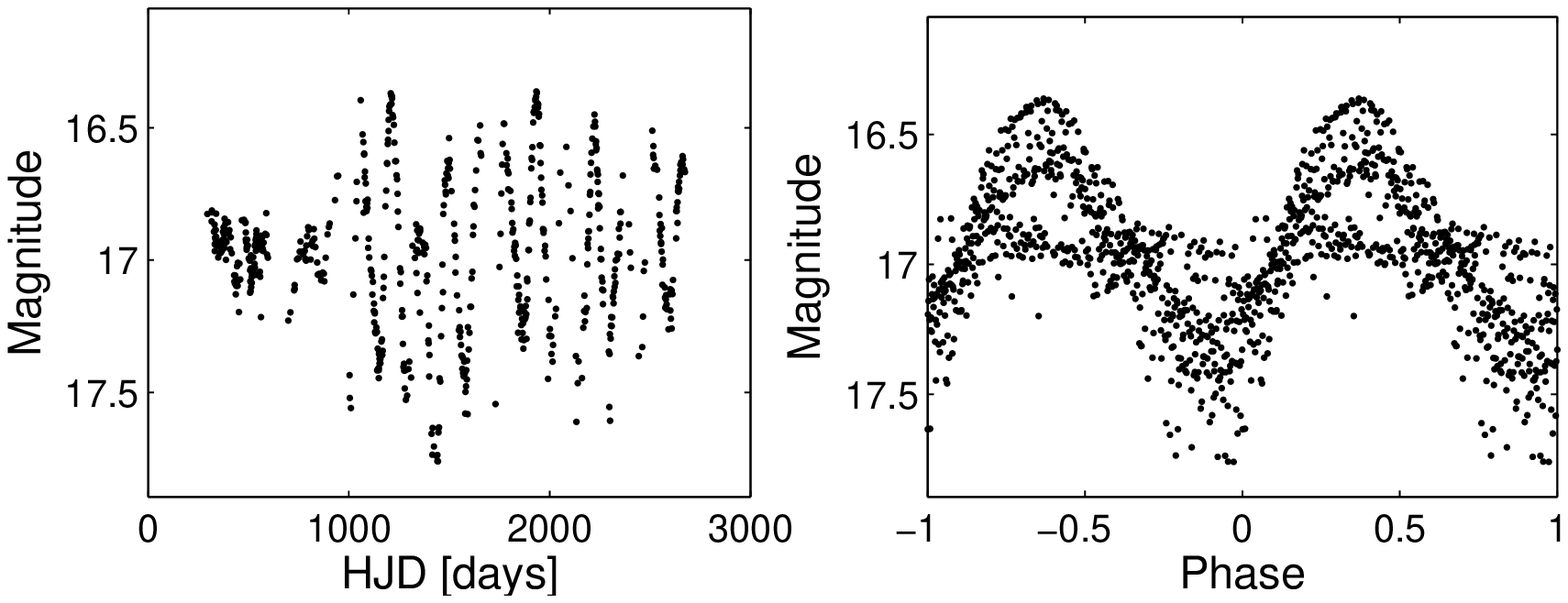}}  
	\caption{ These lightcurves are examples of the false positives found in the catalogs. (a) lightcurve lm0090n29655, folded with the detected period of 278 days, is an example of quasi-periodic behaviour.  (b) lightcurve lm0091l19300, folded with the detected period of 264 days, is mounted on a polynomial trend in the mean. (c) lightcurve lm0090n6107, folded with the detected period of 144 days, varies in amplitude across the time span.  }
\end{figure}

It is important to note that there are periodic behaviors that are not captured in the proposed synthetic lightcurve set. Examples of these are periodicities mounted on polynomial trends, objects with more than one oscillation period, objects that are not periodic in the whole time span and objects whose oscillations amplitude change irregularly or following a modulation pattern, such as semi-regular and irregular LPVs. These cases are considered as non-periodic during the inspection. Examples of these cases are shown in Figures \ref{fig-exfp1}, \ref{fig-exfp2} and \ref{fig-exfp3}, which correspond to false positives found in field lm009. Currently the proposed method is not able to discriminate quasi-periodicities and other irregular periodics.
%[PP: ARE THOSE FP OR FNs?]
%[PH: I'm sure that those examples are False positives. If you mean quasi-periodics in general, then they were considered as non-periodic so they are either FP or TN, not FN]

\begin{table}[th]
	\begin{center}
	\caption{\label{tab:summaryFields} Characteristics of selected fields.}  
	\begin{tabular}{l c c c}
			\hline
	Field	&Number of lightcurves	&Average N & Average SNR\\ \hline
	lm009	&109,802	&548	&1.628 \\
	lm012	& 95,010	&447	&0.959 \\	
	sm001	& 92,666	&830	&1.505 \\
	\hline
	\end{tabular}
	\end{center} 
\end{table}

\begin{table*}[th]
	\begin{center}
	\caption{Results in the selected EROS-2 survey fields.  }  	
	\begin{tabular}{l c c c c c c c}
	\hline
	Field	&$\tilde{N}_p$	&FP	&Prec. [\%]	&FN	&Recall [\%]	&Observed $N_p$	& Synthetic $N_p$\\ \hline
	lm009	&1160	&41	&96.47	&66	&94.43	&1185	&1189	\\
	lm012	&718 	&30	&95.82	&51	&93.10	&739	&743	\\	
	sm001	&1564	&69	&95.59	&99	&93.79	&1594	&1637	\\
	\hline
	\end{tabular}
	\label{tab:resFields}
	\end{center} 
\end{table*}

\subsection{Results on EROS-2 LMC and SMC fields}

%28,800,099 LMC
%4,064,179 SMC

A total of 32.8 million lightcurves from the EROS-2 survey were processed with the proposed periodicity discrimination pipeline, 28.8 million from the LMC and 4 million from the SMC. Table \ref{tab:summaryMCp} shows the summary of the results for the LMC and SMC. $\tilde{N}_p$ corresponds to the number of lightcurves labeled as periodic by our method. The {\em Discarded} column corresponds to the number of periodic lightcurves that appear twice in the list, due to field overlapping and blending. Column $N_p$ corresponds to an estimation of the true number of periodic variables using the synthetic precision and recall values given in Section \ref{sec:true-number-periodics}.

\begin{figure}
	\centering	\vspace{-0pt}
	\subfloat[]{\label{fig-o1}\includegraphics[scale=0.49]{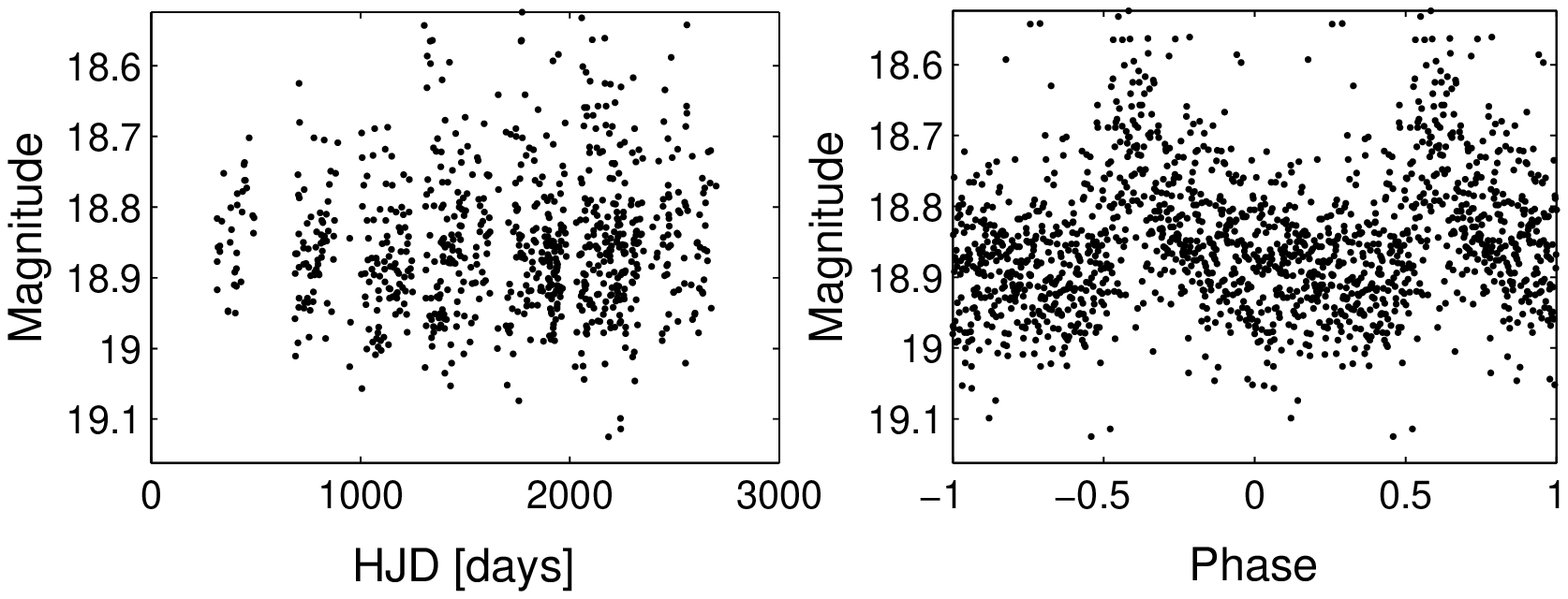}} \\ \vspace{-10pt}
	\subfloat[]{\label{fig-o2}\includegraphics[scale=0.49]{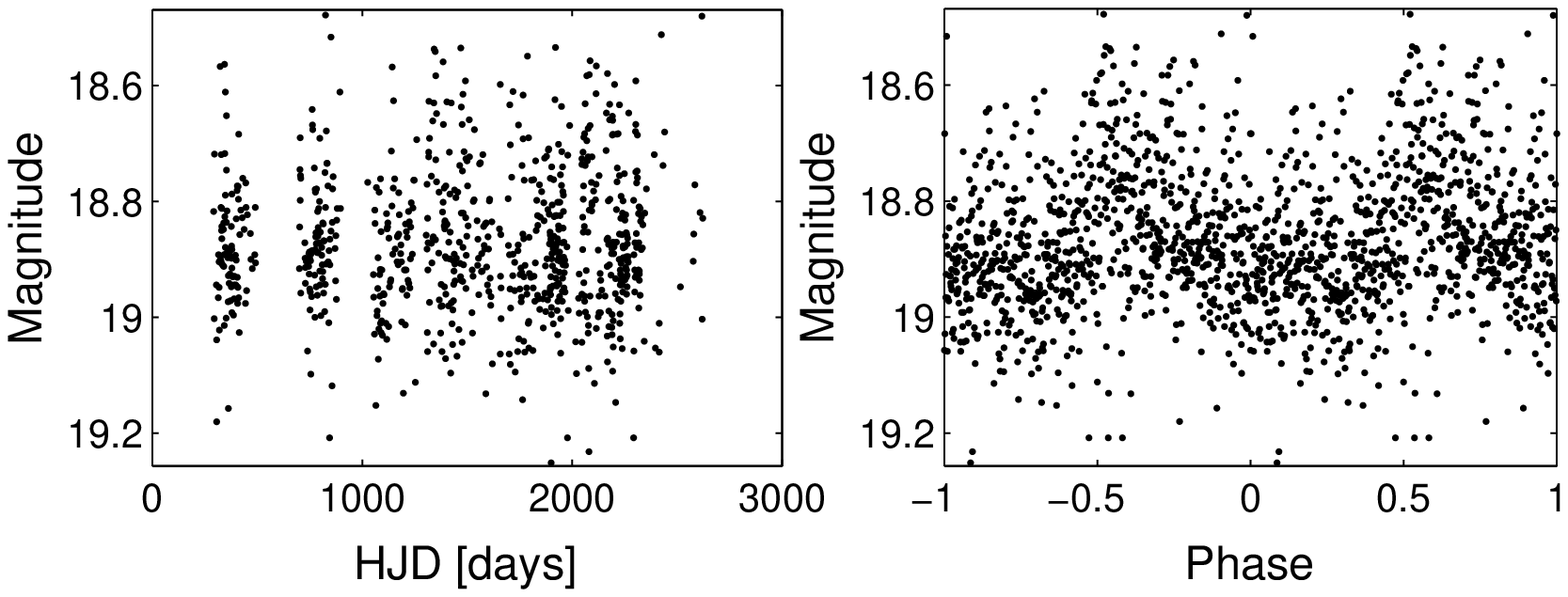}} \\ \vspace{-10pt}
	\subfloat[]{\label{fig-b1}\includegraphics[scale=0.49]{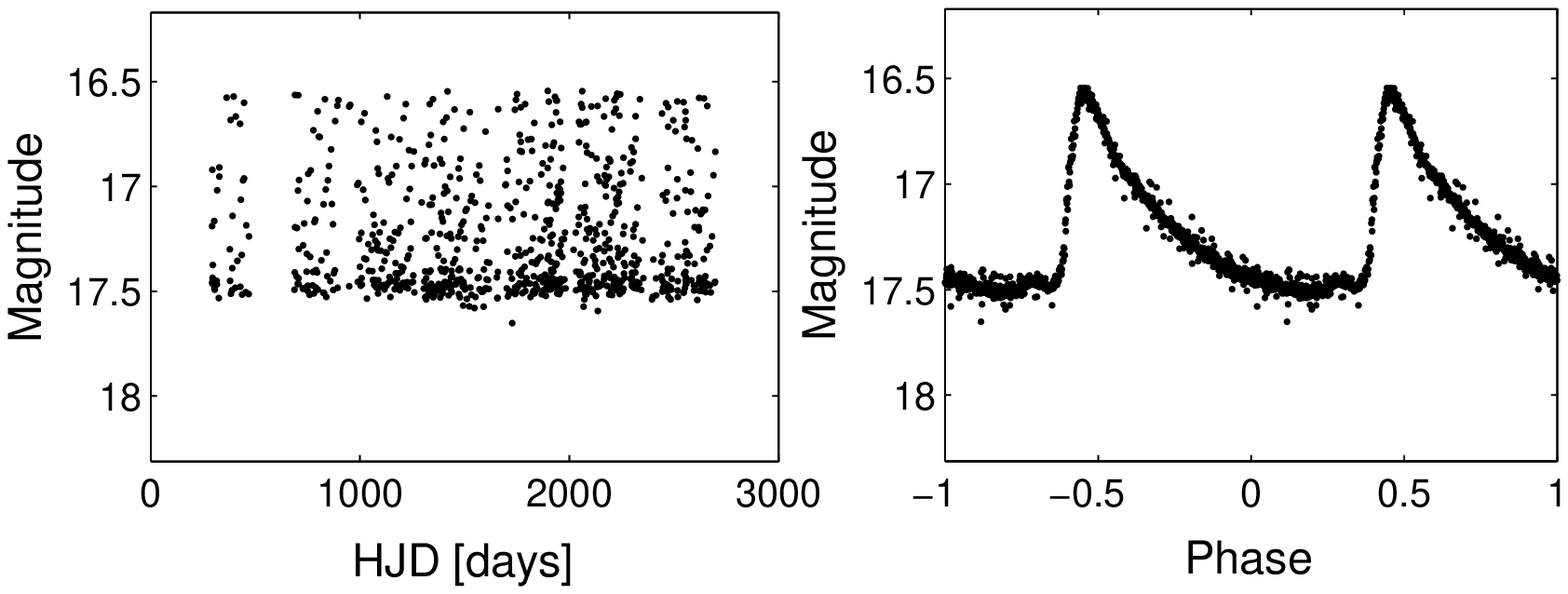}}  \\ \vspace{-10pt}
	\subfloat[]{\label{fig-b2}\includegraphics[scale=0.49]{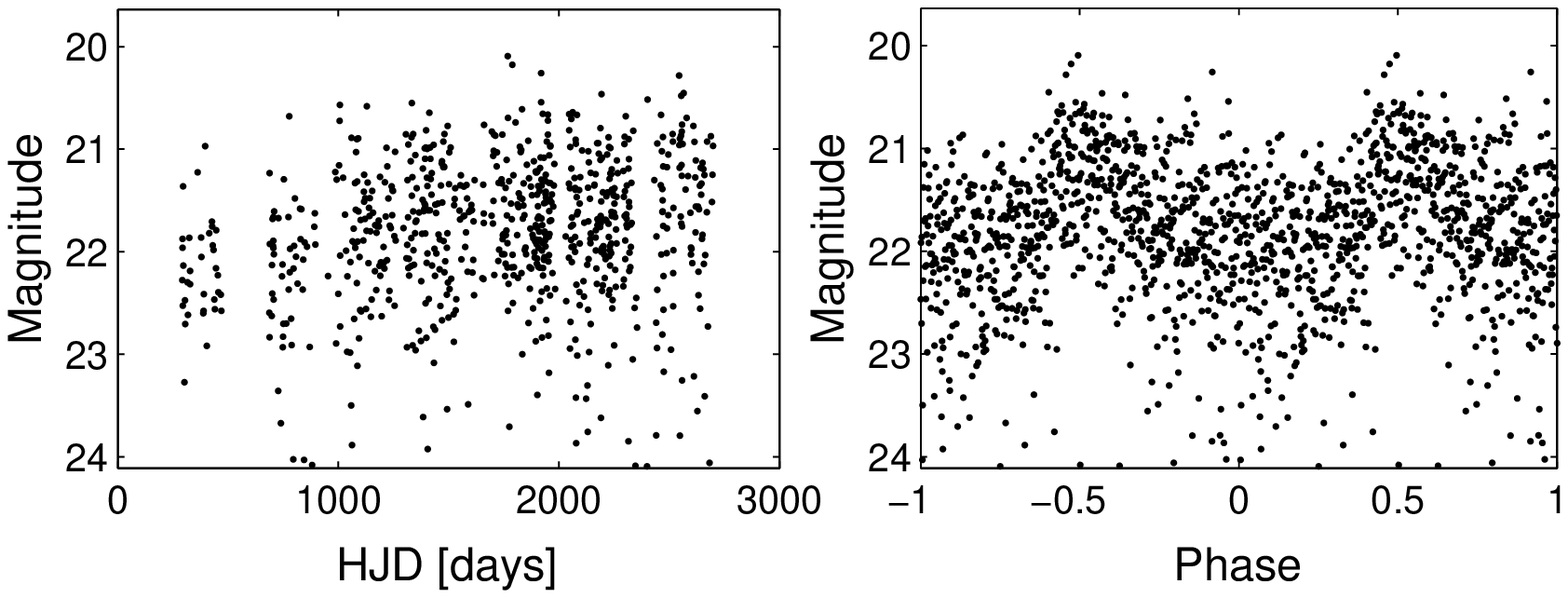}}  \vspace{-0pt}
	\caption{ \label{fig-exob} Examples of overlapping and blending. A period of 2.4796 days is detected for lightcurves sm0077n17908 (a) and sm0010k3199 (b). The angular distance between these lightcurves is 0.5$''$. Their difference in magnitude and CKP value is 0.03 and 0.23, respectively. These lightcurves are associated to a star that is in an overlapped region between fields sm001 and sm007. lightcurves sm0023n10183 (c) and sm0023n10325 (d) are also found to have the same period (1.2535 days), but they reside in the same field. Their angular distance, $\delta$-magnitude and $\delta$-CKP is 4.9$''$, 4.5 and 4.1, respectively. In this case the light from sm0023n10183 (c) introduces a periodicity in its neighbour (d).  }
\end{figure}

To select the `duplicate' lightcurves,  the nearest neighbor for each object in terms of angular distances is firstly identified. If the distance to the nearest neighbor is less than 10$''$ and both objects have the same period, then the lightcurve with the lowest magnitude is added to the discarded set. Using this criterion 2663 pairs of lightcurves are selected from the LMC. From this set 336 correspond to lightcurves that reside in different chips. The average delta magnitude in this set is 0.281 and the average delta CKP is 0.744. Each pair of lightcurves correspond to the same star which appears twice in the survey due to the overlapping in the observational fields. The other 2327 cases correspond to lightcurves that are neighbours in the same chip. The average delta magnitude in this set is 2.15 and the average delta CKP is 3.02, much higher than the previous set. In this set the more luminous star of the pair injects its periodicity in the lightcurve of the less luminous star (blending). Fig. \ref{fig-exob} shows an example of an overlapped pair and blended pair. It is interesting to note that a 72\% of the blended lightcurves are found in the fields within the LMC bar where the star density is the highest,  while the  overlapped lightcurves are equally distributed between bar and non bar fields.
In the SMC 1817 pairs of lightcurves are selected to be discarded. In this case 386 are due to field overlapping and 1431 are due to blending. The average delta magnitude in the overlapped lightcurves is 0.21 and the average delta CKP is 0.78. The average delta magnitude in the blended lightcurves is 2.34 and the average delta CKP is 4.86. The percentage of discarded lightcurves in the SMC is 7.2\% which is higher than the 2.3\% found in the LMC. This again attributed to the fact that SMC seeing is worst than LMC resulting into overlapping PSF which in turn into correlated lightcurves.

Fig \ref{fig-mapLMC} shows a map of the 88 fields of the LMC. The shaded fields correspond to the LMC bar. The percentage of periodic lightcurves is shown for each field  below its name. The fields corresponding to the LMC bar have a higher percentage of periodics. The percentage of periodics tends to drop the further the field is from the LMC bar. Fig \ref{fig-mapSMC} shows a map of the 10 fields of the SMC where the same pattern is apparent. Because the cores of the LMCs have older population of stars it is known that one would expect more periodic stars in those regions.

A grand total of 118,320 and 23,103 periodic lightcurves are found from the LMC and SMC blue channel data, respectively. Using the recall and precision from the training dataset we estimate that the true number of periodic lightcurves is 121,147 for LMC and 24,855 for the SMC. A 0.42\% of the lightcurves of the LMC are periodic and a 0.61\% of the lightcurves in the SMC are periodic.
%I put the number in the summary table too, the percentage of periodics were updated using the estimations of true number of periodics. 

\begin{figure}
	\centering
	\subfloat[]{\label{fig-clustersLMC}\includegraphics[scale=0.4]{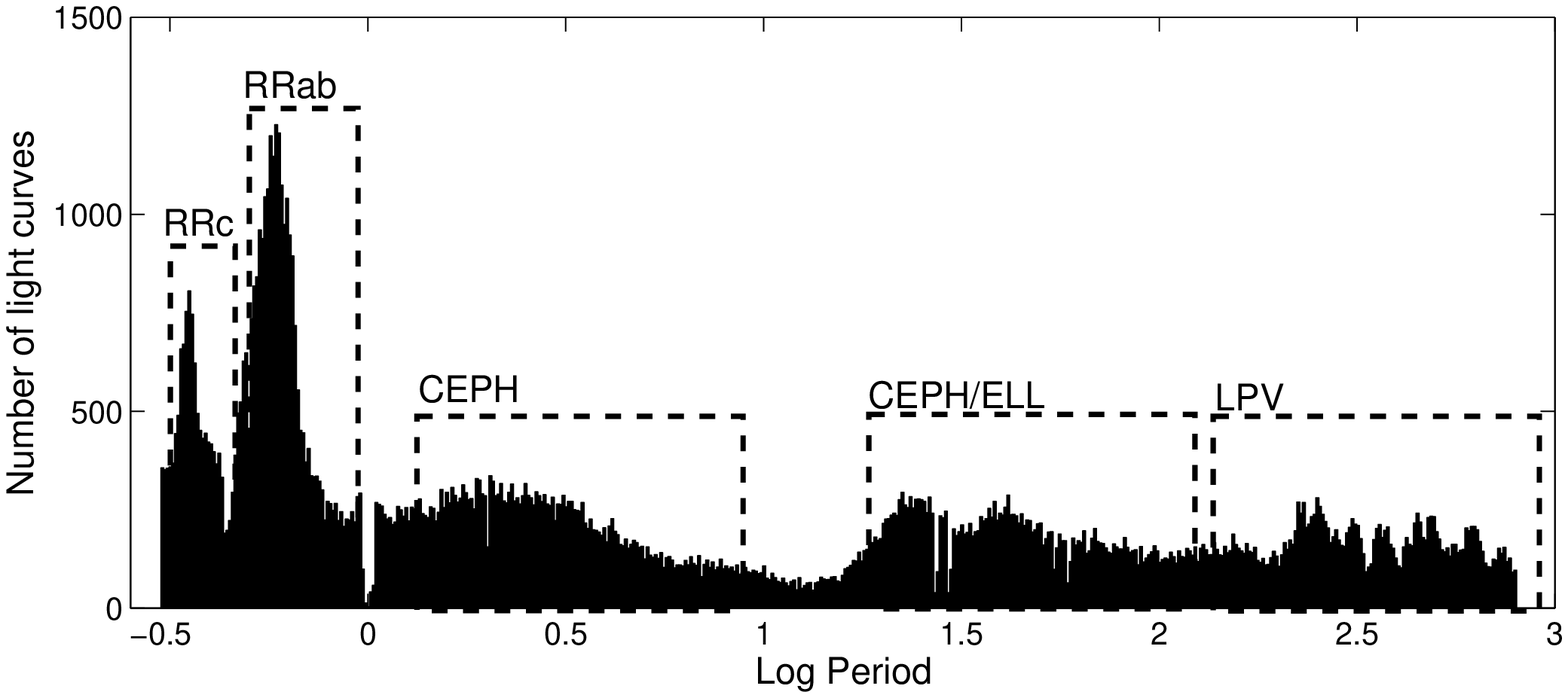} } \\
	\subfloat[]{\label{fig-clustersSMC}\includegraphics[scale=0.4]{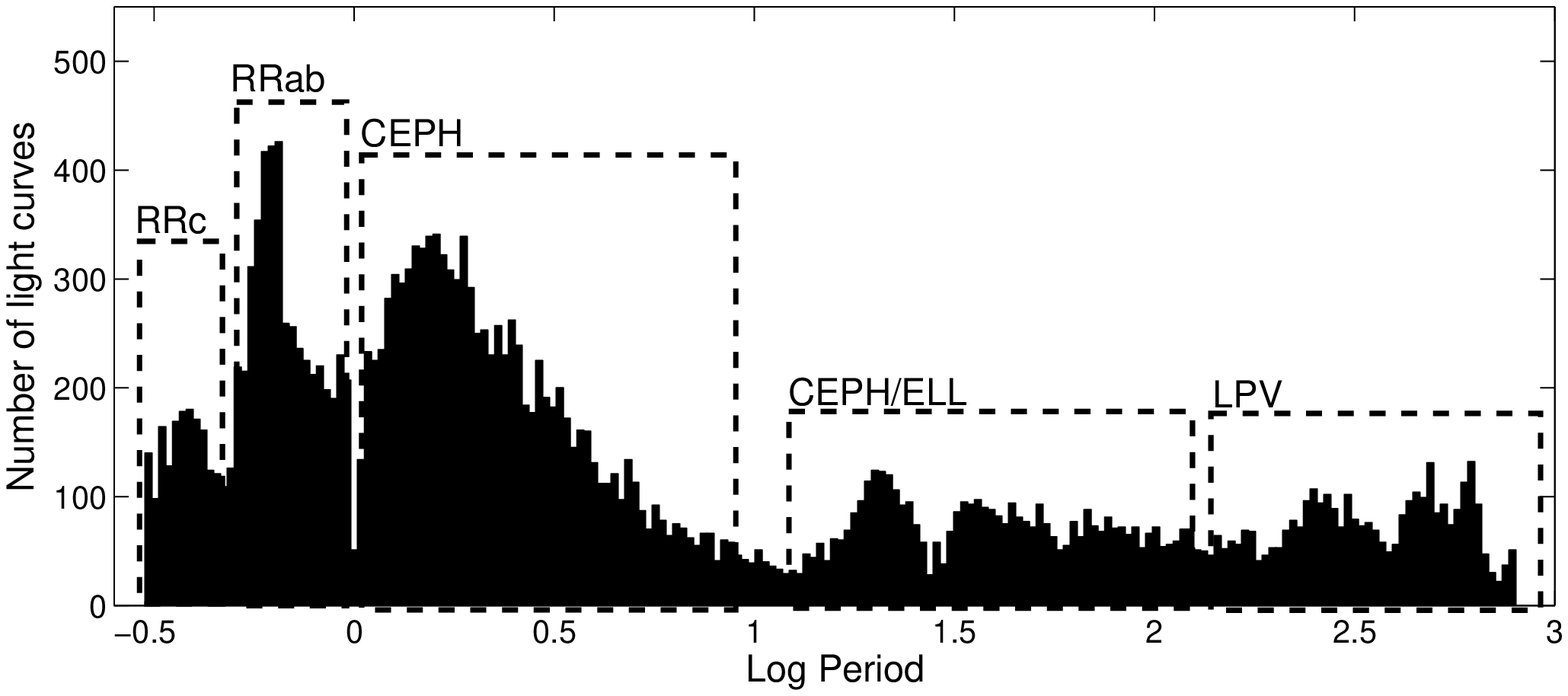} } 	
	\caption{ Histogram of the periods found in the LMC (a) and SMC (b) blue channel data. The regions marked with dotted boxes are associated to clusters of a given type periodic variable star. }
\end{figure}

Fig. \ref{fig-clustersLMC} shows the histogram of the periods found in the LMC blue channel data. Some of the known populations of periodic variables are identified in the histogram. The most notable populations correspond to c-type RR Lyrae (period centered in 0.3 days) and ab-type RR Lyrae (period centered in 0.6 days). These results are consistent with the RR Lyrae period histogram from the MACHO survey results on the LMC \citep{Cook1995}.

\begin{figure}
	\centering	
	\subfloat[]{\label{cmLMCBLUE} \includegraphics[scale=0.48]{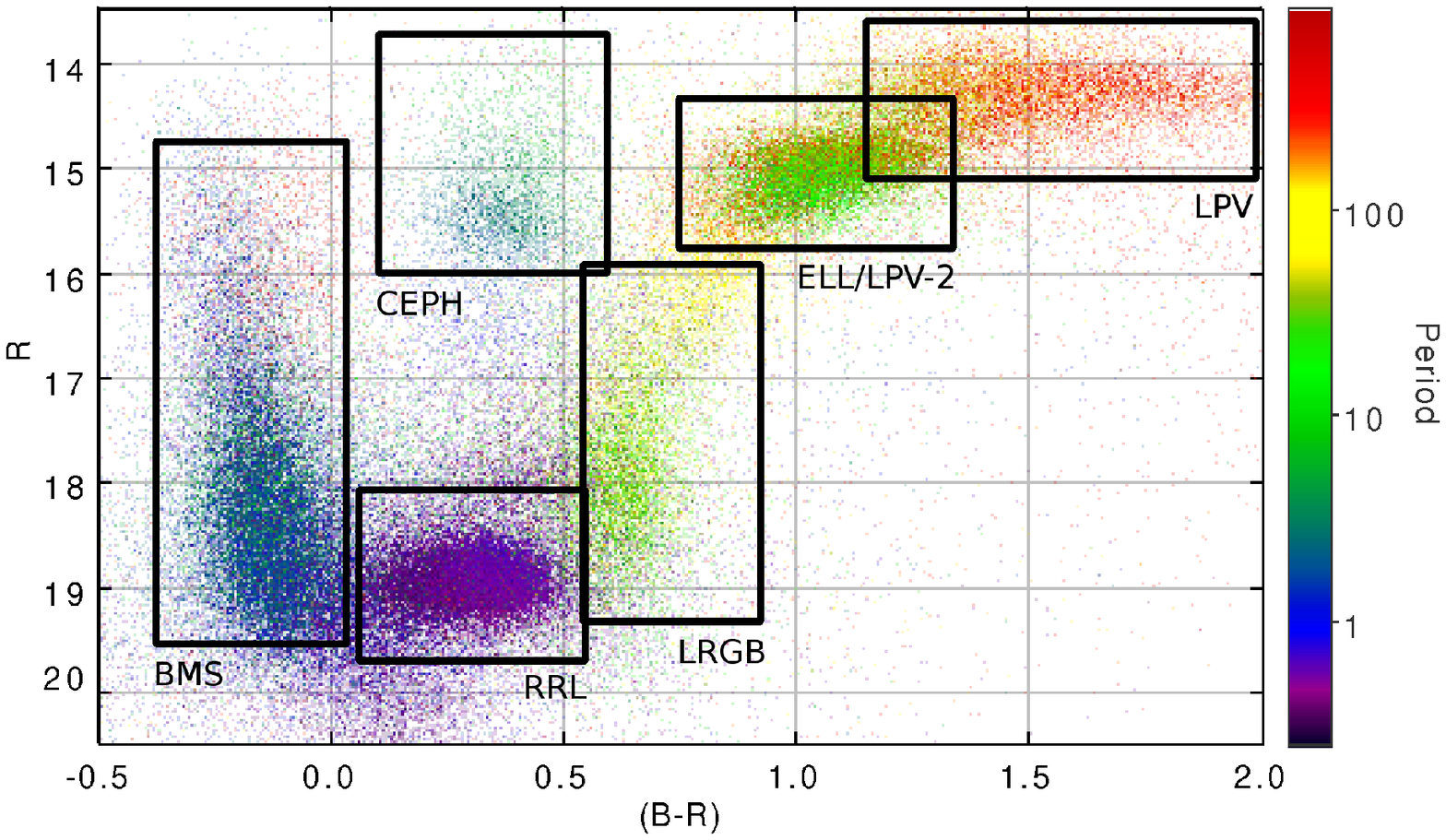}  } \\
	\subfloat[]{\label{cmSMCBLUE} \includegraphics[scale=0.48]{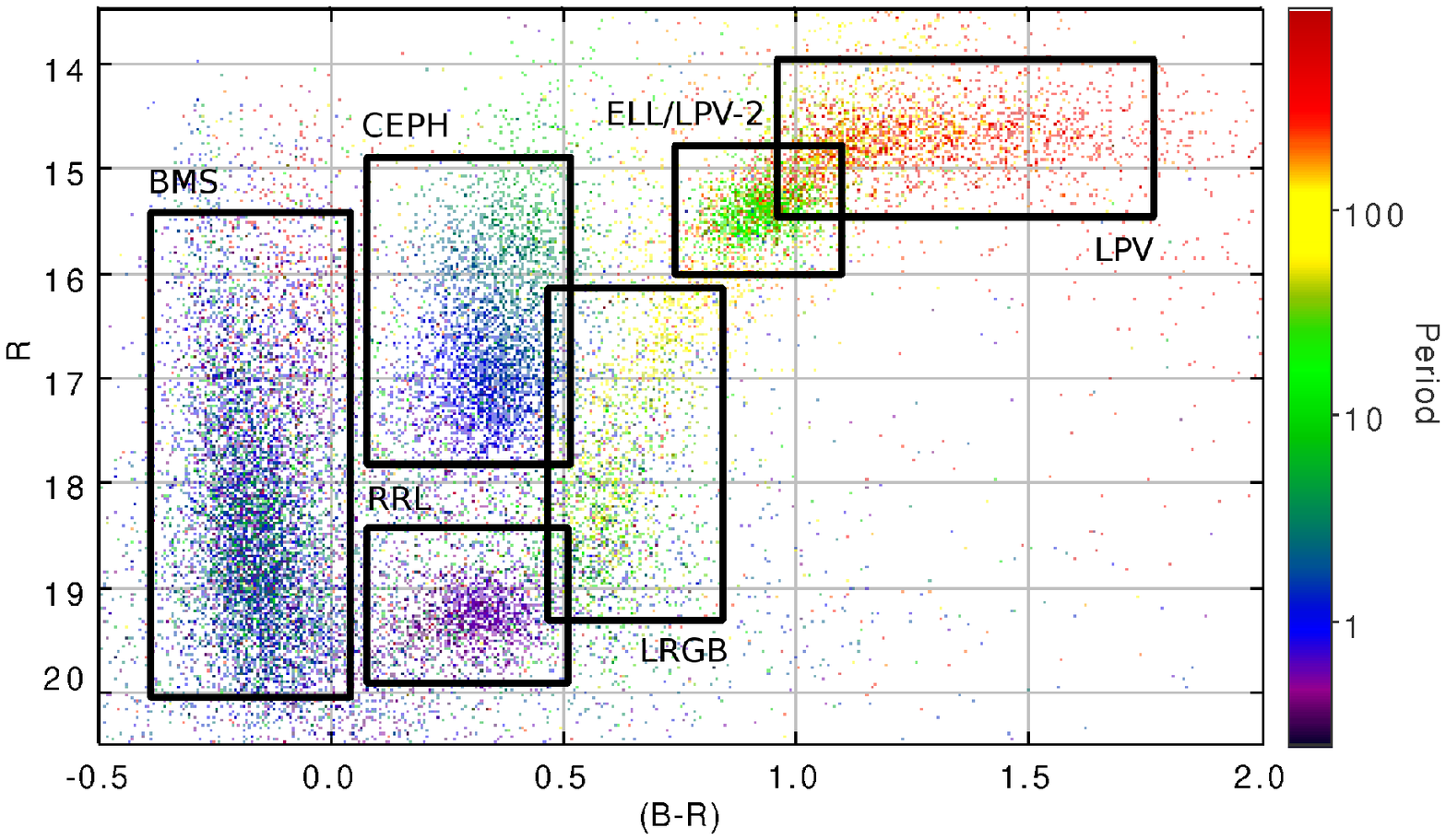} } 	
	\caption{ Color magnitude diagram showing the periodic lightcurves found in the LMC (a) and SMC (b). BMS corresponds to blue main sequence. LRGB corresponds to lower red giant branch. Black boxes mark the location of Cepheid, RR Lyrae, LPV and ellipsoidal variable populations.}
\end{figure}

Fig. \ref{cmLMCBLUE} shows a color magnitude diagram of the periodic lightcurves found in the LMC blue channel. The third axis corresponds to the detected period. The regions of interest are marked with black dotted squares. Examples of the periodic variable stars found in these regions are shown in Fig. \ref{fig-exp-puls} through  \ref{fig-exp-ell}. These results are consistent with the color magnitude diagram of the LMC periodic variables from the OGLE survey \citep{Spano2009}.

Fig. \ref{fig-clustersSMC} and \ref{cmSMCBLUE} show the histogram of periods and the color magnitude diagram of the periodic lightcurves found in the SMC blue channel, respectively. By comparing the histogram and color magnitude diagram with those of the LMC, the following differences arise: the relative size of the Cepheid population is larger in the SMC, the relative size of the c-type RR Lyrae population is larger in the LMC.

\begin{figure}
	\centering	\vspace{-0pt}
	\subfloat[]{\label{fig-ba1}\includegraphics[scale=0.47]{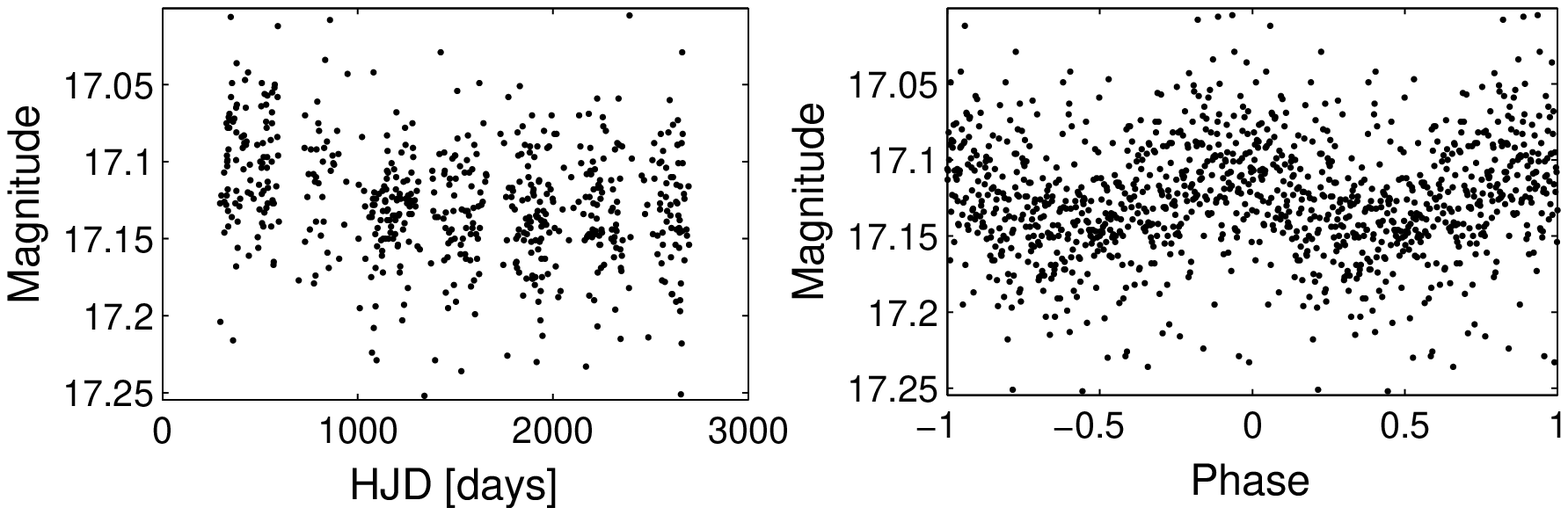}} \\ \vspace{-10pt}
	\subfloat[]{\label{fig-ba2}\includegraphics[scale=0.47]{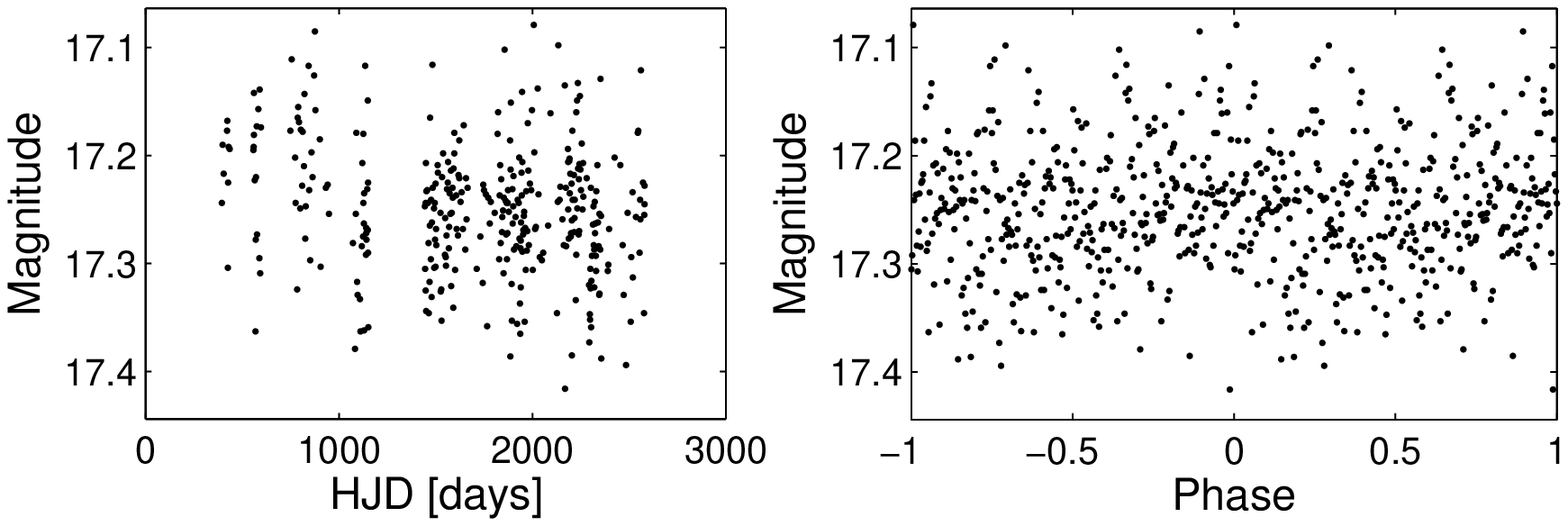}} \\ \vspace{-10pt}
	\subfloat[]{\label{fig-ba3}\includegraphics[scale=0.47]{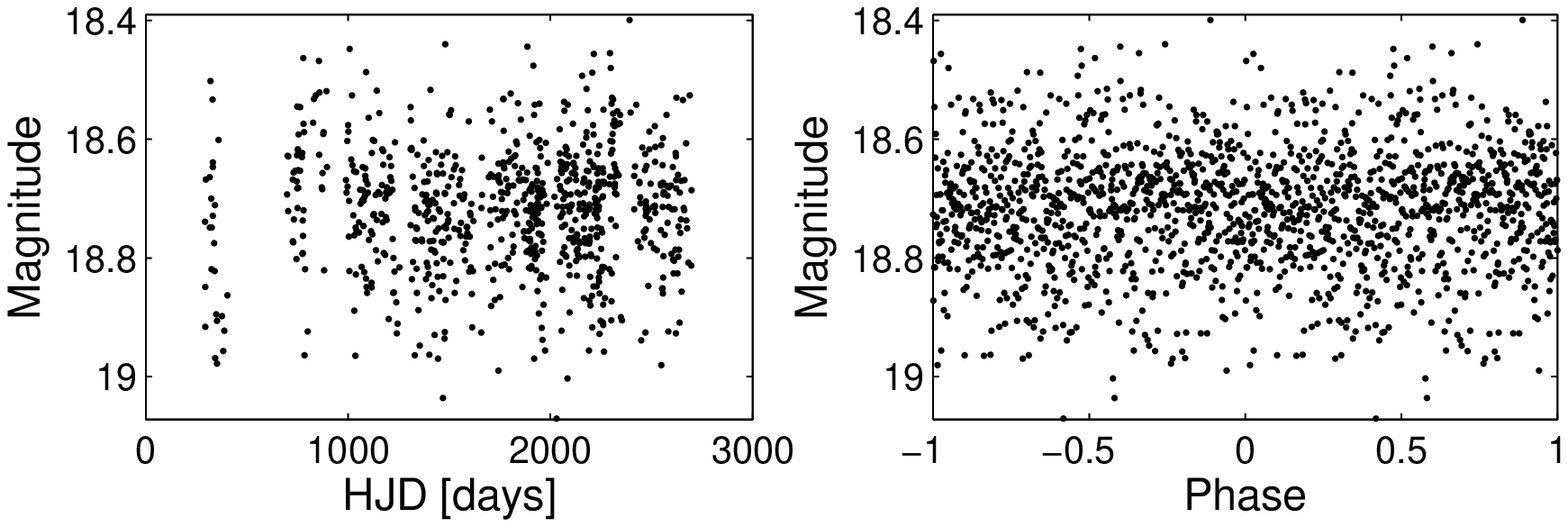}}  \\ \vspace{-10pt}
	\subfloat[]{\label{fig-ba4}\includegraphics[scale=0.47]{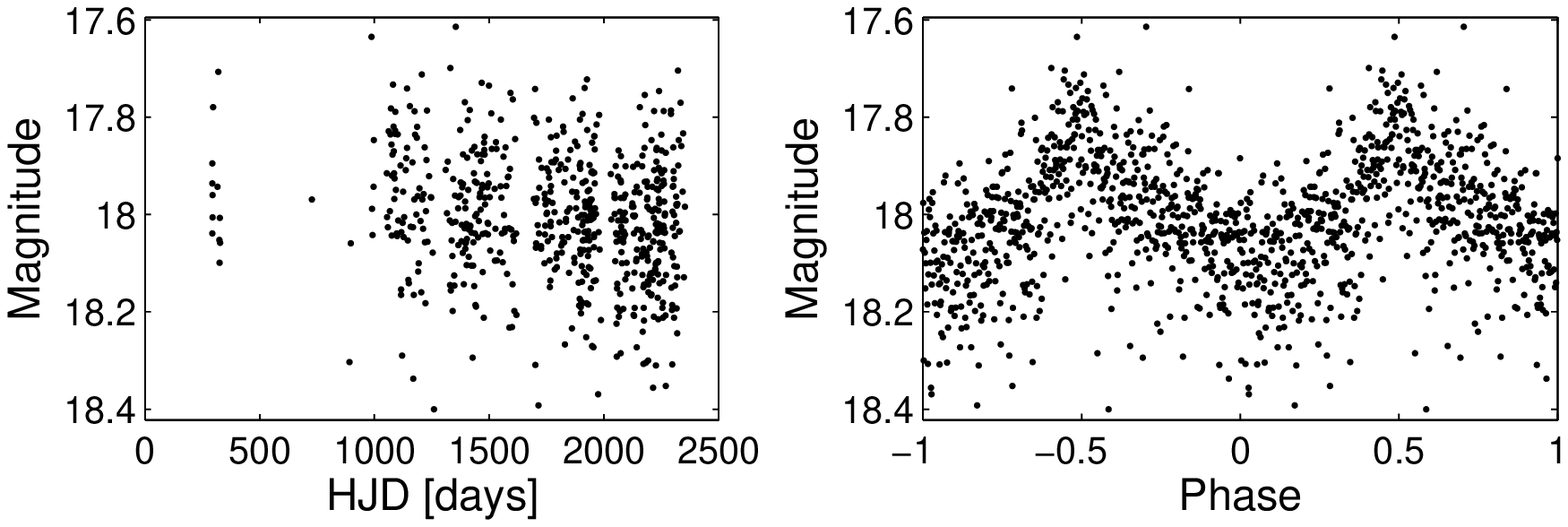}}  \vspace{-0pt}
	\caption{ \label{fig-exchan} Examples of periodic lightcurves detected only in one of the EROS-2 channels. Fig (a) and (b) correspond to lightcurve lm0012k17912. Fig (a) shows the blue channel lightcurve folded with the detected period of 0.48004 days. Using the red channel data no strong periodicity is found. Fig (b) shows red channel lightcurve folded with the 0.48004 days periods. Fig. (c) and (d) correspond to lightcurve sm0010l10270. Fig. (d) shows the red channel data folded with the detected period of 10.4453 days. Using the blue channel data no strong periodicity is found. Fig. (c) shows the blue channel data folded with the period detected in the red channel. }
\end{figure}

The red channel lightcurves are also analyzed for comparison purposes. A grand total of 87,025 and 14,501 periodic lightcurves are collected from the LMC and SMC red channel data, respectively. This represents a decrease of 30\% with respect to the amount of periodics collected from the blue channel. By cross-matching the lists obtained from the blue and red channels in the LMC we found that 68,179 objects appear in both lists, 50,141 objects are found only in the blue channel, and 18,846 objects are found only in the red channel. For the SMC, 12,536 objects appear in both lists, 1,965 appear exclusively in the red and 10,567 appear exclusively in the blue. For a given object the SNR may change between channels as shown in the examples of Fig. \ref{fig-exchan}. By inspecting the histogram of the color $(B - R)_{eros}$ of the EROS-2 lightcurves, it is clear that it is skewed to the blue side. The average color value in the LMC and SMC is 0.46 and 0.31, respectively and therefore the SNR is higher in the blue channel and therefore this  explains why more periodics are found in the blue channel data\footnote{Another reason could be related to the training scheme, in which only blue channel lightcurves where used to create the synthetic database.}.

\begin{table*}[t]
	\begin{center}
	\caption{Periodic lightcurve discrimination results summary on the EROS-2 survey.}  	
	\begin{tabular}{l c c c c c}
	\hline
	&  $N_{LC}$ & $\tilde{N}_p$ & Discarded &  $N_p$ & Periodics [\%] \\ \hline
	LMC & 28,797,305 & 120,983 & 2,663 & 121,147 & 0.42 \\
	SMC & 4,064,179 & 24,920 & 1,817  & 24,855  & 0.61 \\
	\hline
	\end{tabular}
	\label{tab:summaryMCp}
	\end{center} 
\end{table*}

The catalogs are compared with existing periodic variable star catalogs for the LMC and SMC. We first test against the published OGLE catalogs for Cepheids \citep{Soszynski2008-1,Soszynski2010-1}, type II Cepheids \citep{Soszynski2008-2,Soszynski2010-2}, RR Lyrae \citep{Soszynski2009-1,Soszynski2010-3} and LPV \citep{Soszynski2009-2,Soszynski2011} in the LMC and SMC. The OGLE team performed an extent period search using Fourier based methods, analysis of variance and visual inspection. In this test the objective is to reveal how many of the periodic variables reported by the OGLE team can be found in our catalogs and to analyze the discrepancies between the detected periods.  Table \ref{tab:crossmatching1} summarizes the results of the crossmatching. First, for each OGLE object, a nearest neighbor in the EROS catalog is found. Neighbors with a separation larger than 1.5 arcsec are not considered. Column $N_{inEROS}$ corresponds to the number of OGLE objects that were found in the EROS set within the search distance. The OGLE objects that did not have an EROS neighbor were either out of EROS bounds, located on inter-chip EROS zones or located on corrupted EROS chips. Column $N_{match}$ correspond to the number of crossmatched OGLE-EROS objects that appear in our periodic variable catalog. The differences between $N_{inEROS}$ and $N_{match}$ are due to OGLE objects whose CKP is below the periodicity threshold (low SNR light curves). There are cases in which the true period is within the spurious filters areas and was missed in our search. Finally the periods reported by OGLE are compared to the periods found with the our method. The agreement column corresponds to the percentage of lightcurves in which the OGLE period is equal to the period found in our catalog (a 1\% relative error is considered). The multiple column corresponds to the cases in which the reported period is either a multiple, sub-multiple or alias of the OGLE period. The disagreement column corresponds to the cases in which the reported period is not related to the OGLE period.

\begin{figure*}
	\centering	\vspace{0pt}
	\subfloat[]{\label{fig:crossmatch_example11}\includegraphics[scale=0.55]{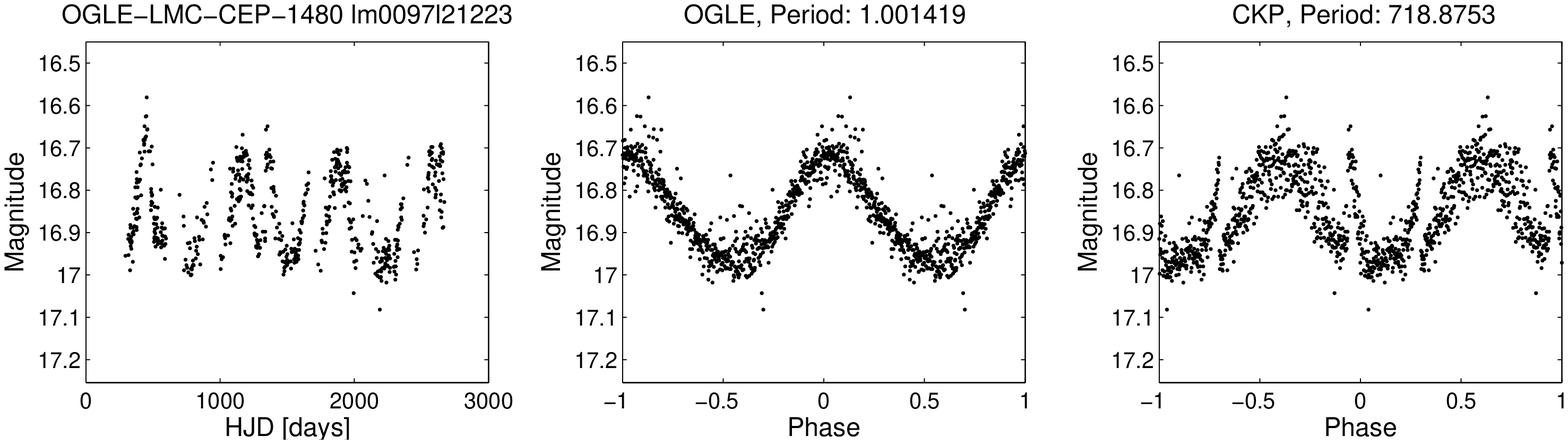}} \\ \vspace{-5pt}
	\subfloat[]{\label{fig:crossmatch_example12}\includegraphics[scale=0.55]{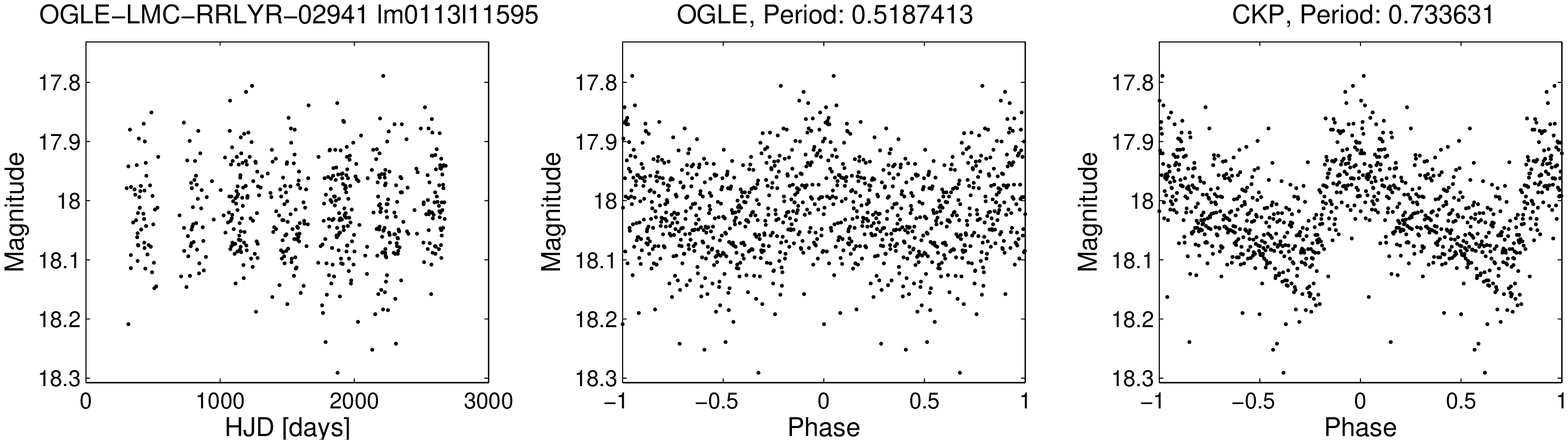}} \\ \vspace{-5pt}
	\subfloat[]{\label{fig:crossmatch_example13}\includegraphics[scale=0.55]{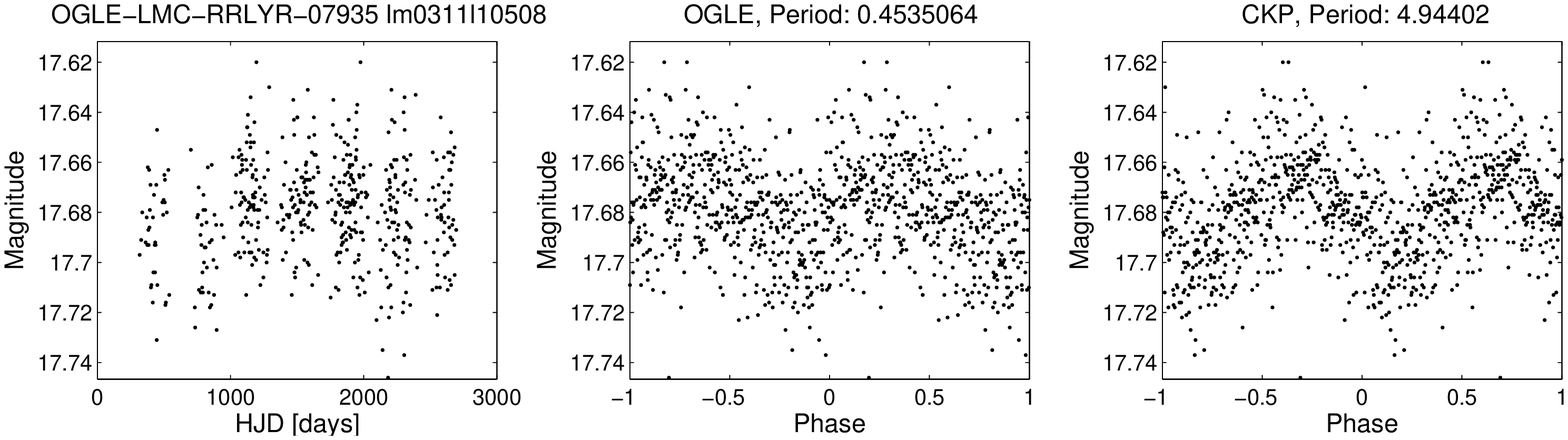}}  \\ \vspace{-5pt}
	\subfloat[]{\label{fig:crossmatch_example14}\includegraphics[scale=0.55]{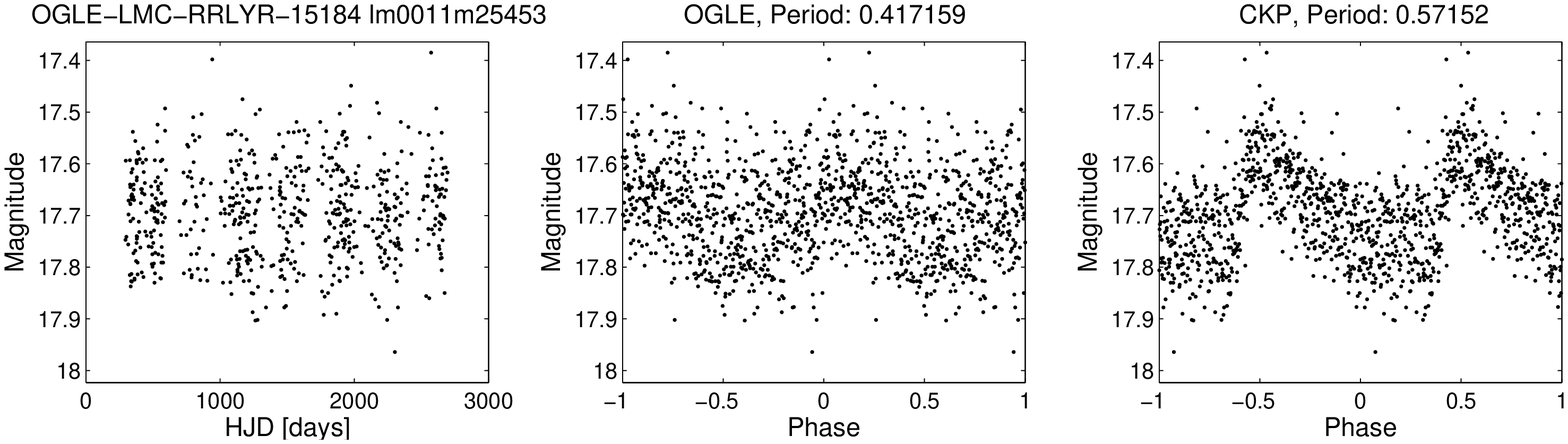}}  \vspace{-5pt}
	\caption{ \label{fig:crossmatch_example1} Light curves in which the reported period is in disagreement with the OGLE period. The EROS and OGLE labels, along the periods are shown in the title of each light curve.  }
\end{figure*}

There is a high level of agreement between the reported and OGLE periods for Cepheids, type II Cepheids and RR Lyrae classes, in both the LMC and SMC. The periods labeled as multiples were visually inspected. In these cases the OGLE period is the correct period, but it was not found by the proposed method because it was either below 0.3 days or filtered in the spurious period rejection stage. Examples of the lightcurves in which the reported period is in disagreement with the OGLE period are shown in Fig. \ref{fig:crossmatch_example1}.

%Explain what we do when OGLE has many periods!: we use the one with lowest error to ours

\begin{figure*}
	\centering	\vspace{0pt}
	\subfloat[]{\label{fig:crossmatch_example21}\includegraphics[scale=0.55]{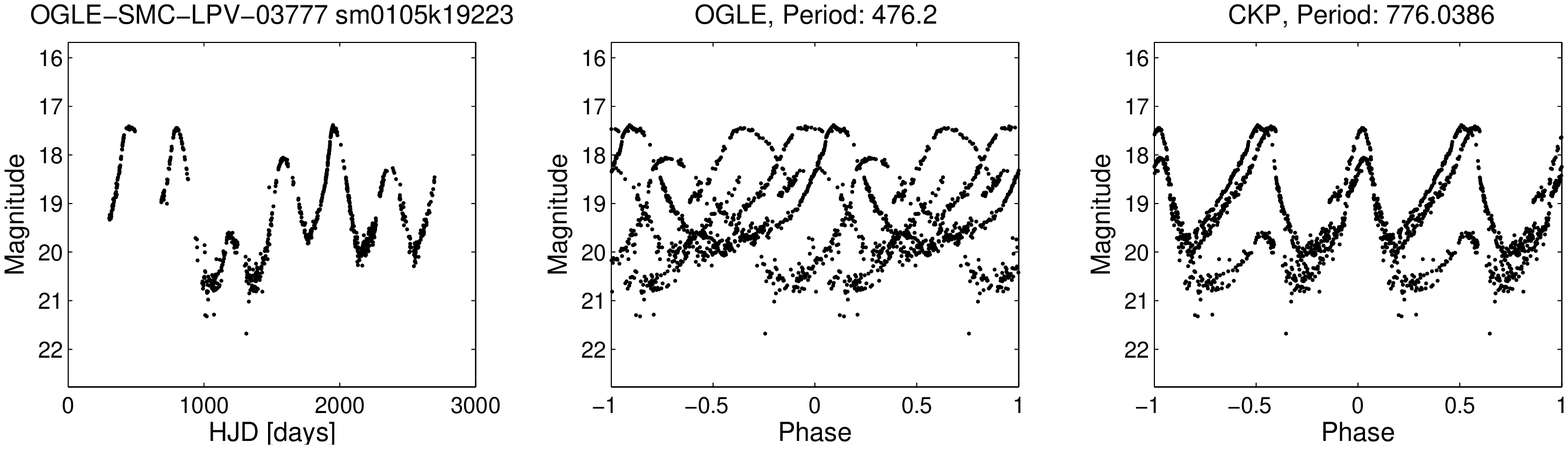}} \\ \vspace{-5pt}
	\subfloat[]{\label{fig:crossmatch_example22}\includegraphics[scale=0.55]{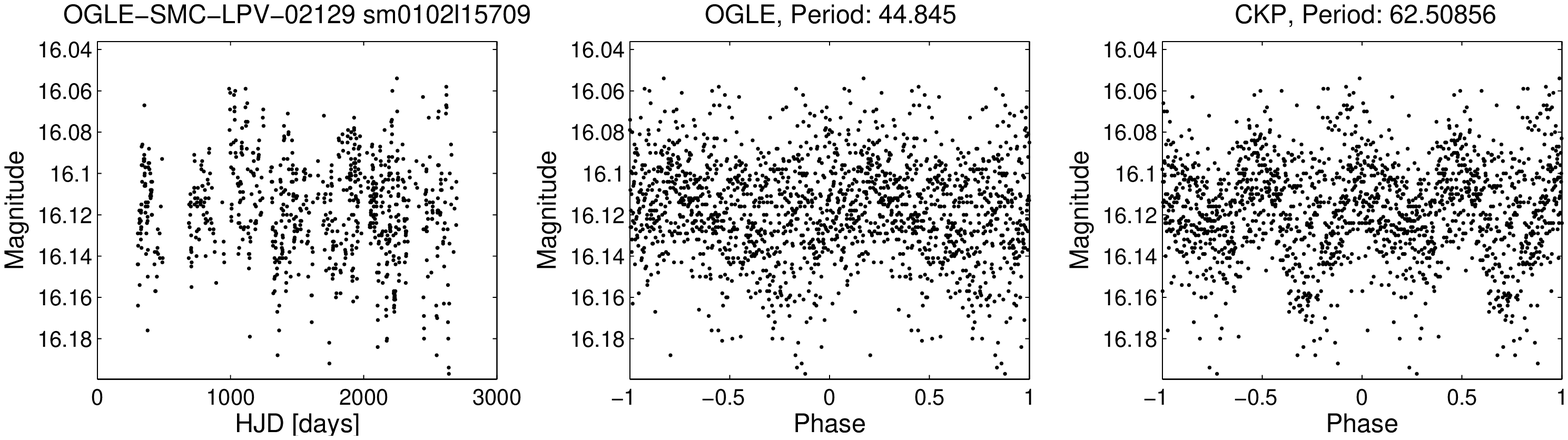}} \\ \vspace{-5pt}
	\subfloat[]{\label{fig:crossmatch_example23}\includegraphics[scale=0.55]{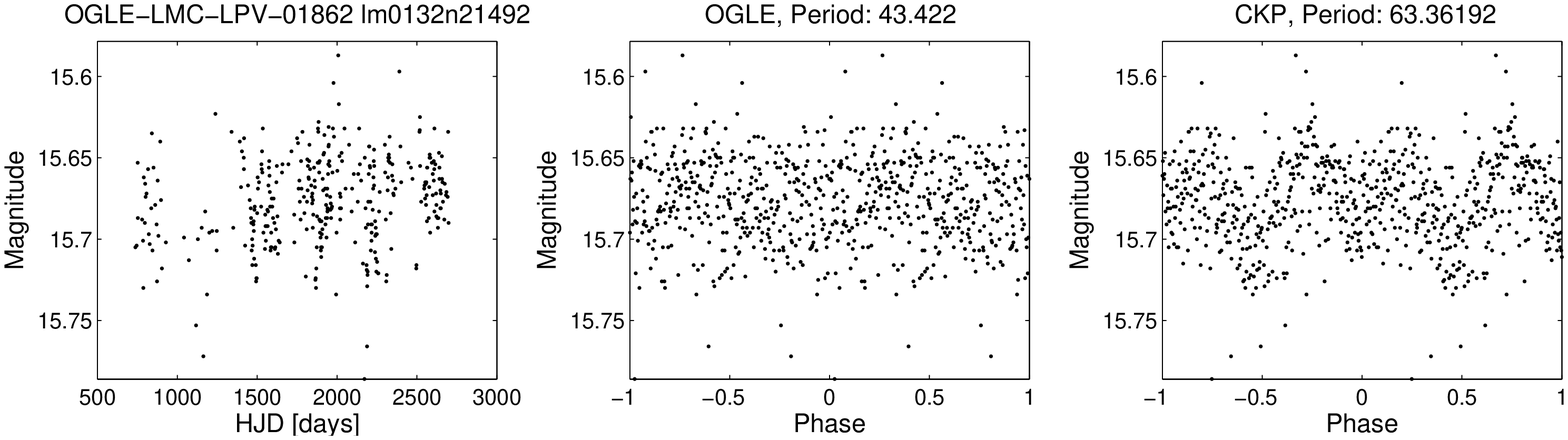}}  \\ \vspace{-5pt}
	\subfloat[]{\label{fig:crossmatch_example24}\includegraphics[scale=0.55]{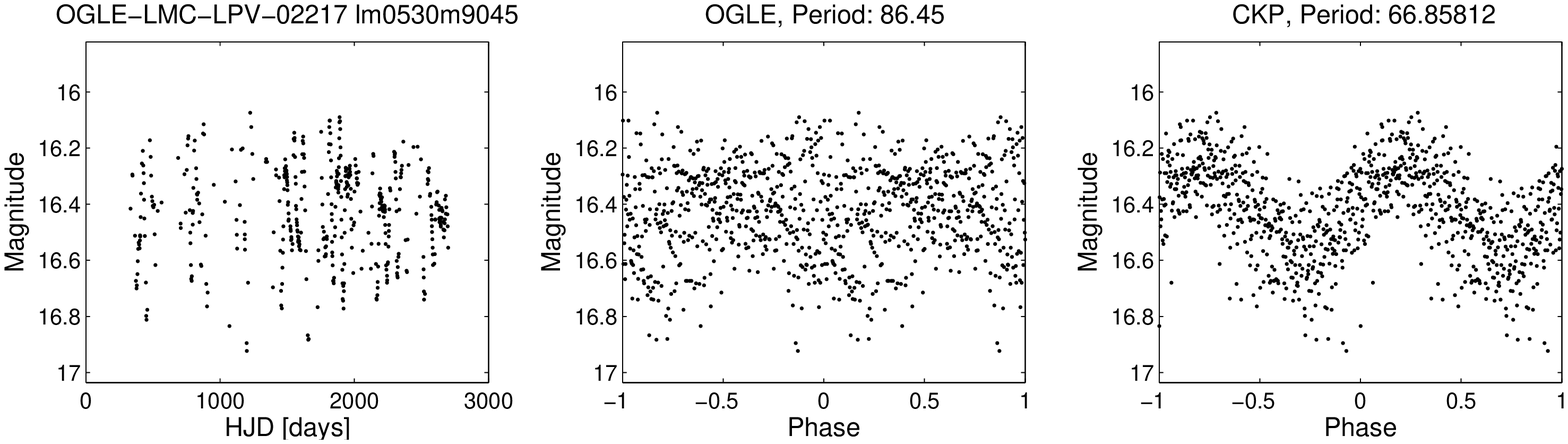}}  \vspace{-5pt}
	\caption{ \label{fig:crossmatch_example2} Examples of LPVs in which the reported period is in disagreement with the OGLE period. The EROS and OGLE labels, along the periods are shown in the title of each light curve.} %According to OGLE, Figures (a) and (b) corresponds to the OSARG class, Fig. (c) is a semiregular variable and Fig. (d) is a Mira variable }
\end{figure*}

For the LPV class the difference between $N_{inEROS}$ and $N_{match}$ is larger than in other classes (i.e. more objects with CKP below periodicity threshold). This is expected as the LPVs are known to suffer from irregularities that affect their period. Additionally, the level of agreement between periods is lower than the other classes. Fig. \ref{fig:crossmatch_example2} shows examples of disagreeing periods in the LPV class.

%PH: Some examples with disagreeing periods do not manifest the period we found in the OGLE light curve (I-band)!!!! Are these guys truly the same object? Can we blame the channel frequencies? I check the R-channel of the OGLE light curves too, but that channel has too few points to get a clear folded lightcurve. 

\begin{figure}
	\centering	\vspace{-0pt}
	\subfloat[]{\label{fig:crossmatch_example31}\includegraphics[scale=0.45]{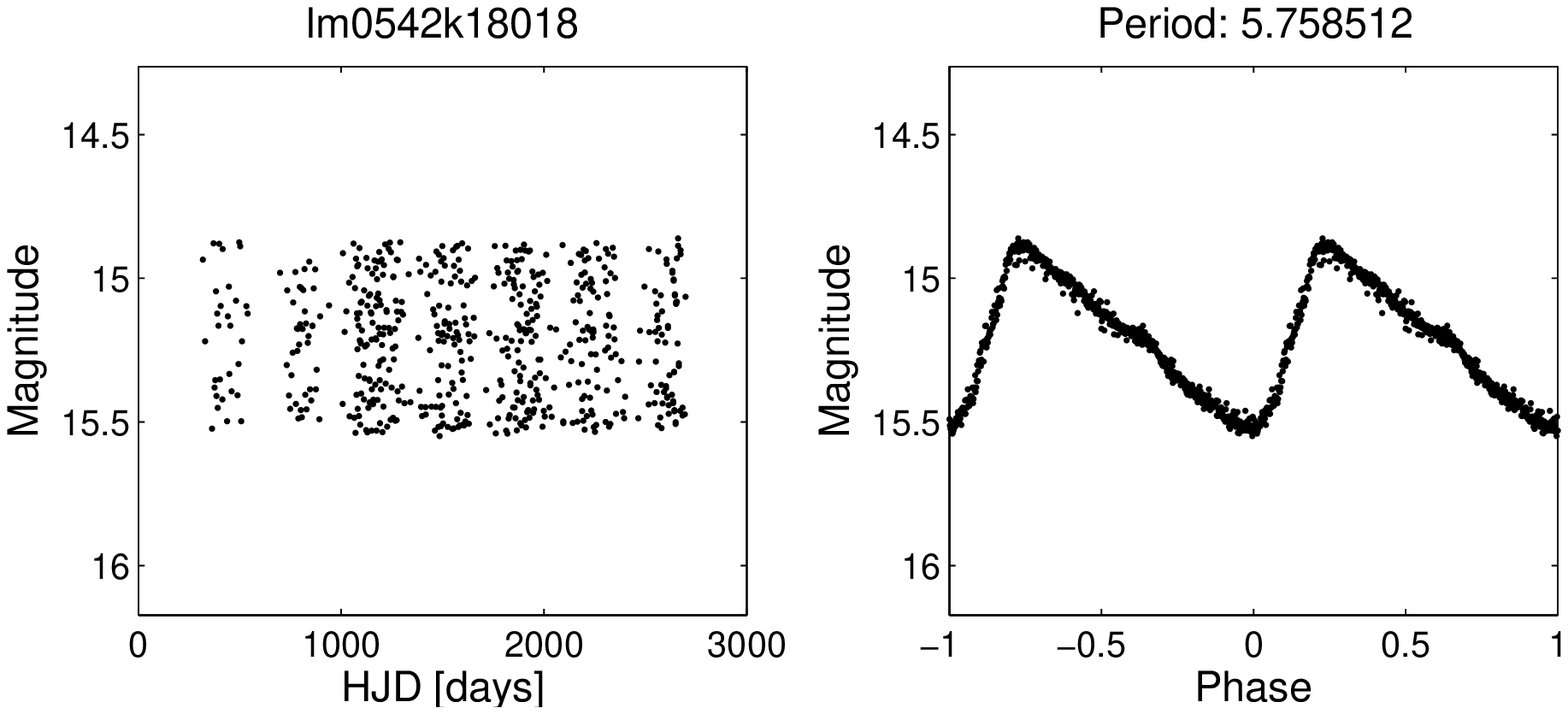}} \\ \vspace{-5pt}
	\subfloat[]{\label{fig:crossmatch_example32}\includegraphics[scale=0.45]{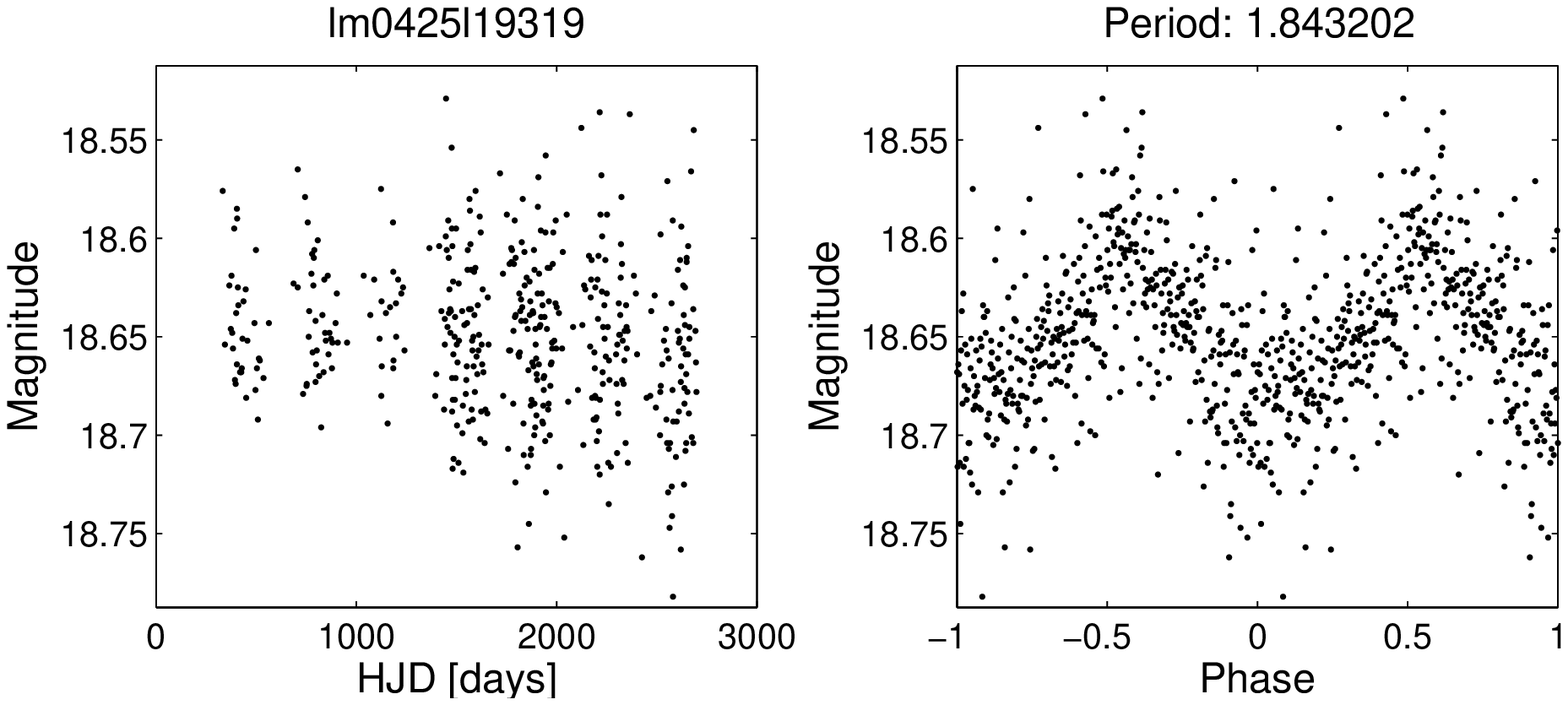}} \\ \vspace{-5pt}
	\subfloat[]{\label{fig:crossmatch_example33}\includegraphics[scale=0.45]{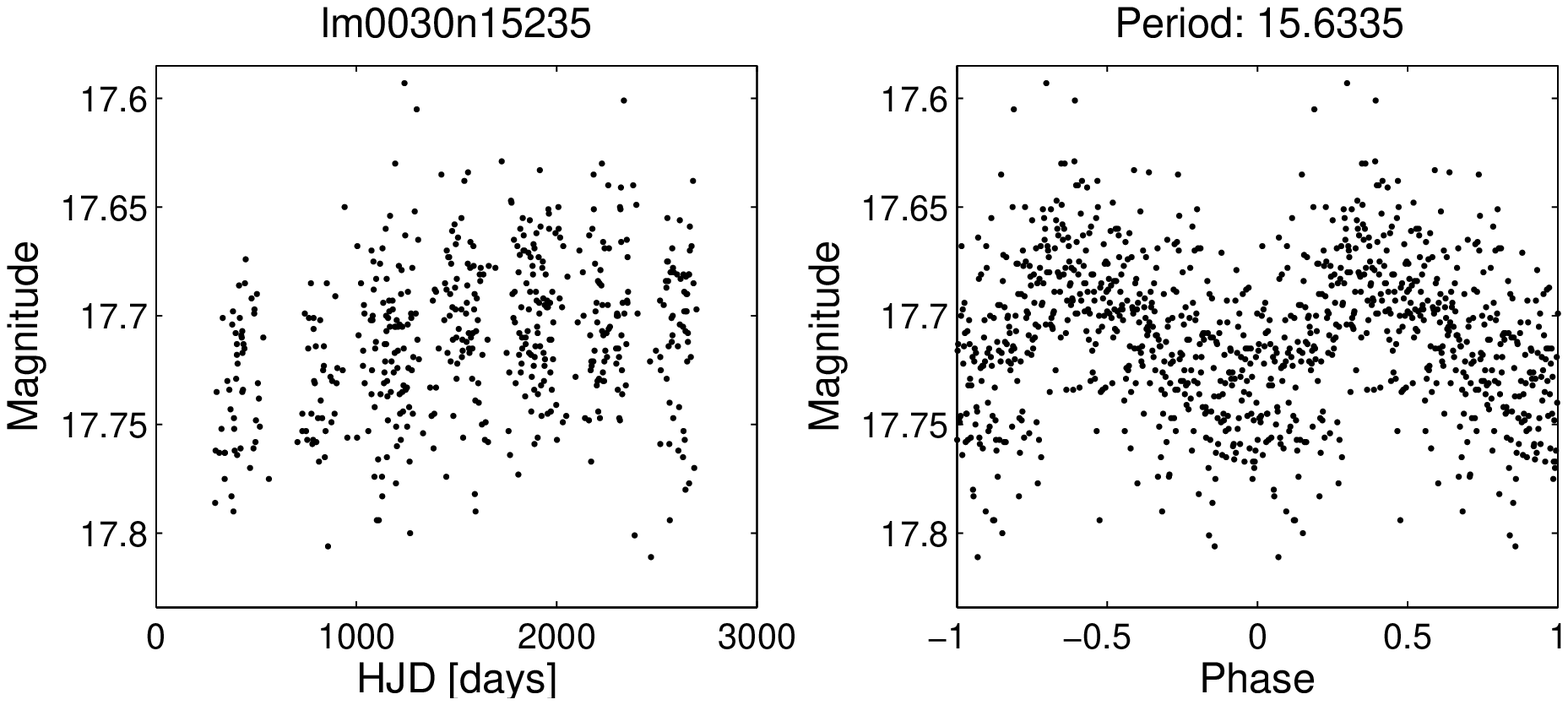}}  \\ \vspace{-5pt}
	\subfloat[]{\label{fig:crossmatch_example34}\includegraphics[scale=0.45]{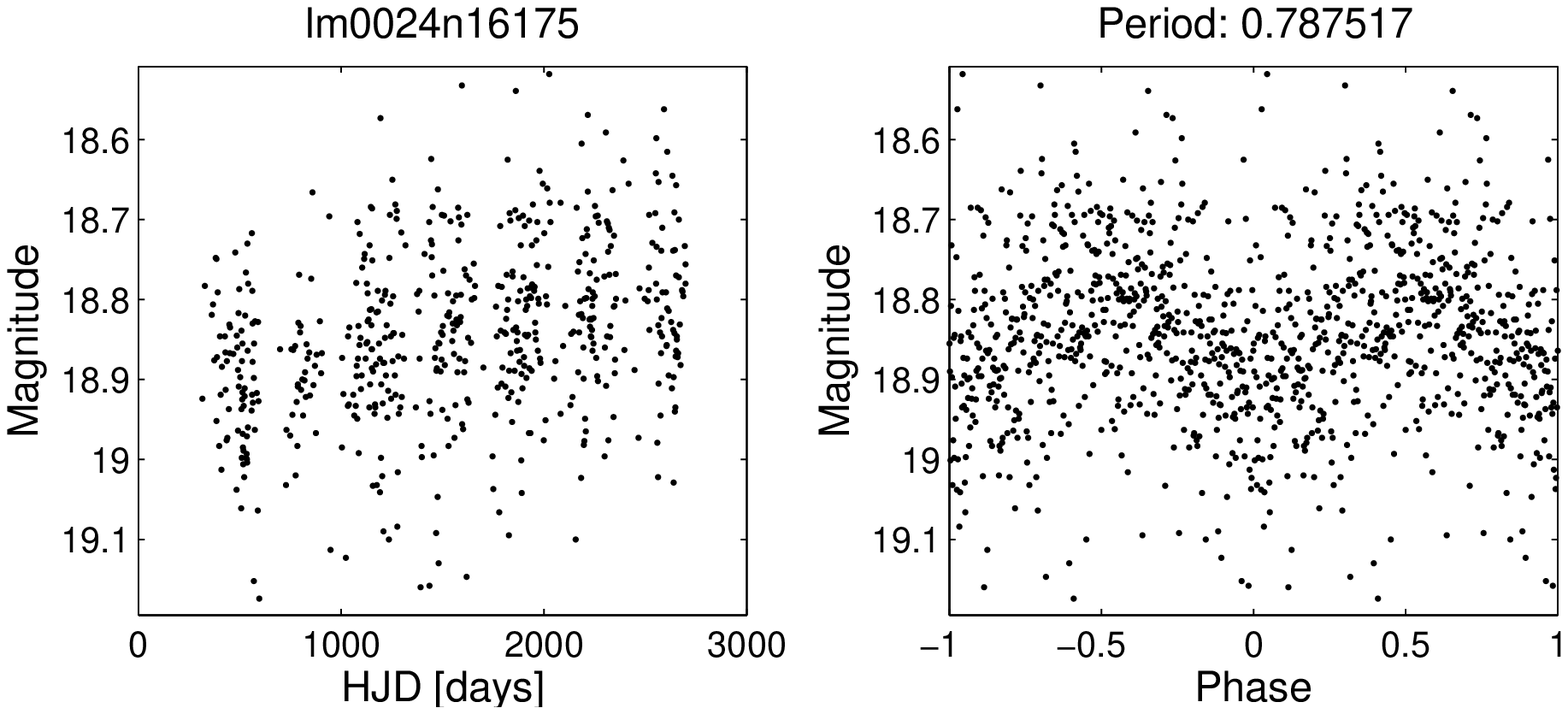}}  \vspace{-5pt}
	\caption{ \label{fig:crossmatch_example3} Examples of periodic light curves not found by OGLE. Fig (a) corresponds to a Cepheid variable with high SNR not found by OGLE. The majority of these light curves have a low CKP value which translates roughly to low SNR.  Figures (b), (c) and (d) are low SNR examples.  }
\end{figure}

There are 80,304 objects in our periodic catalog that do not have a neighbor from the OGLE periodic variable catalogs (within 2.5 arcsec).  Some of these objects may have not been surveyed by the OGLE project, or they could belong to classes with currently not available catalogs such as eclipsing binaries. A 60\% of these light curves have a low CKP value which translates roughly to low SNR. This could indicate that the proposed method is more sensitive than the method used by the OGLE team. Fig. \ref{fig:crossmatch_example3} shows examples of periodic light curves found in the EROS catalog that do not appear in the OGLE catalogs. 

%PH: The thing is there are still plenty of cepheids in these 80k. Shall we show some examples and talk about this? Of course these cepheids may be out of the OGLE survey bounds, or be inter OGLE chips or be on dead OGLE chips, so I'm not quite sure what to discuss and how to present this...

%PH: is it worth showing 4 light curves out of 80k? shall we been talking about this issue?

The periodic variable catalogs are also compared to the lists of beat Cepheids found in the EROS-2 data by \citet{Marquette2009}. The catalog contain Cepheids pulsating on their fundamental and first overtone (F/FO) and first and second overtone (FO/SO), respectively. The periods were obtained using a combination of Fourier decomposition, Analysis of Variance and visual inspection. The results are summarized in Table \ref{tab:crossmatching2}. There are eight cases that do not appear in our catalog due to their CKP value being below the threshold. In the remaining 409 cases, only three cases show disagreement with the reported period. The one case in which the period is not a multiple of the EROS-2 period was shown in Fig. \ref{fig:crossmatch_example11}.

%\begin{table*}[t]
%	\begin{center}
%	\caption{Crossmatching with OGLE periodic variable catalogs in the LMC and SMC.}  	
%	\begin{tabular}{l c c c c c c}
%	\hline
%	OGLE catalog&  $N_{catalog}$ & $N_{inEROS}$ & $N_{match}$ & Agree [\%] & Multiple [\%]& Disagree [\%]\\ \hline
%	OGLE-LMC-CEPH	&3,375	&2,727	&2,711	&2,679	&28	&4	\\
%	OGLE-LMC-t2CEPH	&203	&161	&148	&140	&6	&2	\\
%	OGLE-LMC-RRLyr	&24,906	&18,092	&17,272	&15,903	&1,167	&202	\\ %according to EROS 'isVar' only 12309 are variable
%	OGLE-LMC-LPV	&91,995	&74,960	&20,430	&15,767	&407	&4,256	\\ %SRV 50% disagreement!, MIRA 10%, OSARG, 15%
%	OGLE-SMC-CEPH	&4,630	&3,413	&3,395	&3,371	&19	&5	\\
%	OGLE-SMC-t2CEPH	&43	&30	&30	&28	&1	&1	\\
%	OGLE-SMC-RRLyr	&2,475	&1,392	&1,360	&1,329	&23	&8	\\
%	OGLE-SMC-LPV	&19,384	&14,103	&4,413	&3,103	&113	&1,197	\\
%	\hline
%	\end{tabular}
%	\label{tab:crossmatching1}
%	\end{center} 
%\end{table*}

\begin{table*}[t]
	\begin{center}
	\caption{Crossmatching with OGLE periodic variable catalogs in the LMC and SMC.}  	
	\begin{tabular}{l c c c c c c}
	\hline
	OGLE catalog&  $N_{catalog}$ & $N_{inEROS}$ & $N_{match}$ & Agree [\%] & Multiple [\%]& Disagree [\%]\\ \hline
	OGLE-LMC-CEPH	&3,375	&2,727	&2,711	&98.8	&1.0	&0.2	\\
	OGLE-LMC-t2CEPH	&203	&161	&148	&94.6	&4.1	&1.3	\\
	OGLE-LMC-RRLyr	&24,906	&18,092	&17,272	&92.0	&6.8	&1.2	\\ %according to EROS 'isVar' only 12309 are variable
	OGLE-LMC-LPV	&91,995	&74,960	&20,430	&77.2	&2.0	&20.8	\\ %SRV 50% disagreement!, MIRA 10%, OSARG, 15%
	OGLE-SMC-CEPH	&4,630	&3,413	&3,395	&99.3	&0.6	&0.1	\\
	OGLE-SMC-t2CEPH	&43		&30		&30		&93.4	&3.3	&3.3	\\
	OGLE-SMC-RRLyr	&2,475	&1,392	&1,360	&97.7	&1.7	&0.6	\\
	OGLE-SMC-LPV	&19,384	&14,103	&4,413	&70.3	&2.6	&27.1	\\
	\hline
	\end{tabular}
	\label{tab:crossmatching1}
	\end{center} 
\end{table*}

\begin{table*}[t]
	\begin{center}
	\caption{Crossmatching with EROS-2 beat Cepheid catalogs for the LMC and SMC.}  	
	\begin{tabular}{l c c c c c}
	\hline
	Beat Cepheids catalog&  $N_{catalog}$ & $N_{match}$ & Agree [\%] & Multiple [\%]& Disagree [\%]\\ \hline
	F/FO pulsation	&115	&109	&100.0	&0.0	&0.0	\\
	FO/SO pulsation	&302	&300	&99.0	&0.66	&0.33	\\
	\hline
	\end{tabular}
	\label{tab:crossmatching2}
	\end{center} 
\end{table*}

\section{Beyond CKP}
\subsection{Multimodes}
It is known that periodic stars exhibit multimode oscillations which  is manifested in the morphology of the lightcurves. Despite the fact that the methodology presented in this paper was not designed to find multimodes, we have explored the multimodes in a two level search approach. For each periodic lightcurve the prime lightcurve $P_0$ is used to `remove' the periodic signal. This procedure is known as whitening and is performed as follows:
\begin{enumerate}
\item Fold the light curve with $P_0$.
\item Obtain a template of the periodicity by smoothing the folded light curve using a moving average of 30 samples.
\item Subtract the template from the folded light curve. 
\item Rearrange the light curve samples to their original time order.
\end{enumerate}
If the whitened lightcurve is found to be periodic with period $P_1$, that  is not multiple/sub-multiple or alias of $P_0$, then the light curve is selected as a dual mode candidate. Subsequent oscillation modes can be found by repeating the procedure above. 

%
%\begin{enumerate}
%
%
%\item Evaluate the light curve with the pipeline described in Section \ref{pipeline}. If the light curve is non-periodic, discard it. If the light curves is periodic, save the period as $P_0$. 
%\item Subtract the periodicity found in the previous step from the light curve. This procedure is known as whitening and is performed as follows
%\begin{enumerate}
%\item Fold the light curve with $P_0$. Order the magnitudes following the newly obtained phases.
%\item Obtain a template of the periodicity by smoothing the folded light curve using a moving average of 30 samples.
%\item Subtract the template from the folded light curve. 
%\item Rearrange the light curve samples to their original time order.
%\end{enumerate}
%\item Evaluate the whitened light curve with the pipeline described in Section \ref{pipeline}. If the light curve is non-periodic, discard it. If the light curve is periodic, save the period as $P_1$. If $P_1$ is not multiple/sub-multiple or alias of $P_0$, then the light curve is selected as a dual mode candidate. Subsequent oscillation modes can be found by repeating the procedure from step 2, using the whitened light curve.
%\end{enumerate}

\begin{figure}
	\centering	
	\subfloat[]{\includegraphics[scale=0.37]{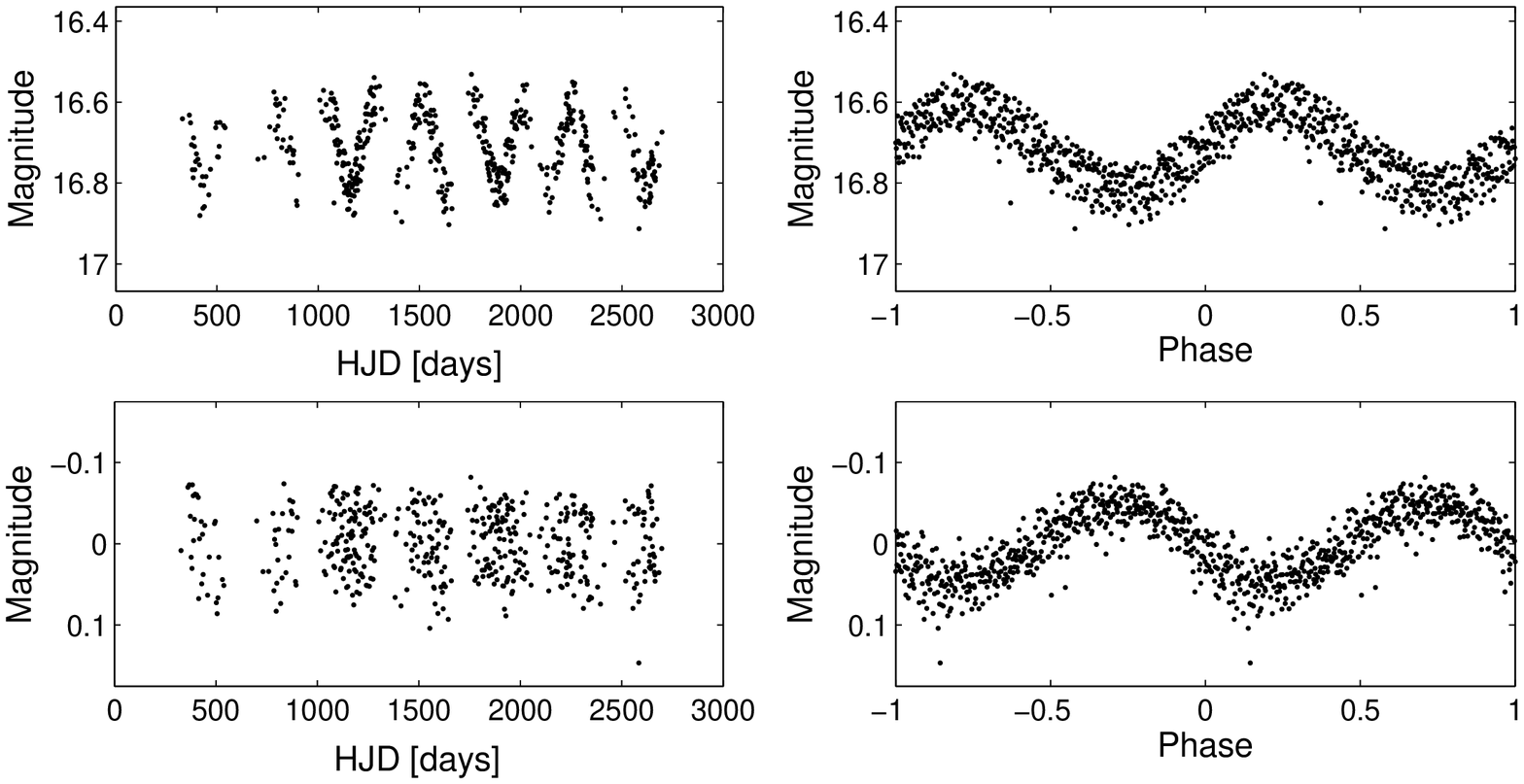}}  \\
	\subfloat[]{\includegraphics[scale=0.37]{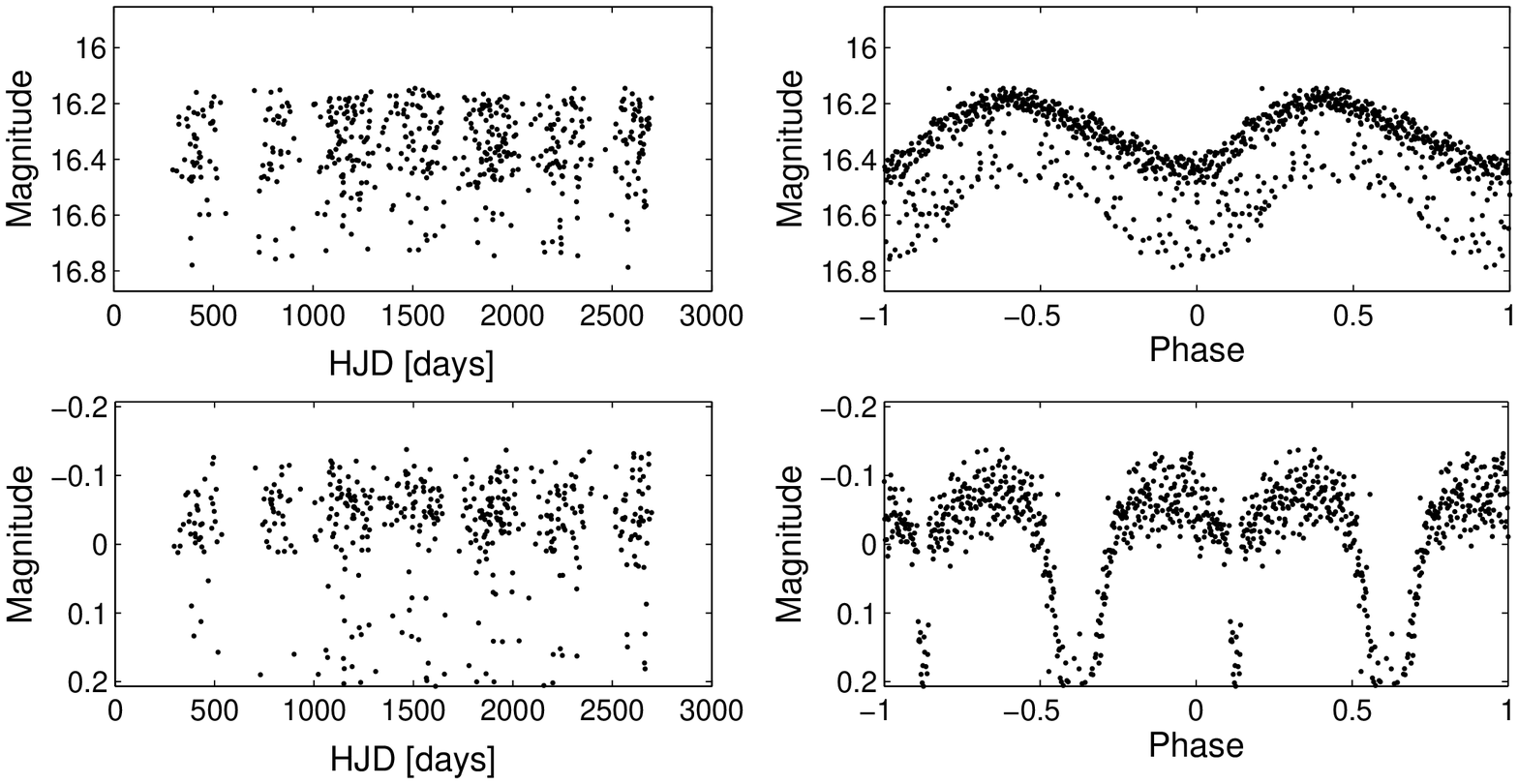}} \\
	\caption{ \label{fig-exp-MM1} Light curves lm0356k24082 (a) and lm0100m7313 (b) are selected as dual mode candidates. On each plot, the first and second rows correspond to the original and whitened light curve, respectively. In (a) the original light curve is folded with $P_0 = 244.06$ days. The whitened light curve is folded with $P_1 = 3.6399$ days. In (b) the original light curve is folded with $P_0 = 6.3419$ days. The whitened light curve is folded with $P_1 = 84.19$ days. }
\end{figure}

\begin{figure}
	\centering	
	\includegraphics[scale=0.4]{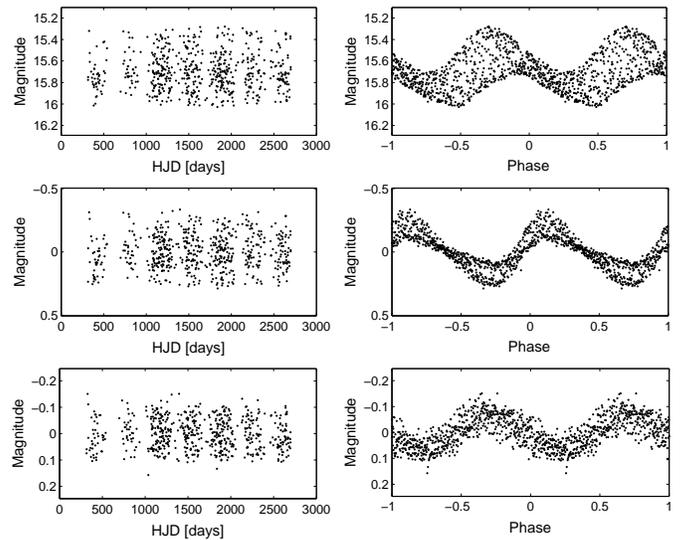}
	\caption{ \label{fig-exp-MM2} Light curve lm056518888 is selected as a triple mode candidate. In the plot the first, second and third rows correspond to the original, first whitened and second whitened light curves, respectively.  The original light curve is folded with the detected period $P_0 = 2.4725$ days. The first whitened light curve is folded with $P_1 = 3.4455$ days. The second whitened light curve is folded with $P_2 = 1.4395$ days.  }
\end{figure}

This procedure is applied on 34,000 periodic light curves from the LMC with CKP values above 2.0 \footnote{We only selected the most prominent periodic lightcurves}. From this set 1165 light curves are selected as dual mode candidates. After evaluating the double mode candidates, 116 are found to have a third oscillation mode. Examples of dual mode and triple mode candidates are shown in Figures \ref{fig-exp-MM1} and \ref{fig-exp-MM2}, respectively. The lists of double and triple mode candidates can be found at \href{http://timemachine.iic.harvard.edu}{http://timemachine.iic.harvard.edu}. 

\begin{figure}
	\centering	
	\includegraphics[scale=0.47]{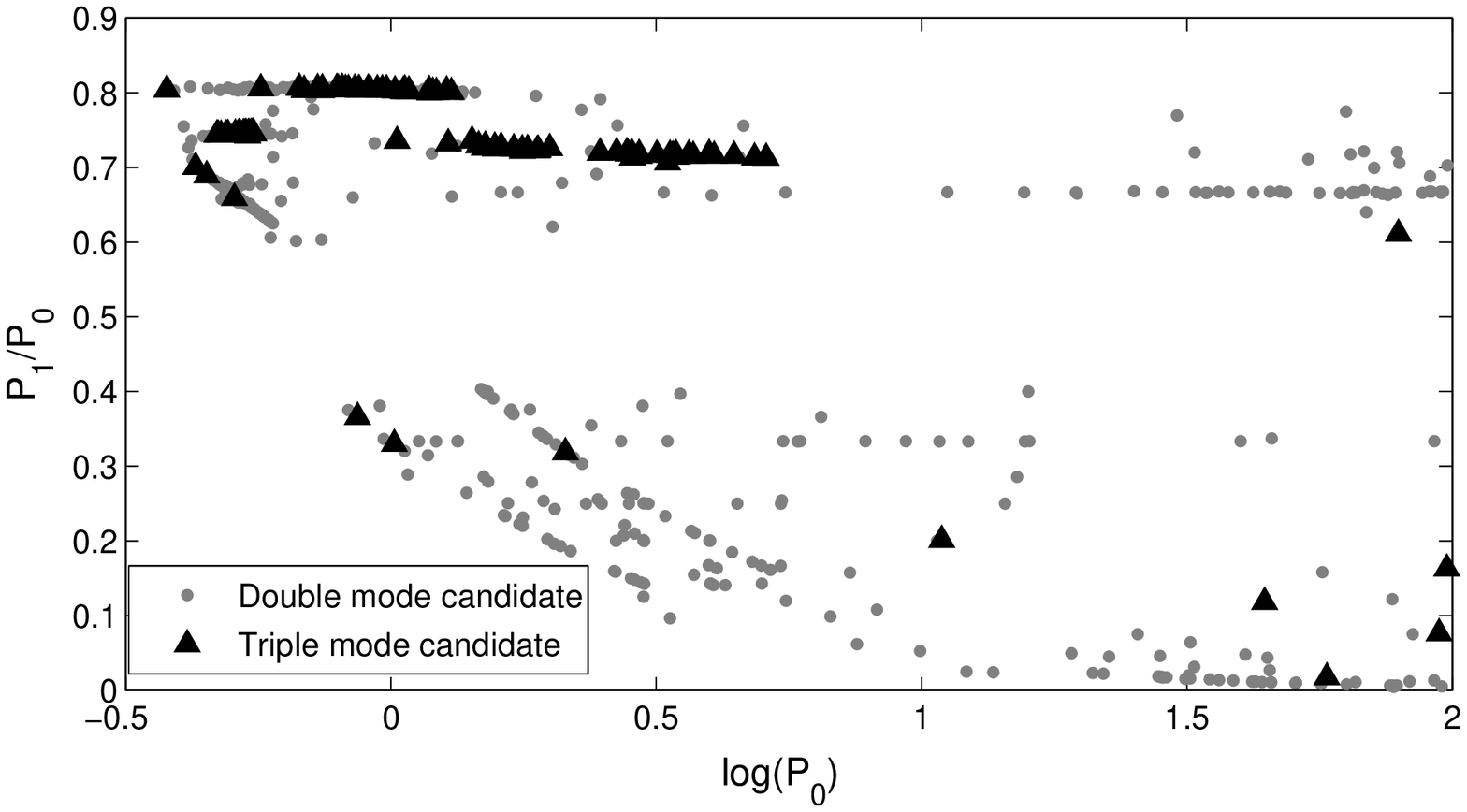}
	\caption{ \label{fig-exp-MM3} Petersen diagram of the 1165 dual mode candidates found in the LMC. The triangles mark the location of 116 triple mode candidates. Clear structures arise in the diagram.  }
\end{figure}

Fig. \ref{fig-exp-MM3} shows a Petersen diagram of the 1165 light curves selected as dual modes candidates. The triangles in the plot mark the 116 light curves in which a third mode was found. The periods are sorted so that $P_0 >P_1$ in all cases. The triple mode candidates occupy two horizontal lines at period ratios of 0.72 and 0.8. These values are close to the known ratios associated to the first and second overtones \citep{Moskalik2012}. A prominent horizontal line appears at $P_1/P_0 \sim 2/3$ for fundamental periods above 10 days. According to \citet{Smolec2012} this ratio is associated to the period doubling phenomenon. Another interesting feature, shown in the lower left part of the diagram, are two curves that follow an inversely proportional relationship between the period ratio and fundamental period.

\subsection{Odd periodic stars}
\begin{figure}
	\centering	
	\subfloat[]{\includegraphics[scale=0.3]{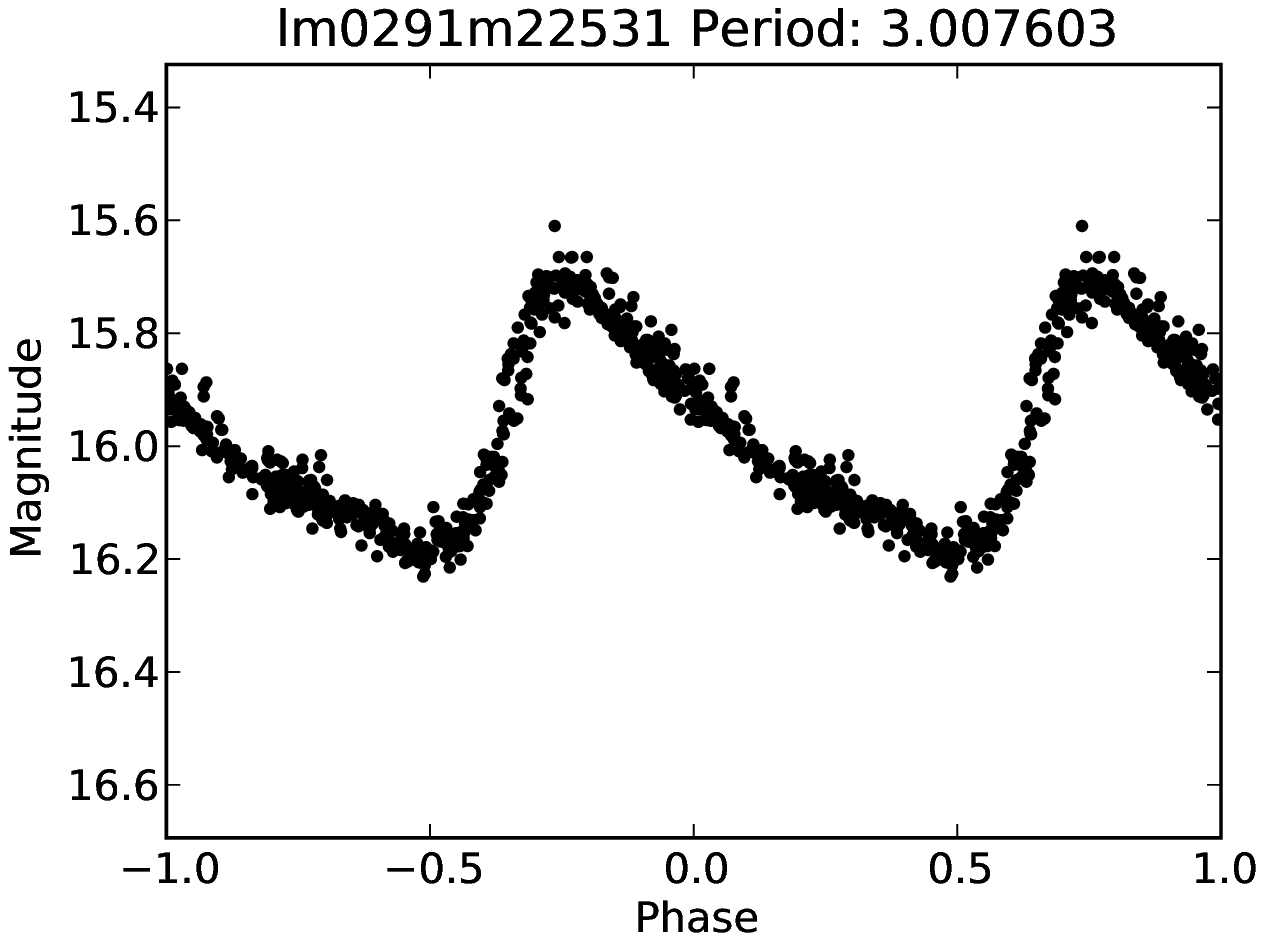}} 
	\subfloat[]{\includegraphics[scale=0.3]{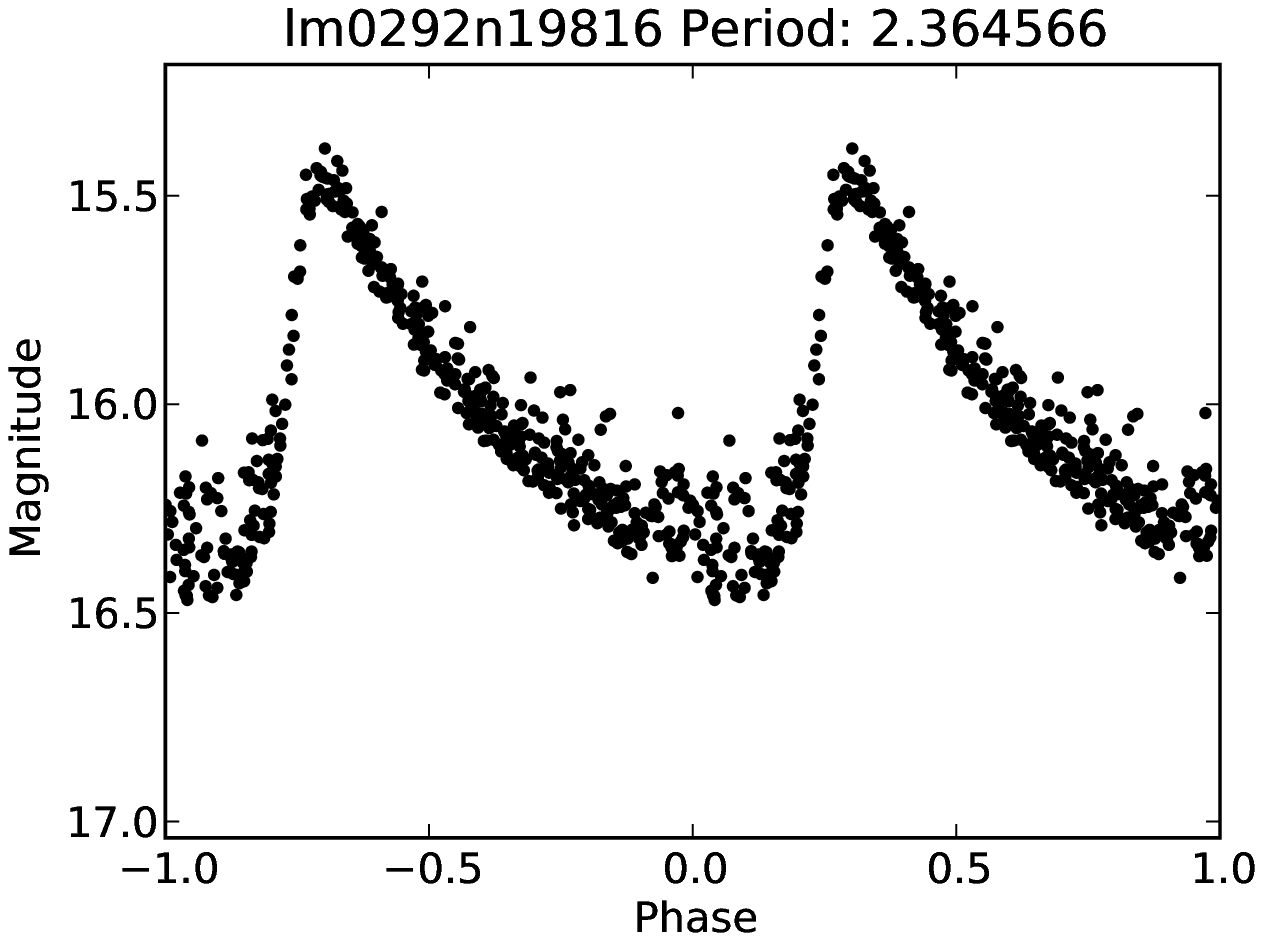}} \\
	\subfloat[]{\includegraphics[scale=0.3]{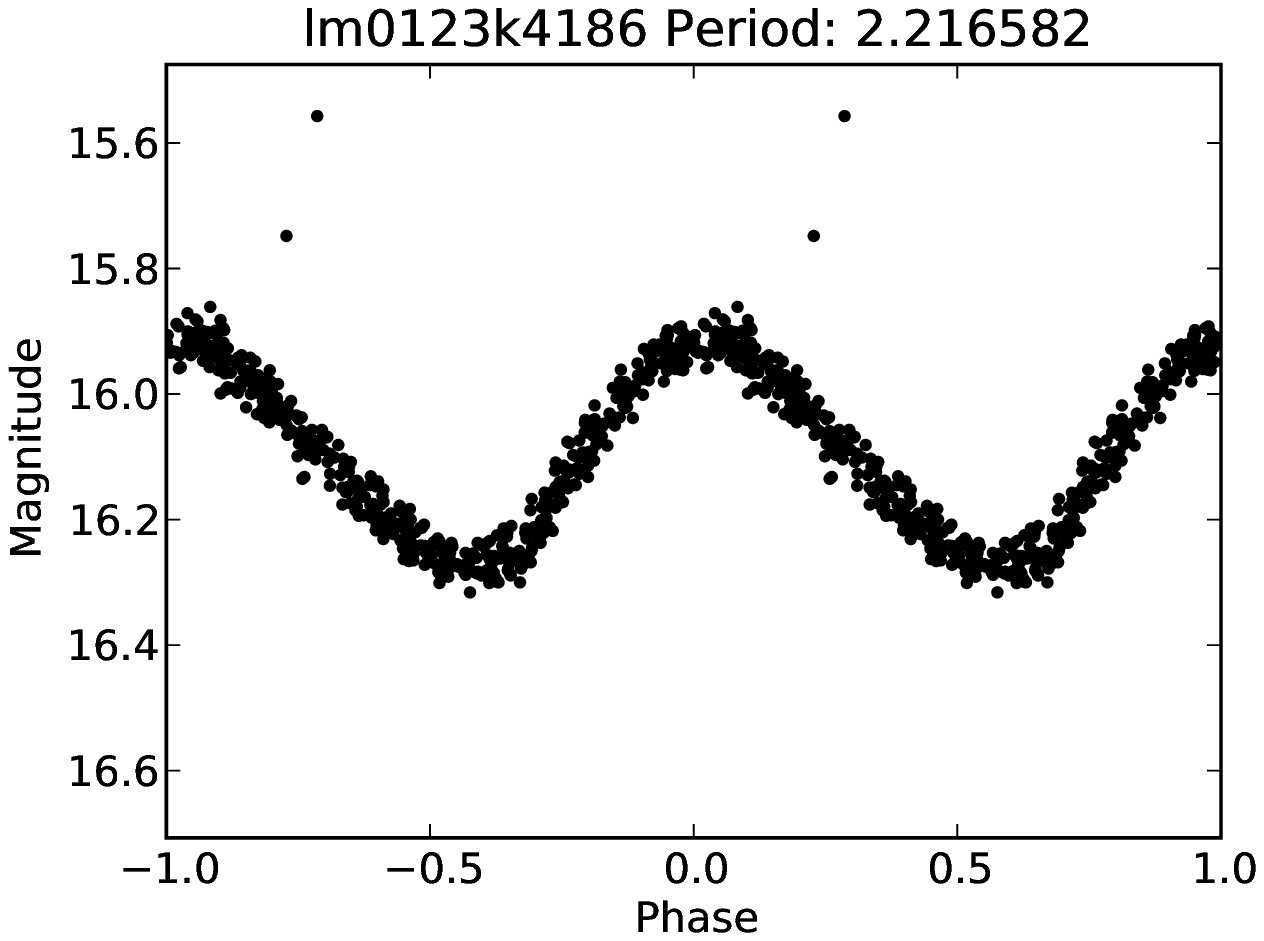}}  
	\subfloat[]{\includegraphics[scale=0.3]{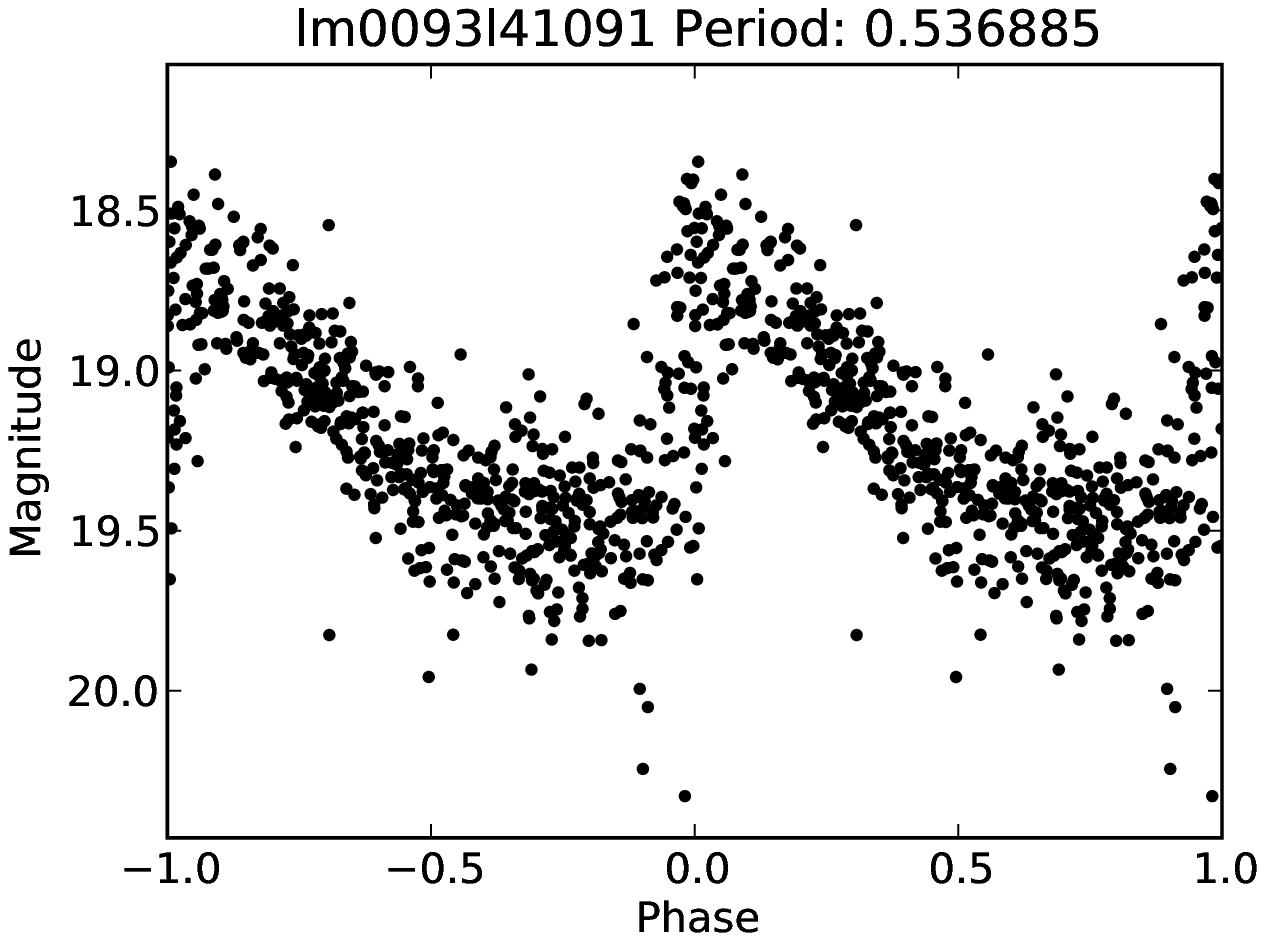}} \\
	\subfloat[]{\includegraphics[scale=0.3]{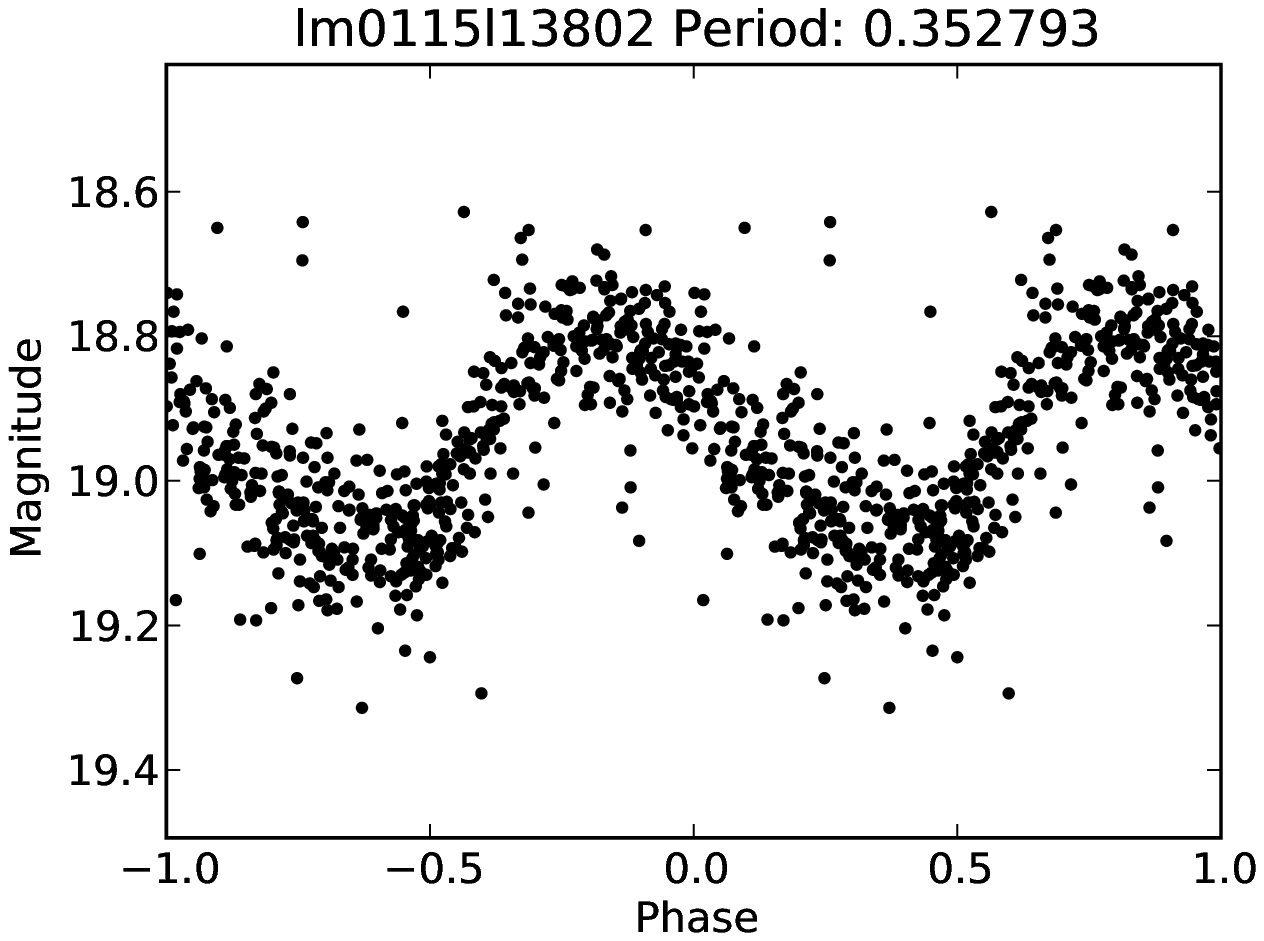}} 
	\subfloat[]{\includegraphics[scale=0.3]{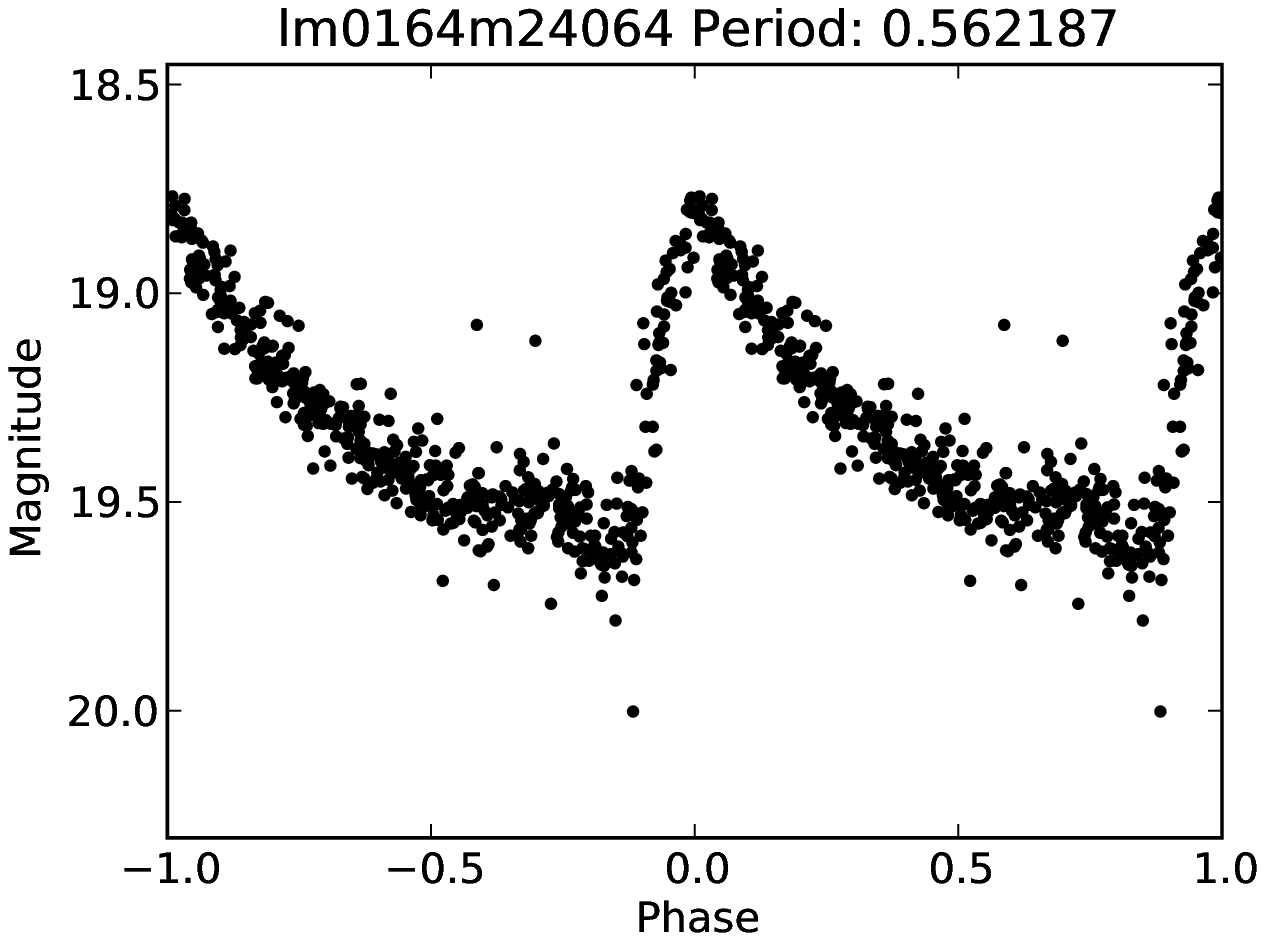}}  
	\caption{ \label{fig-exp-puls} Examples of EROS-2 periodic lightcurves folded with their estimated period. (a), (b) and (c) are Cepheids taken from the CEPH cluster (see Fig. \ref{cmLMCBLUE}). (d), (e) and (f) are RR Lyrae taken from the RRL cluster. (d) and (f) are examples of RRab class stars. (e) is an example of an RRc class star.}
\end{figure}

\begin{figure}
	\centering	
	\subfloat[]{\includegraphics[scale=0.3]{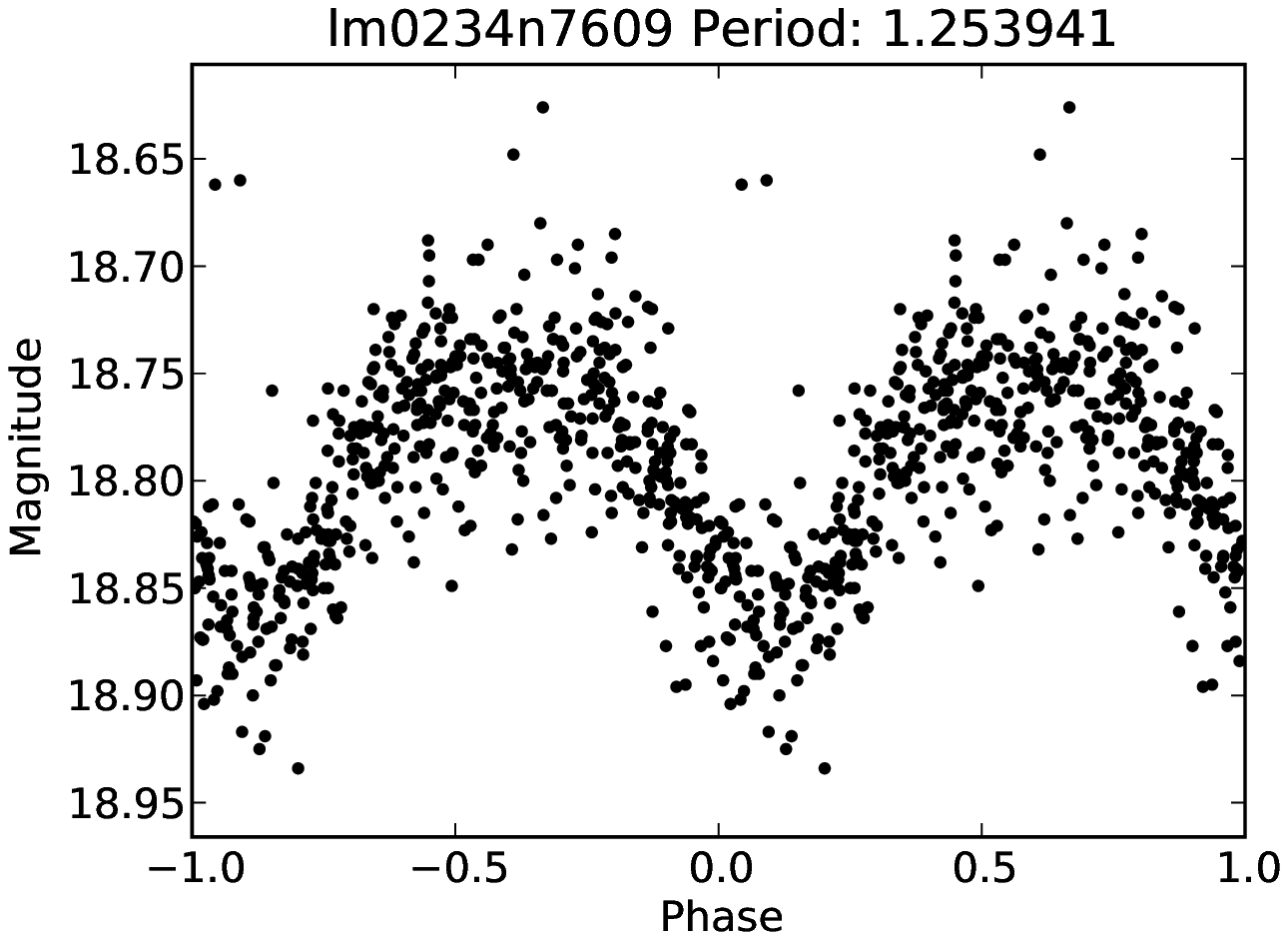}} 
	\subfloat[]{\includegraphics[scale=0.3]{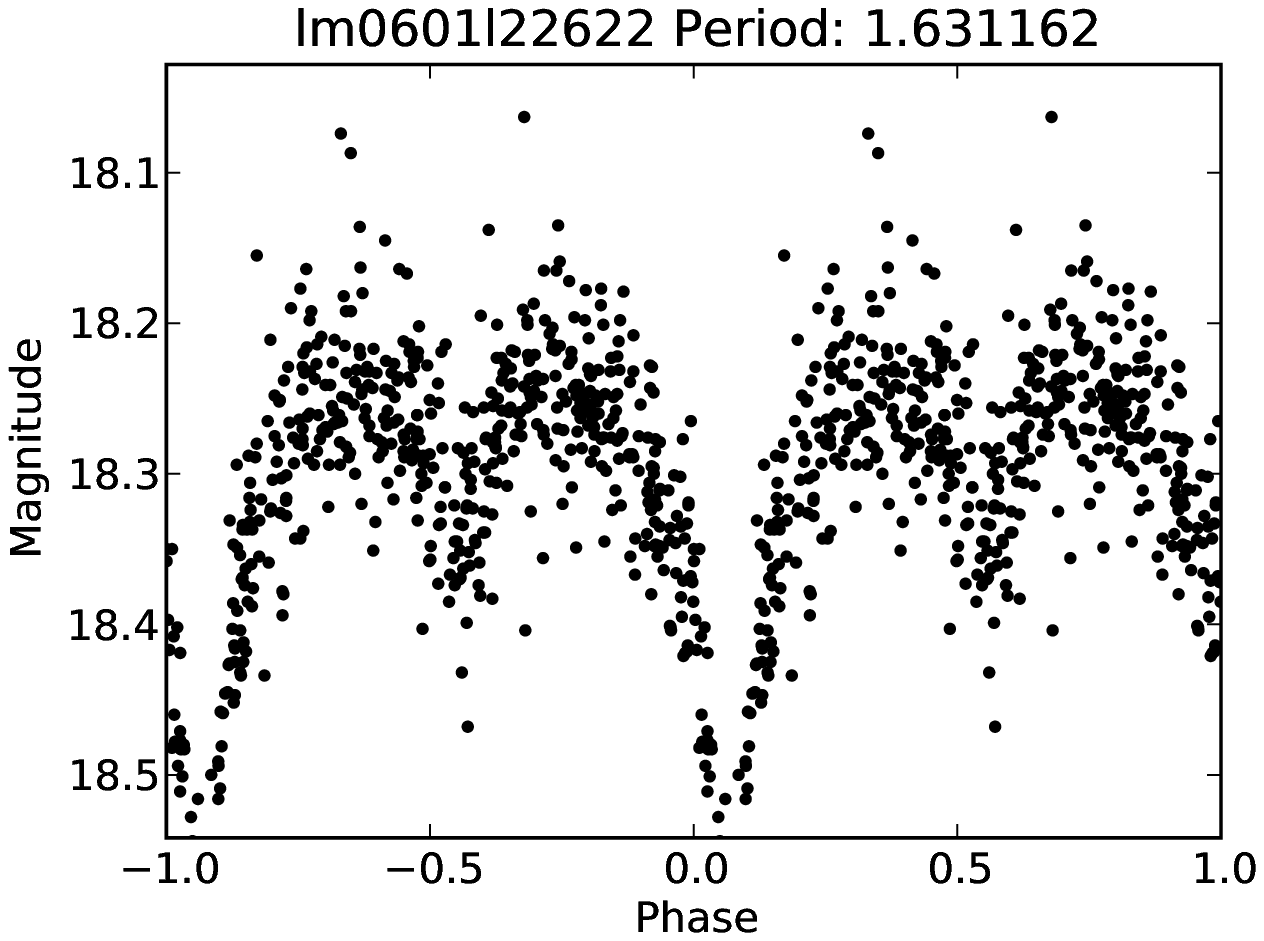}} \\
	\subfloat[]{\includegraphics[scale=0.3]{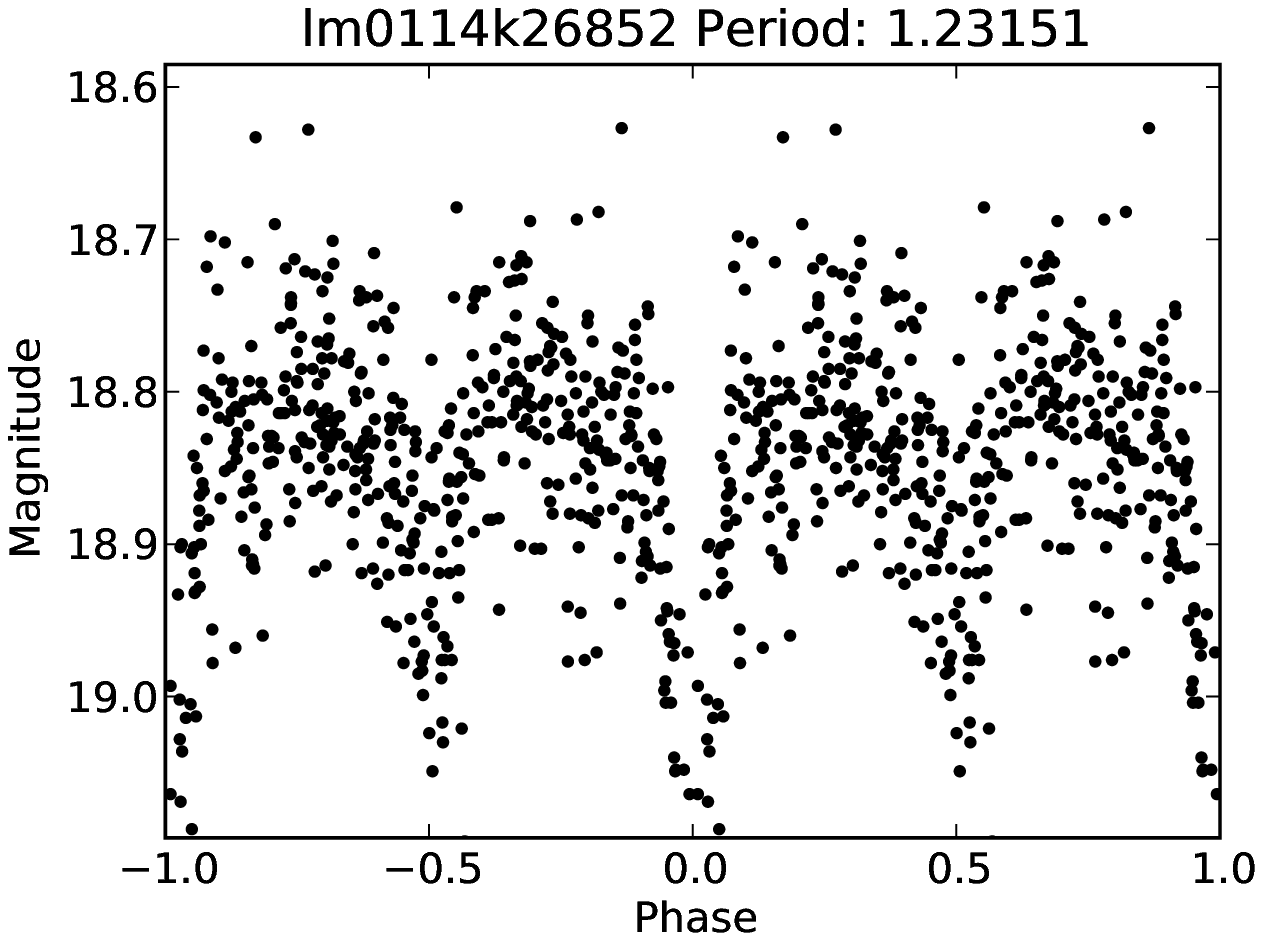}} 
	\subfloat[]{\includegraphics[scale=0.3]{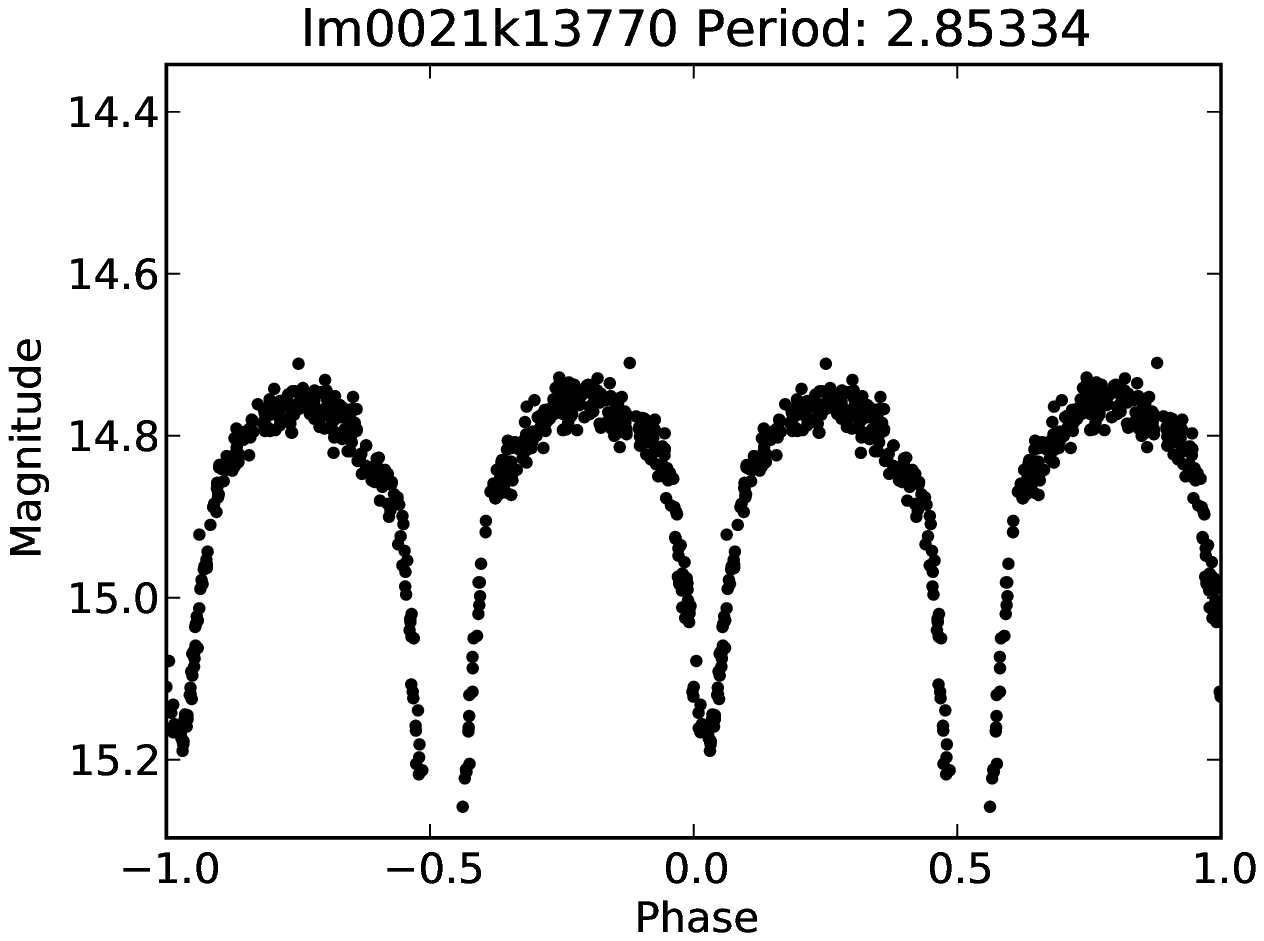}} \\
	\subfloat[]{\includegraphics[scale=0.3]{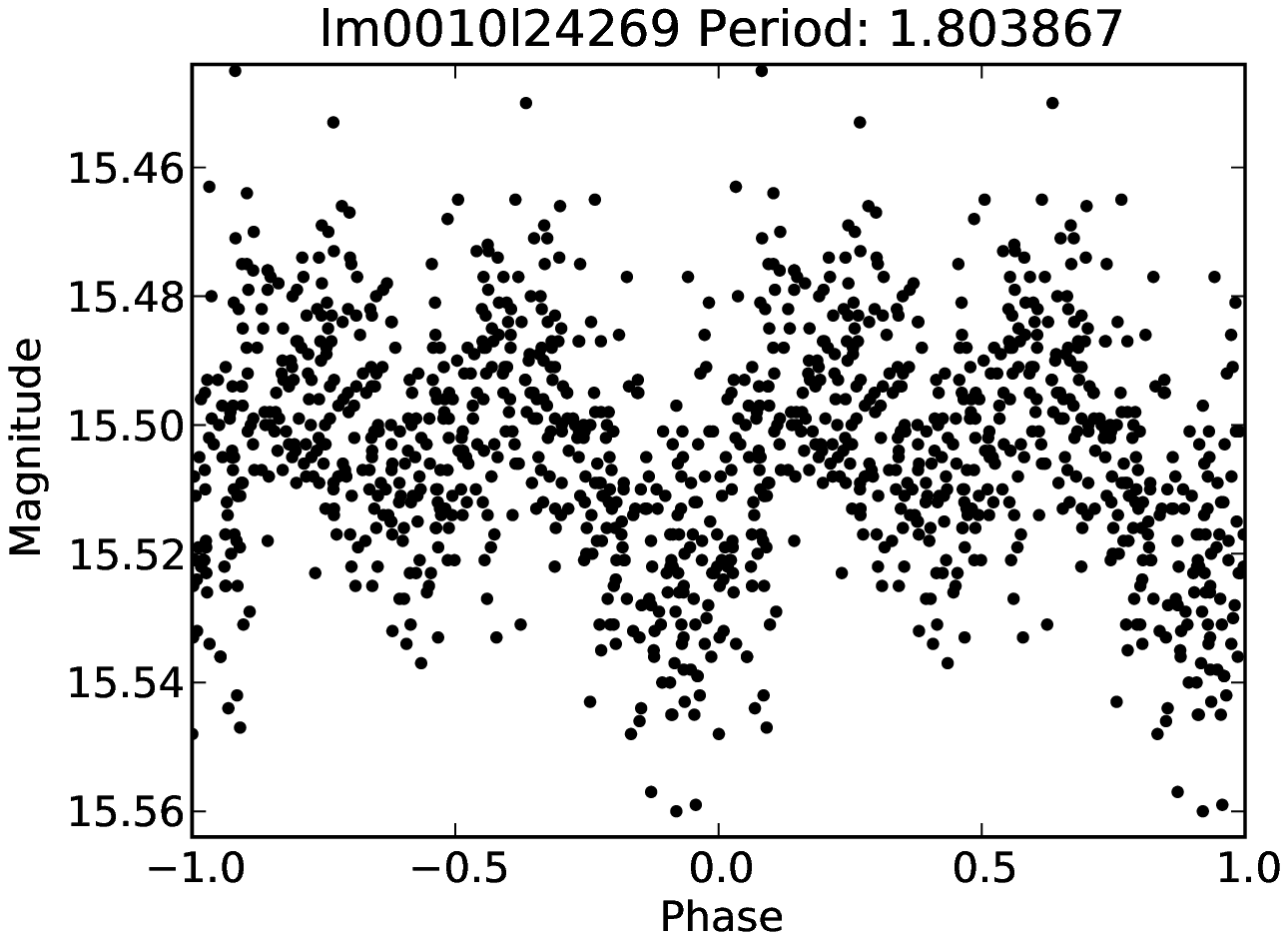}} 
	\subfloat[]{\includegraphics[scale=0.3]{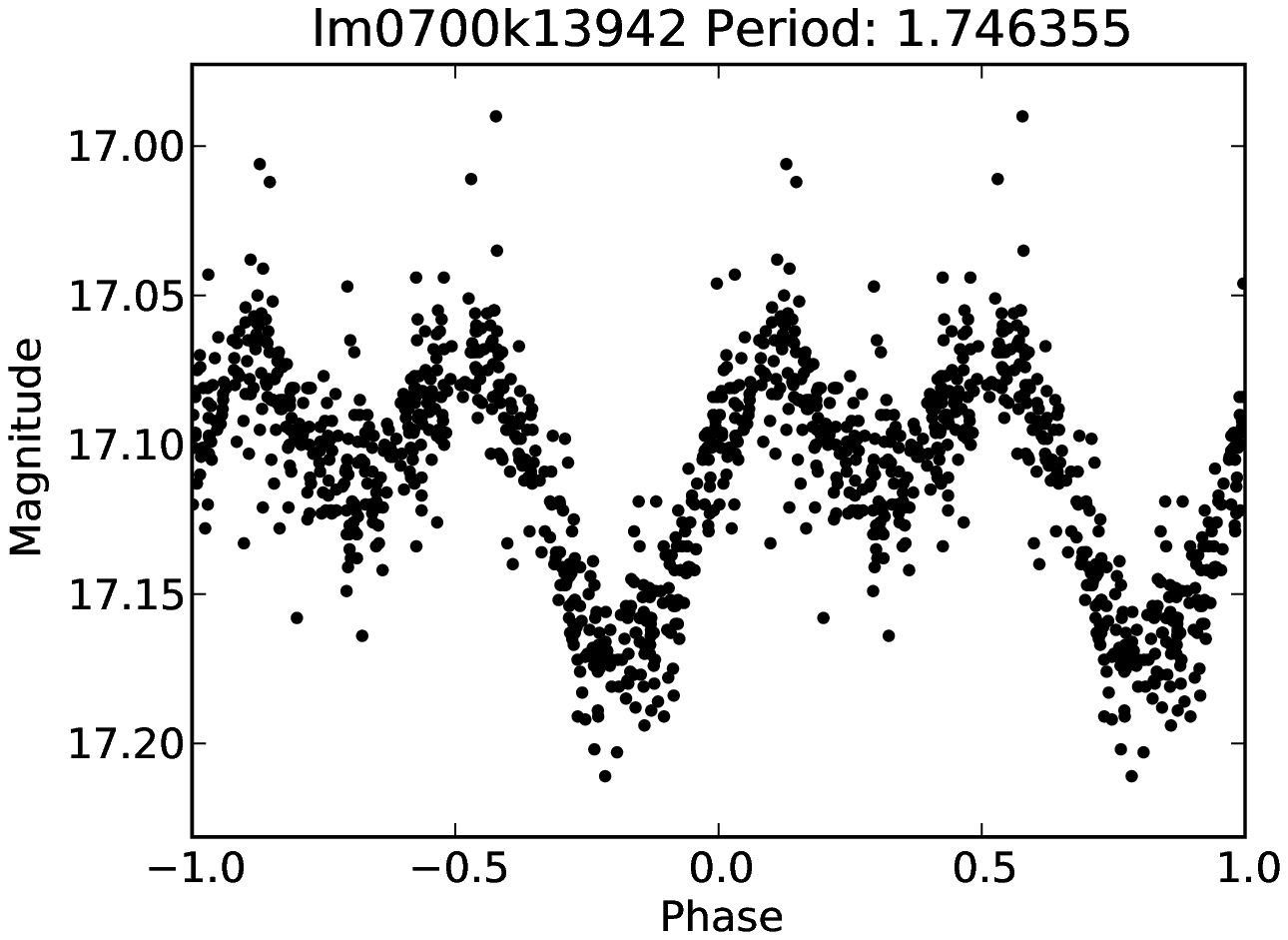}}  
	\caption{ \label{fig-exp-eb} Examples of EROS-2 periodic lightcurves folded with their estimated period. These lightcurves corresponds to eclipsing binary stars found in the blue main sequence (see Fig. \ref{cmLMCBLUE}). }
\end{figure}

\begin{figure}
	\centering	
	\subfloat[]{\includegraphics[scale=0.3]{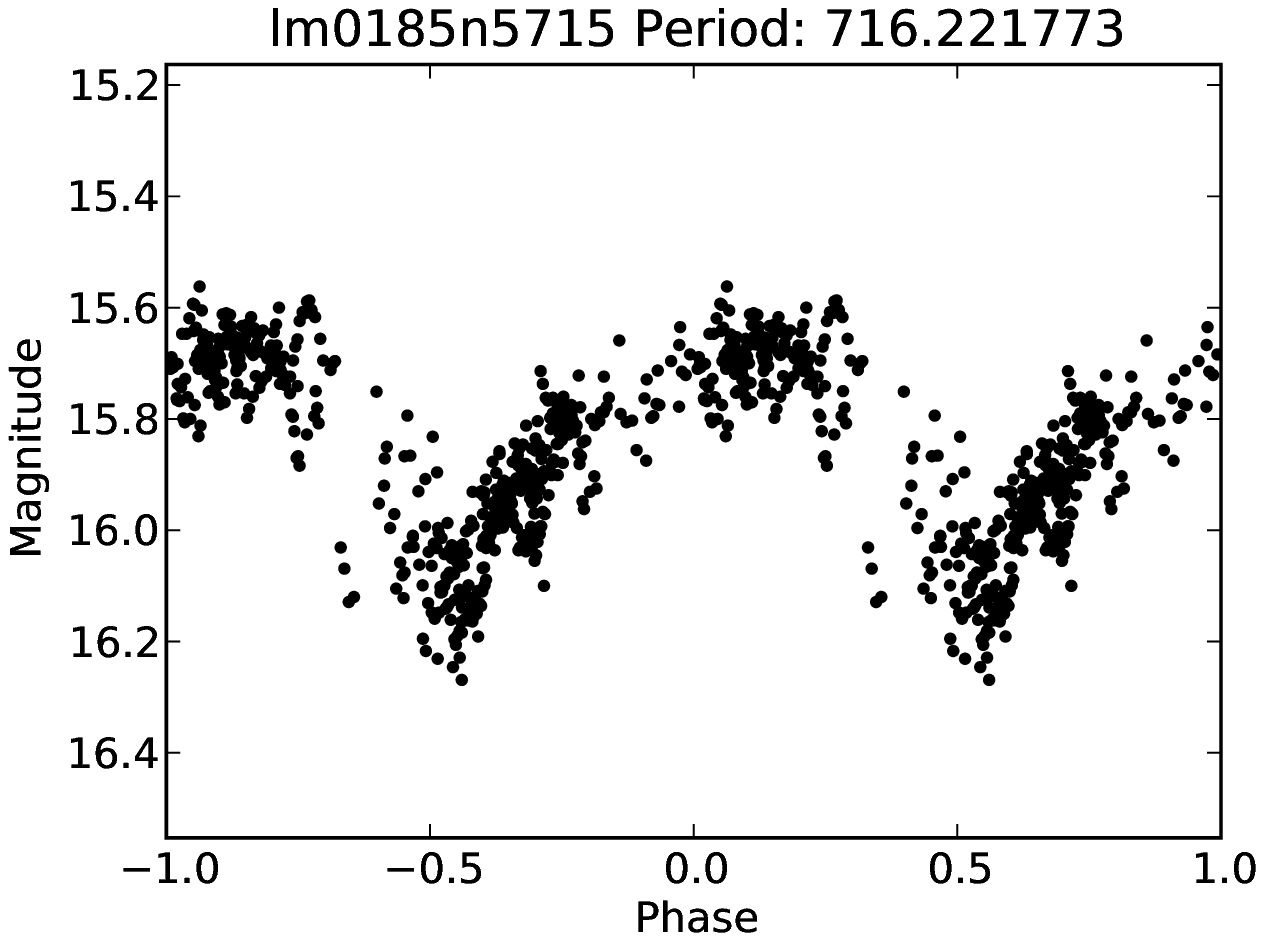}} 
	\subfloat[]{\includegraphics[scale=0.3]{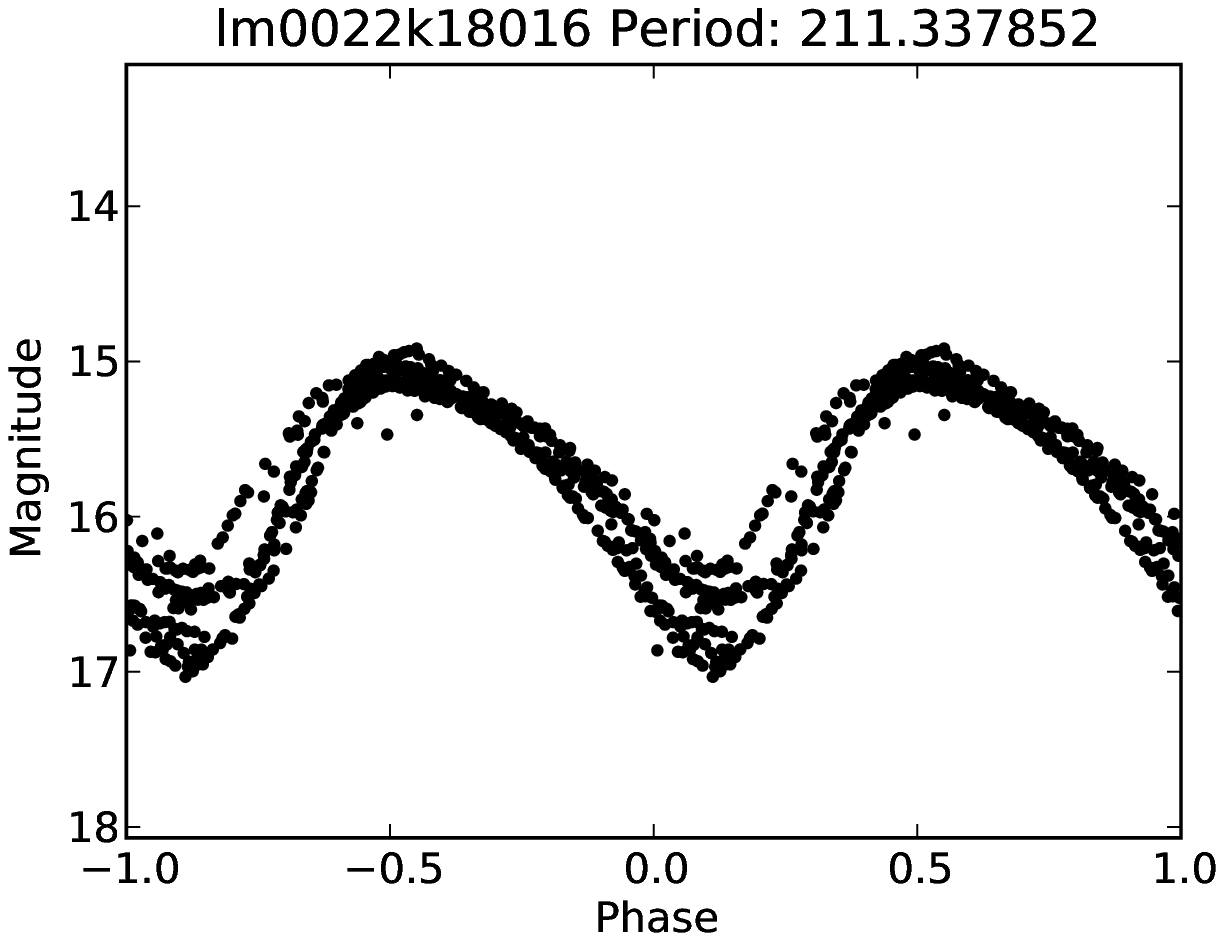}} \\
	\subfloat[]{\includegraphics[scale=0.3]{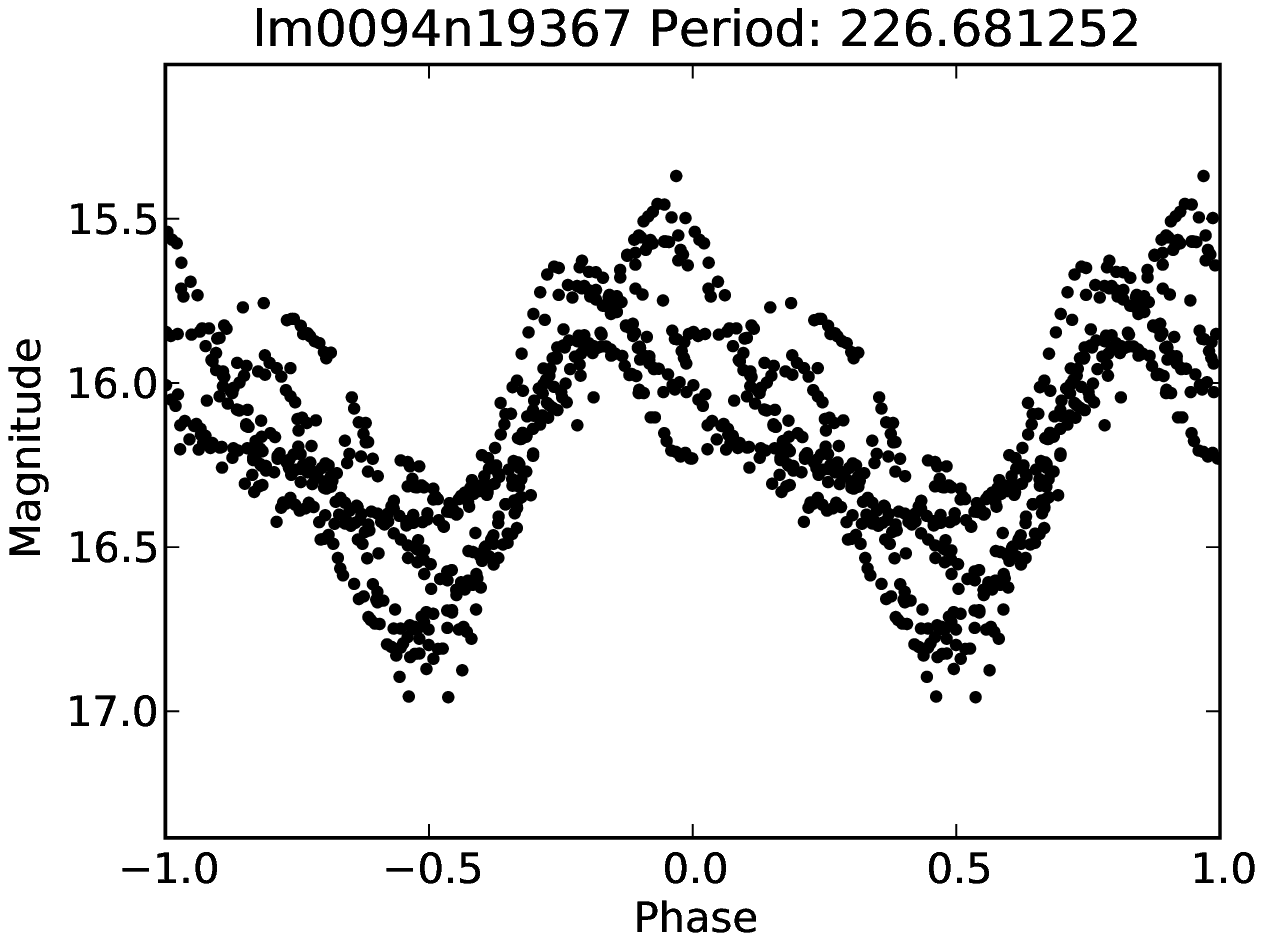}}  
	\subfloat[]{\includegraphics[scale=0.3]{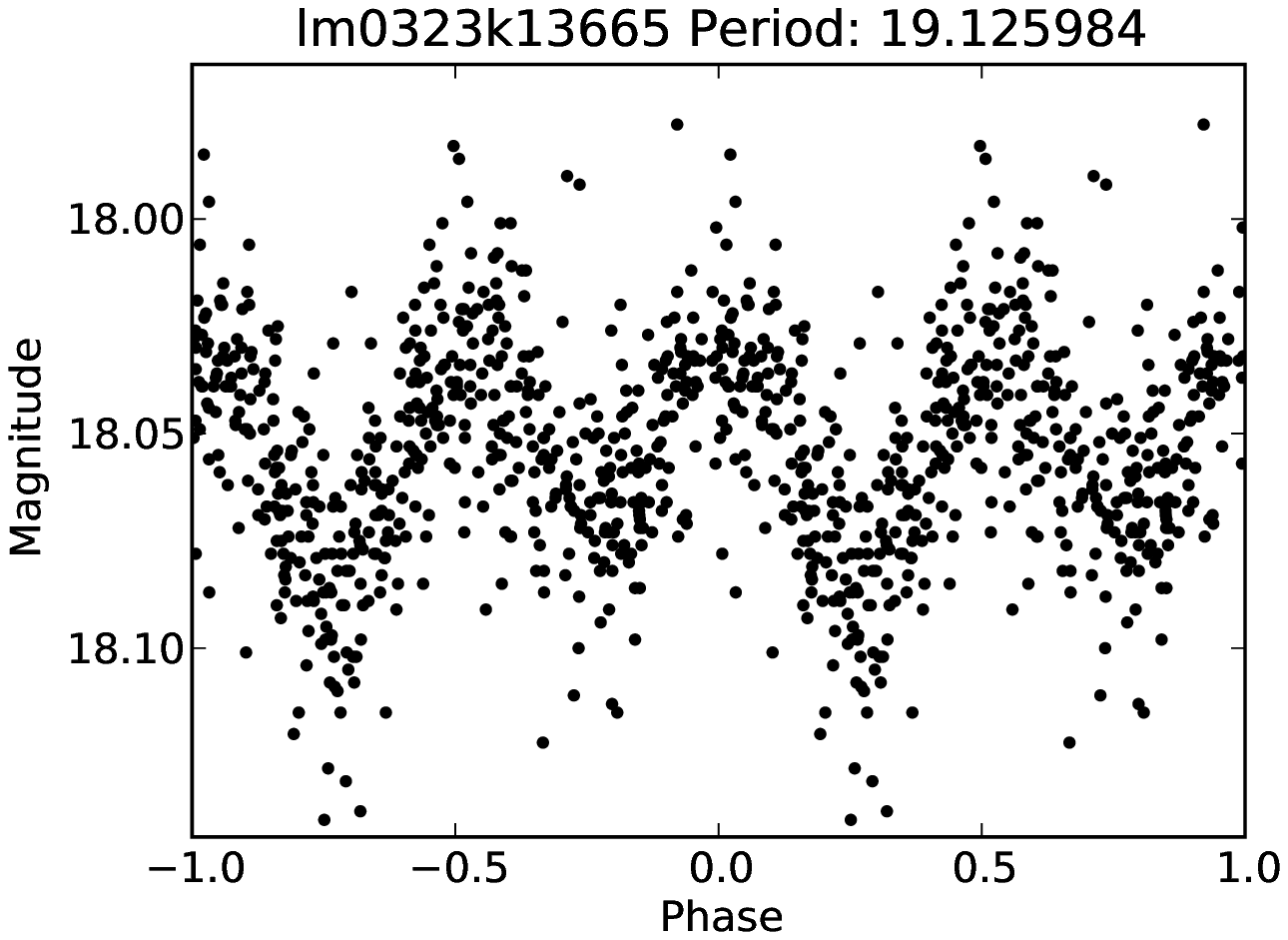}} \\
	\subfloat[]{\includegraphics[scale=0.3]{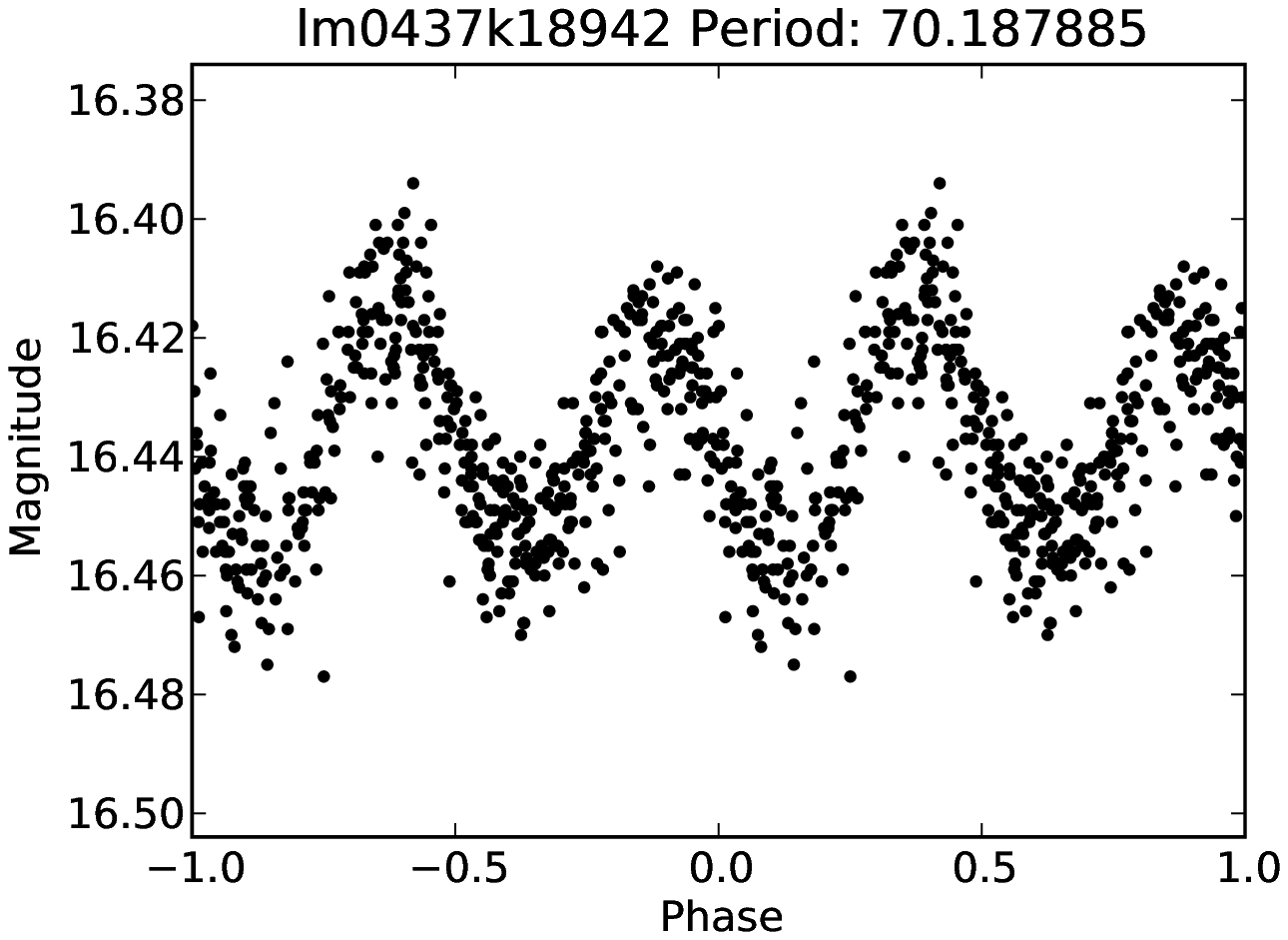}} 
	\subfloat[]{\includegraphics[scale=0.3]{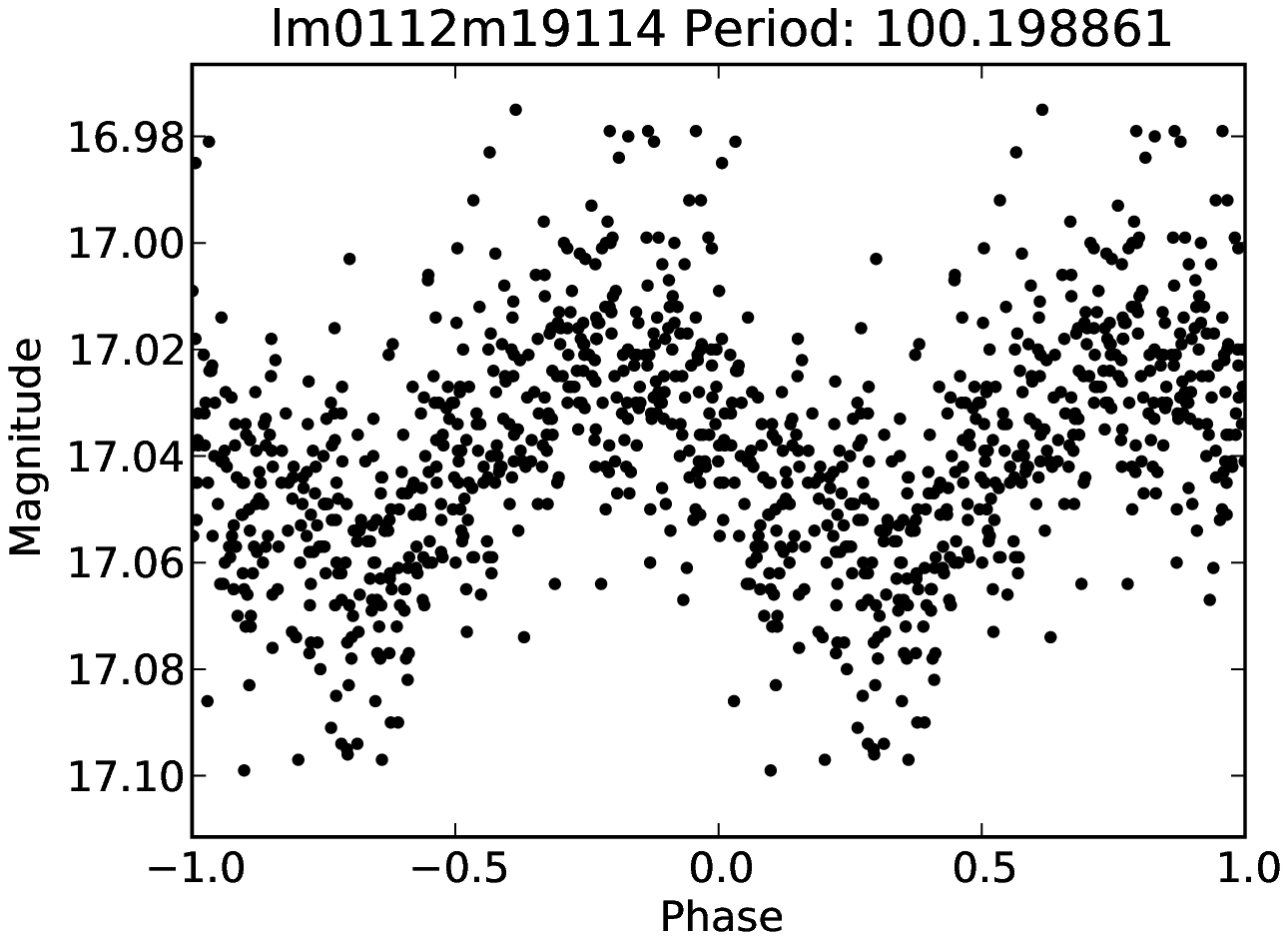}}  
	\caption{ \label{fig-exp-lpv} Examples of EROS-2 periodic lightcurves folded with their estimated period. (a), (b) and (c) correspond to long period variables found in the LPV cluster  (see Fig. \ref{cmLMCBLUE}). (d), (e) and (f) correspond to periodic variable stars found in the lower red giant branch. }
\end{figure}

\begin{figure}
	\centering	
	\subfloat[]{\includegraphics[scale=0.3]{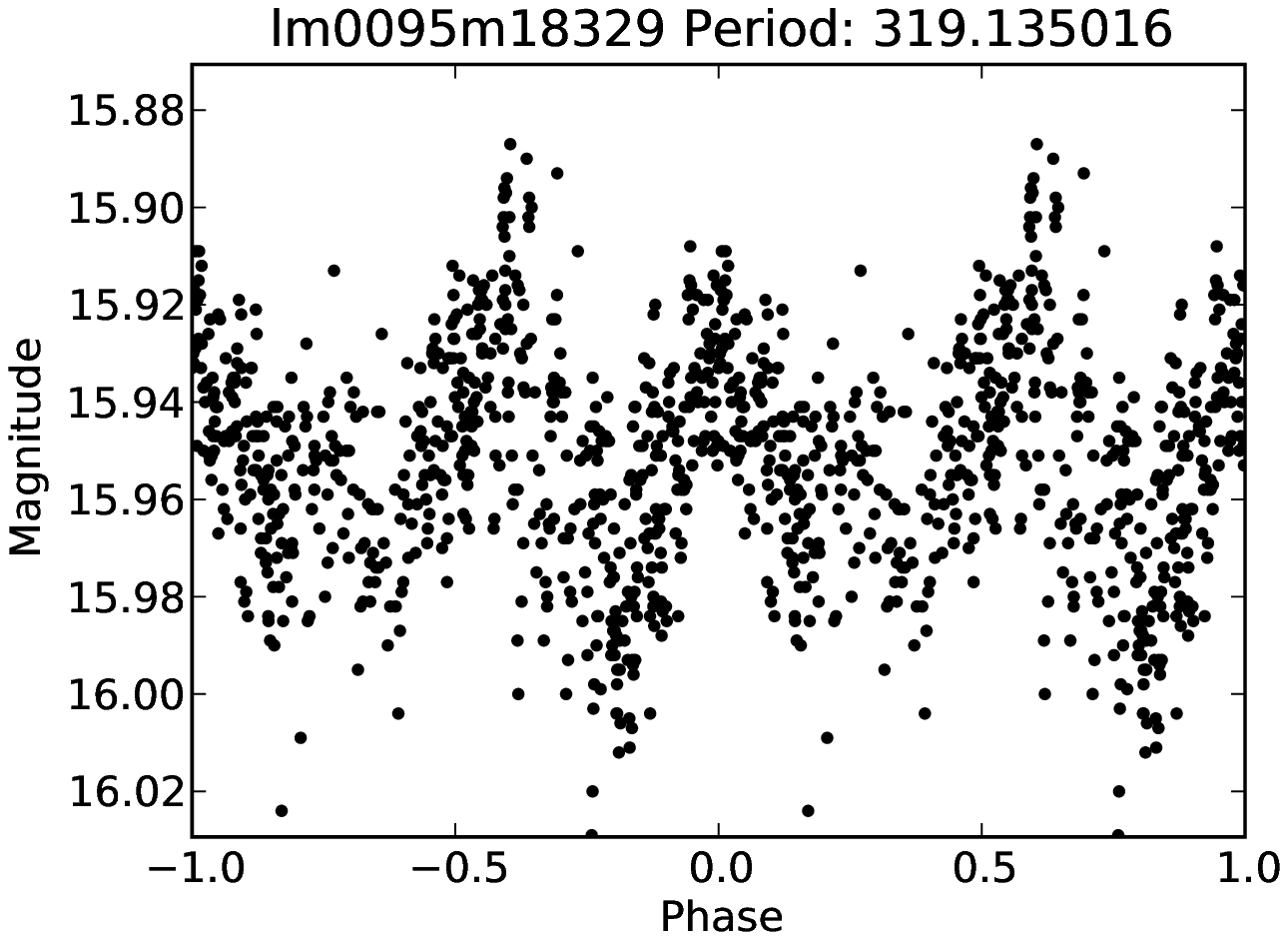}} 
	\subfloat[]{\includegraphics[scale=0.3]{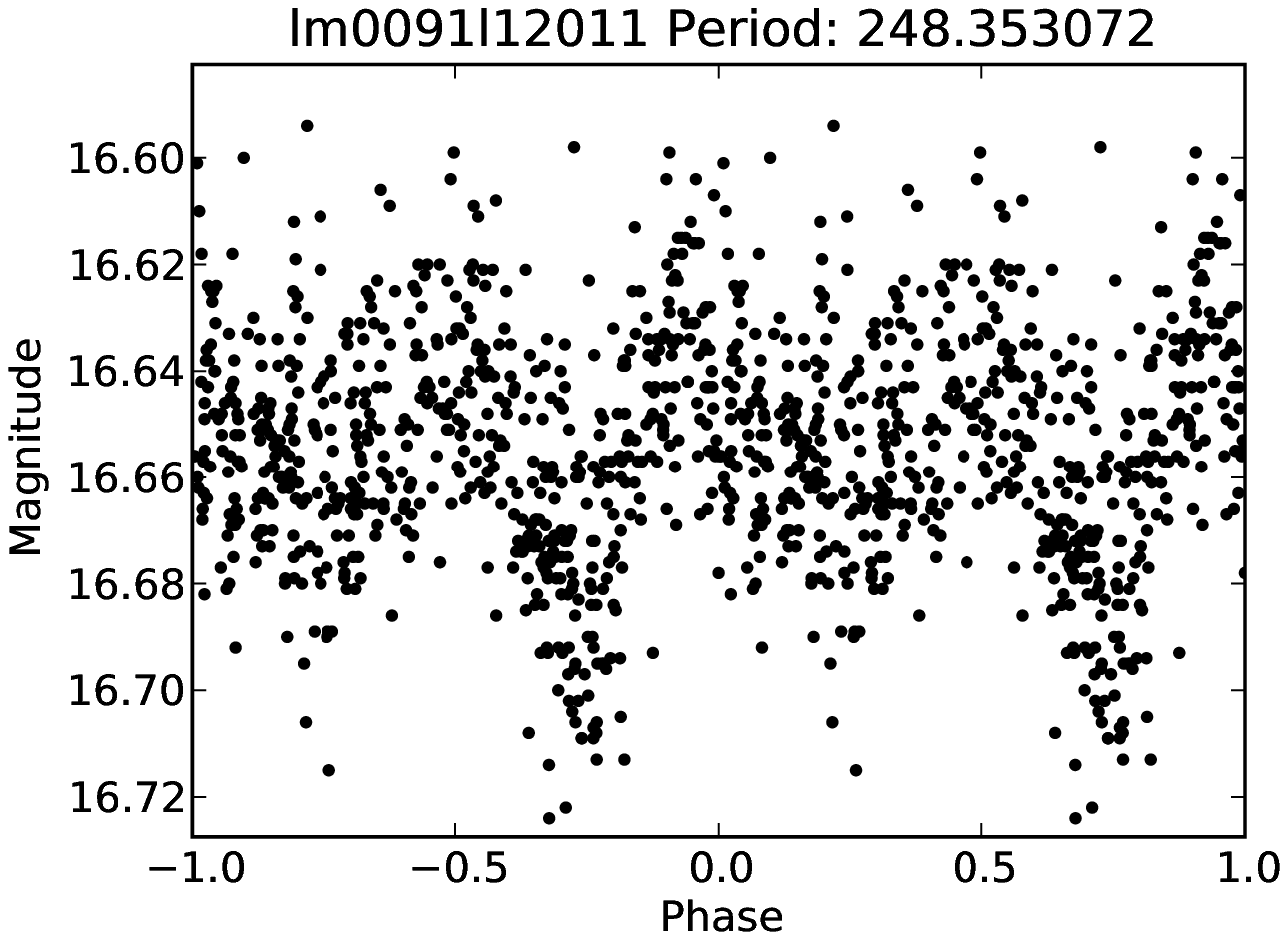}} \\
	\subfloat[]{\includegraphics[scale=0.3]{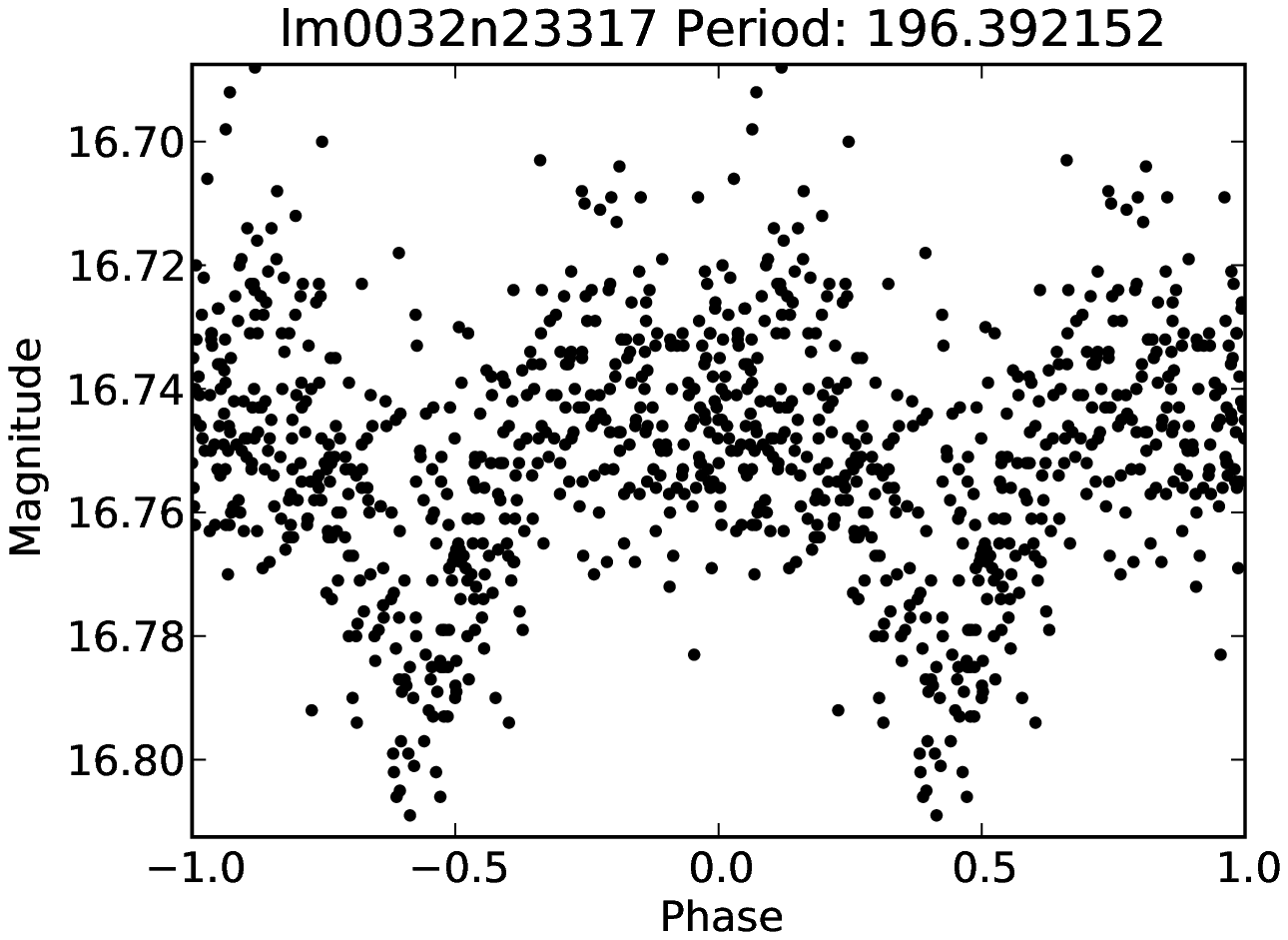}}  
	\subfloat[]{\includegraphics[scale=0.3]{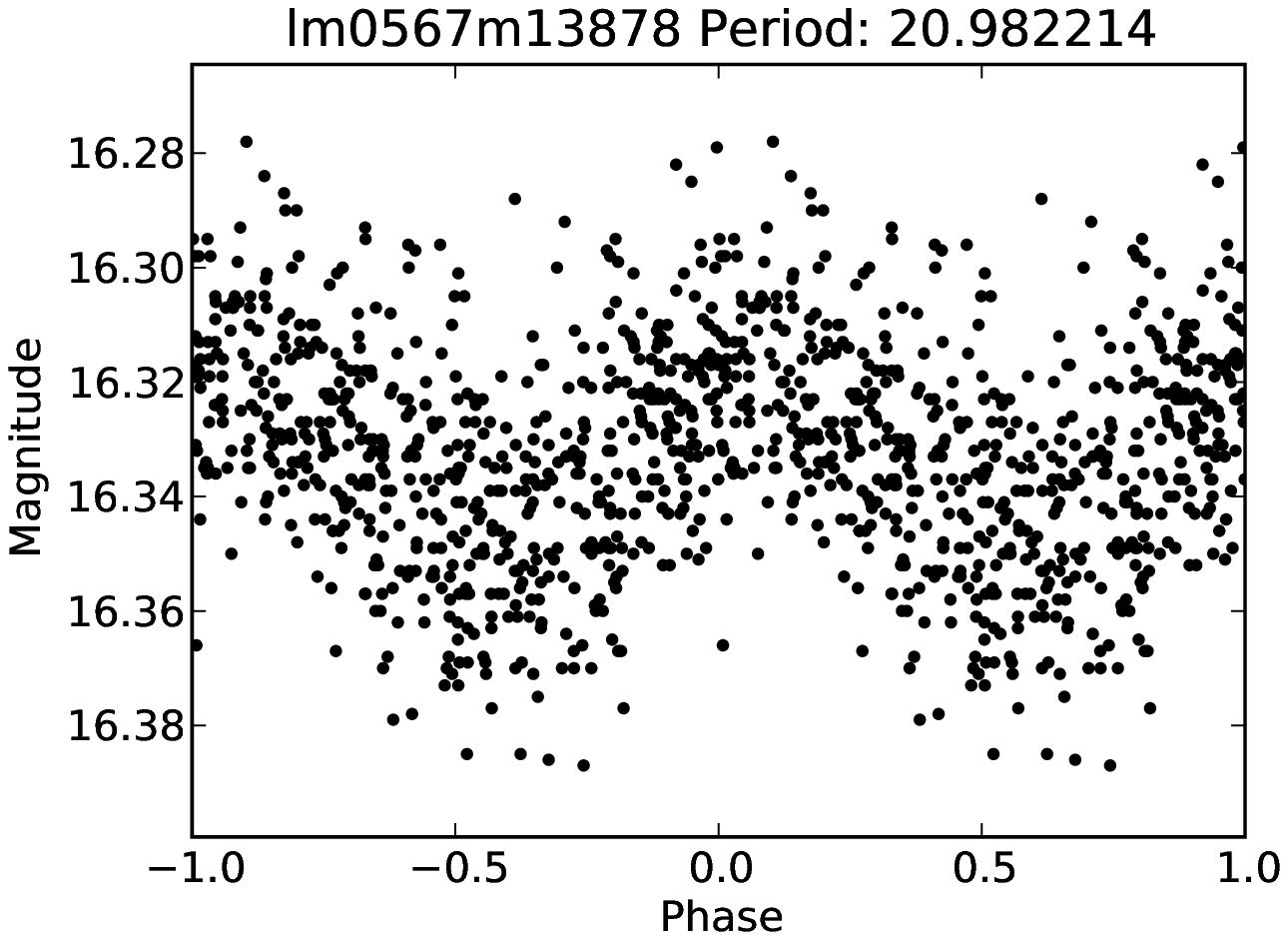}} \\
	\subfloat[]{\includegraphics[scale=0.3]{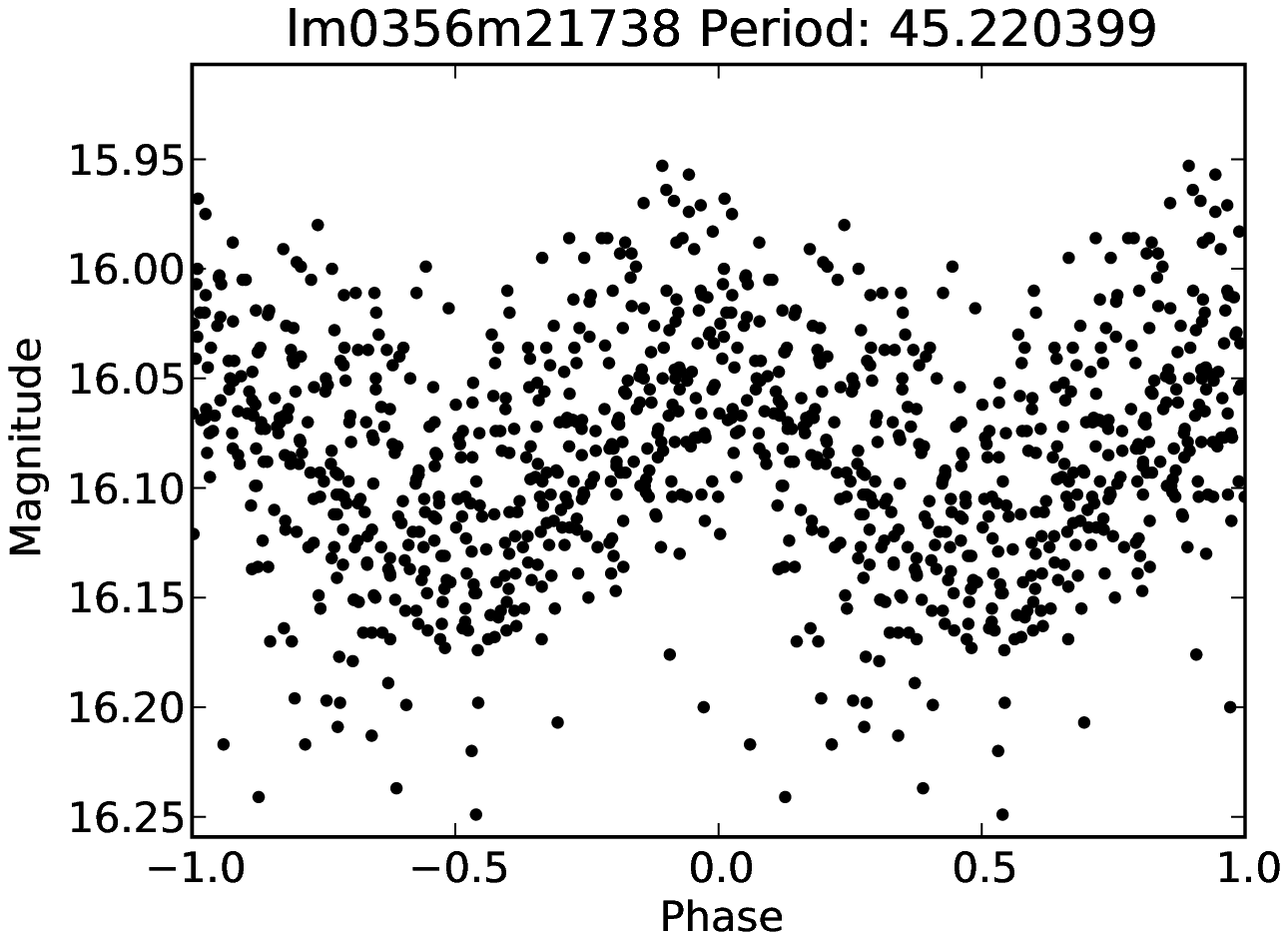}} 
	\subfloat[]{\includegraphics[scale=0.3]{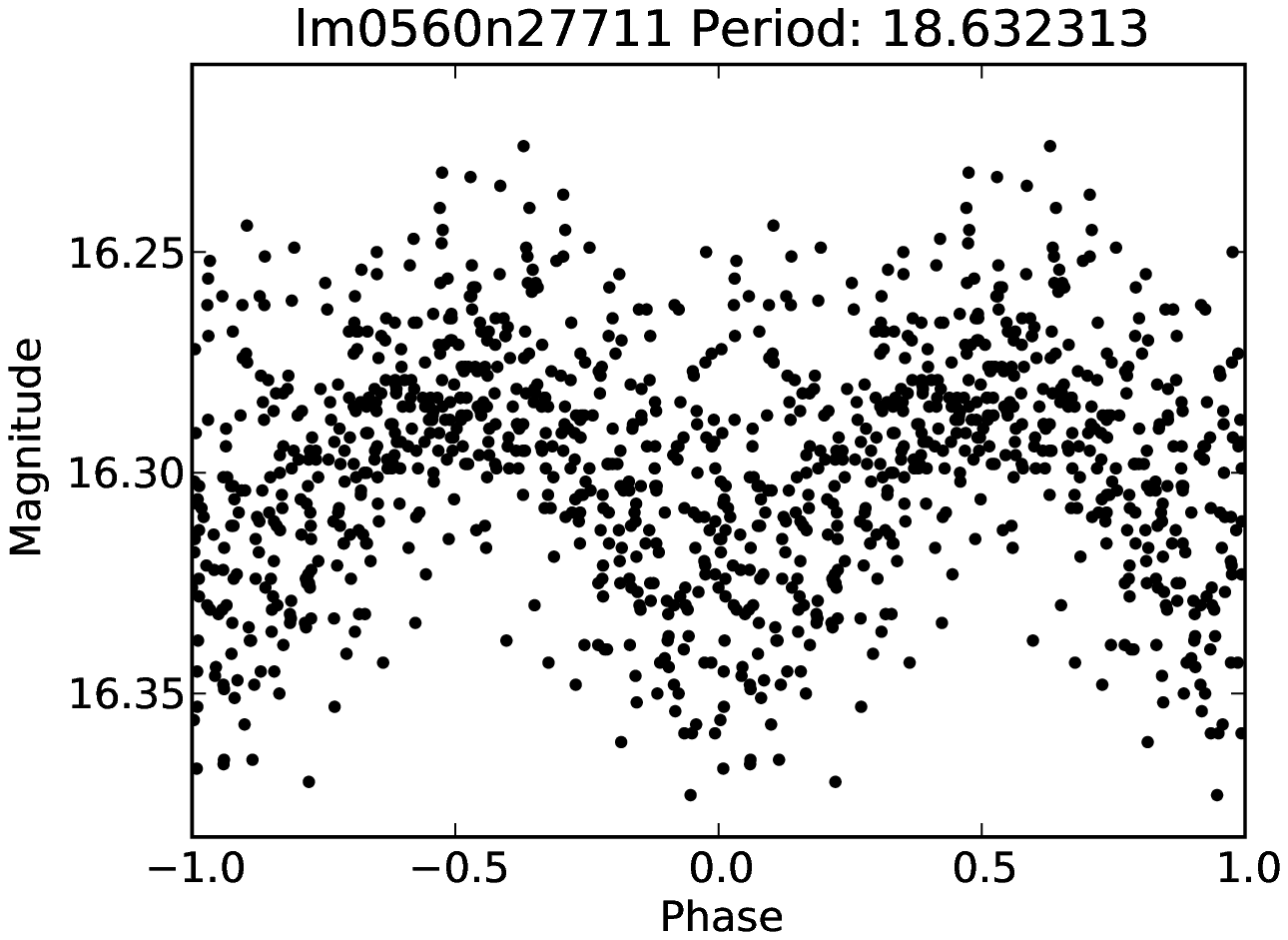}}  
	\caption{ \label{fig-exp-ell} Examples of EROS-2 periodic lightcurves folded with their estimated period. (a), (b) and (c) correspond to long period variables found in the LPV-2 cluster (see Fig. \ref{cmLMCBLUE}. (d), (e) and (f) correspond to ellipsoidal variables found in the ELL cluster ). }
\end{figure}

\begin{figure}
	\centering	
	%\subfloat[]{\includegraphics[scale=0.55]{odd/odd1.eps}} \hspace{5pt}
	%\subfloat[]{\includegraphics[scale=0.55]{odd/odd8.eps}} \\
%\subfloat[]{\includegraphics[scale=0.55]{odd/odd10.eps}} \\
	%\subfloat[]{\includegraphics[scale=0.55]{odd/odd3.eps}}  \hspace{5pt}
	%\subfloat[]{\includegraphics[scale=0.55]{odd/odd2.eps}} \\
	%\subfloat[]{\includegraphics[scale=0.55]{odd/odd5.eps}} \hspace{5pt}
	\subfloat[]{\includegraphics[scale=0.55]{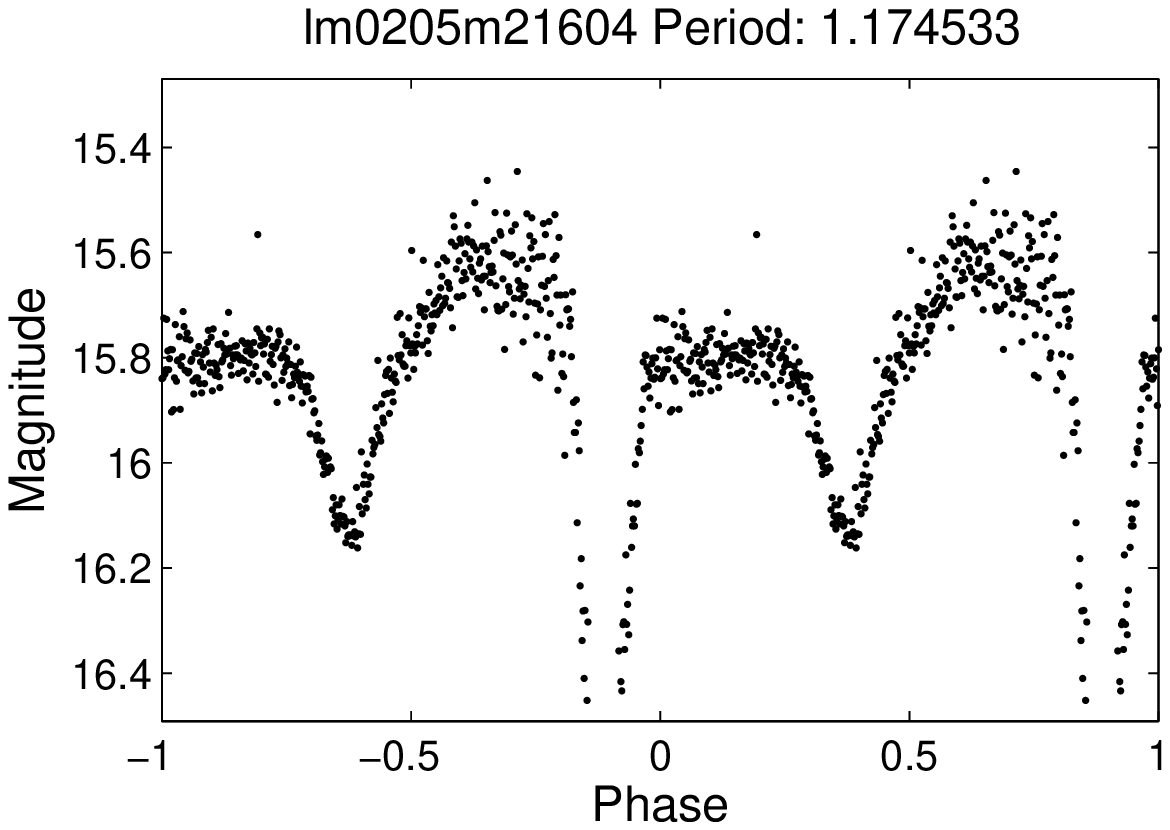}}  \hspace{5pt}
	\subfloat[]{\includegraphics[scale=0.55]{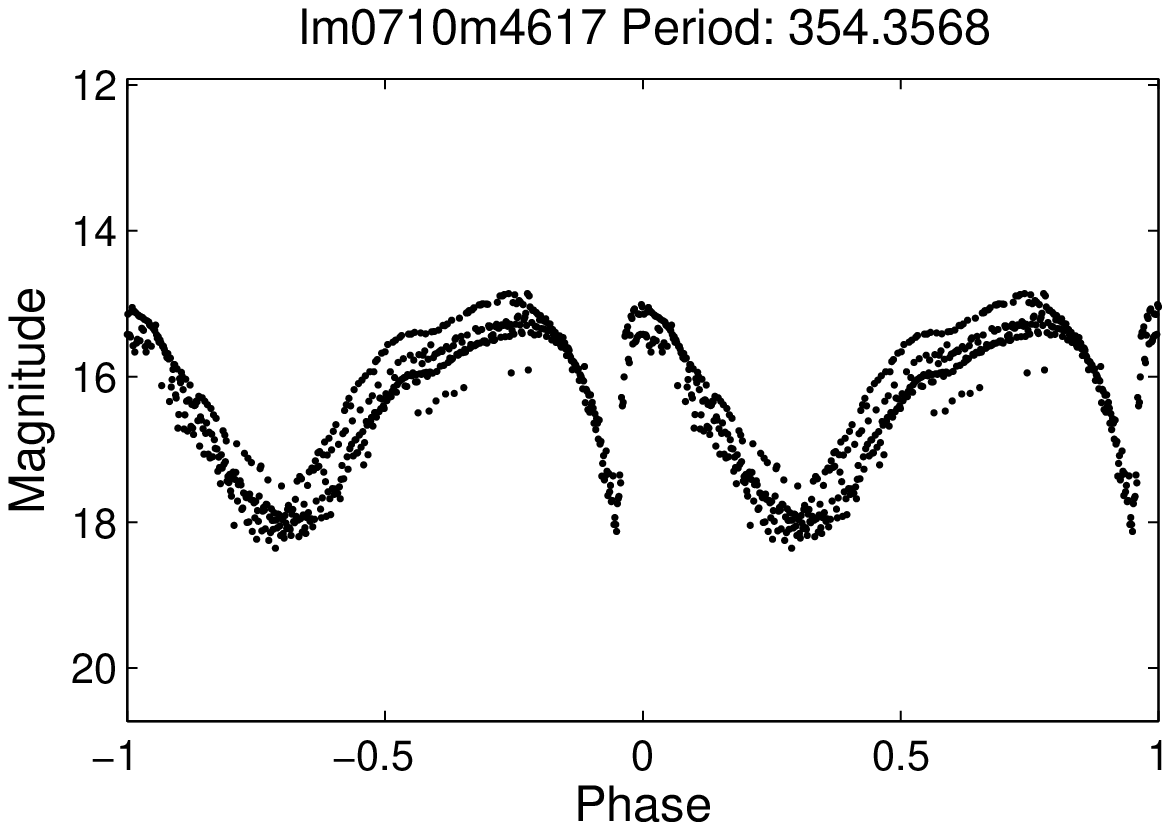}}  
	\caption{ \label{fig-exp-odd1} Examples of EROS-2 periodic lightcurves folded with their estimated period. A priori these objects cannot be attributed to any known class.}
\end{figure}

The method presented here is a not a classification method and therefore the method does not distinguish between types of periodic variables. Most of the periodic objects found in this work can be classified to known classes as it is clearly shown in Figures \ref{fig-exp-puls}, \ref{fig-exp-eb}, \ref{fig-exp-lpv} and \ref{fig-exp-ell}. It is also expected that there should or could be stars with periodic behavior that does not fall in one of the known categories. It is the scope of a different paper to identify those rare or novel phenomena.  Right here we only present a number of objects that we could not obviously attribute to any known classes or combination of classes. Figure \ref{fig-exp-odd1}  shows two such cases.

\section{Computational issues}

The proposed periodicity discrimination pipeline  has been programmed for computational architectures based on graphical processing units (GPUs). 
%GPGPU (general purpose computing in GPUs) is a new paradigm for highly parallel applications where high-complexity calculations are offloaded to the GPU.
 The implementation is programmed in CUDA \citet{CUDA2012}, which is a variation of C developed by GPU manufacturer NVIDIA.
 % Today, modern CPUs offer up to 240 GFlops (latest Intel Xeon E5 2690 released in march 2012) of numerical performance while modern GPUs offer approximately 4 TFlops (Nvidia GTX 680), equivalent to a super-computer from 10 years ago. Due to this big difference of orders of magnitude, people from the HPC community are moving towards GPU computing \cite{Kindratenko2011}. But, not all applications benefit from GPGPU. Explicit thread and data parallelism must be exploited in order to get the theoretical speed-ups of GPUs over CPUs. Problems that require looping or conditional branching are not well suited for GPGPU. 

\begin{figure}
	\centering
	\includegraphics[scale=0.65]{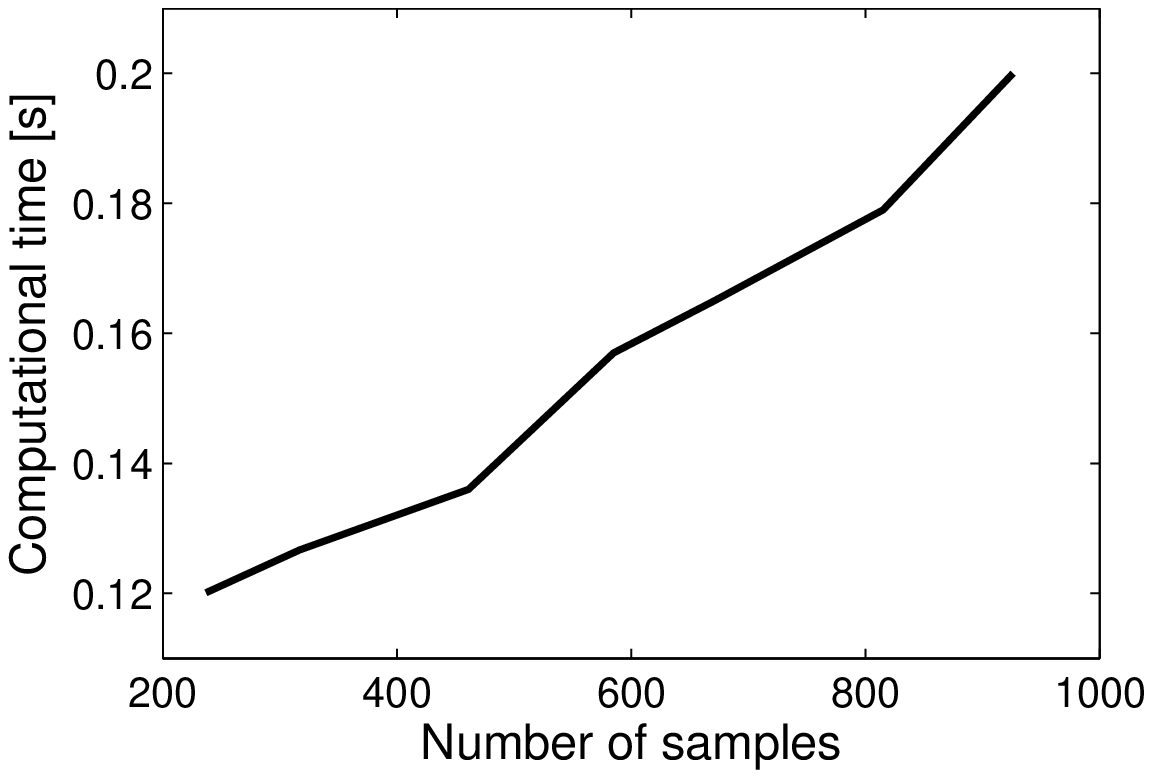} 
	\caption{ Computational time per lightcurve as a function of the number of samples  }
	\label{fig:ctime1}
\end{figure}

To evaluate the CKP metric (Eq. \ref{HP}), one requires the $N(N-1)/2$ interactions between the $N$ samples of the time series\footnote{The kernel matrices given by Eq. \eqref{Gausskernel} and Eq. \eqref{periodickernel} are symmetric, thus only the upper triangular part needs to be computed. The diagonal of the kernel matrices is constant and is omitted from the computations.}. The CKP can be computed efficiently by mapping each of these interactions to a single GPU thread. The final value of the CKP is obtained through a $\log(N)$-step sum reduction performed on the GPU. The computational time required to analyze one lightcurve using our periodic discrimination pipeline is shown in Fig. \ref{fig:ctime1}. These times include the importation and transferring of the lightcurves to the GPU device. Times were measured on a NVIDIA Tesla C2070 GPU. 

The 32.8 million lightcurves from the EROS-2 survey are processed on the NSCA Dell/NVIDIA cluster Forge. Forge is part of the Extreme Science and Engineering Discovery Environment (XSEDE). Forge has a total of 288 NVIDIA Tesla C2070 accelerators distributed on 44 nodes, however the maximum number of nodes that can be used at a time is 26. %The cluster has two queues. The first queue has 18 nodes available with 6 GPUs per node. The second queue has 8 nodes available with 8 GPUs per node. For a given node, $M$ OpenMP threads are launched, where $M$ is the number of available GPUs. Each thread controls one GPU. 
Each GPU process one chip from EROS-2. Table \ref{tab:ctime1} shows the total computational time required to process the 32.8 million lightcurves from the LMC and SMC. These times does not include the time required to transfer the dataset to the cluster nor the time a job is waiting on the queue. 
%Due to sharing of the cluster resources an average of 12 nodes at a time could be used. Using 12 nodes the EROS-2 subset is processed in 18 hours. 

\begin{table}[t]
	\begin{center}
	\caption{Total computational time required to process the 32.8 million EROS-2 lightcurves (LMC plus SMC) on XSEDE Forge cluster. GPUs in all nodes are NVIDIA Tesla C2070.}  	
	\begin{tabular}{l c c c}
	\hline
	Hardware &  Computational time \\ \hline
	Using 1 GPU &  52.2 days\\
	Using 6 GPUs (1 node) &  8.71 days \\
	Using 12 nodes (6 GPUs/node) &  17.41 hours \\
	Using all available nodes & 7.28 hours \\
	\hline
	\end{tabular}
	\label{tab:ctime1}
	\end{center} 
\end{table}

%--------------------------- Conclusion and acknowledgement  ----------------------

\section{Conclusion} \label{sec:conclusion}

We presented and described a fully automated pipeline for periodic light curve discrimination. The method is based on the CKP, a robust information theoretic metric that discriminates periodic behavior by analyzing the similarities between  lightcurve samples. The method is computational efficient; the pipeline takes 0.16 seconds to discriminate if a light curve is periodic or not. The 32.8 million light curves were processed using a GPU cluster in less than 24 hours. This suggests that with few additional optimizations and up-to-date hardware the methods may scale well for modern and larger light curve databases.

The periodicity discrimination pipeline was tested on light curves from the EROS-2 survey. The methods were calibrated using synthetic time series that preserve the characteristics of EROS-2 light curves. The calibration procedure is general and it could be applied to other astronomical time series databases easily. In total 32.8 million light curves from the LMC and SMC were processed finding a grand total of 121,147 and 24,855 periodic variables in the LMC and SMC, respectively. The results obtained are consistent in terms of period distribution and localization of the periodic variables in the color-magnitude diagram. The observed results suggest that the periodic variable catalogues generated by our method could be use to find multimode variables and periodic variables that do not fall in any known category. It is also hinted that higher order analysis, such as stellar classification and clustering may be carried out straight-forwardly using the provided periods.

Using the synthetic dataset and visually inspecting a small subset of the dataset, we were able to 
characterize the completeness and efficiency of the pipeline. We infer that 0.5\% of the lightcurves with SNR$>0.5$
are periodic.

Future work involves quasi-periodic and semi-regular behavior discrimination, more in-depth analysis of non-stationarities (trends) and developing more general kernel size selection schemes.

\section{Acknowledgement}

This work was funded by CONICYT-CHILE under grant FONDECYT 1110701 and 1140816, and its Doctorate Scholarship program. Pablo Est\'evez acknowledges support from the Ministry of Economy, Development, and Tourism's Millennium Science Initiative through grant IC12009, awarded to The Millennium Institute of Astrophysics, MAS.

%[PP: ADD IACS AND RESONANCE]
%PH: how about these?

The authors would like to thank the Harvard Institute for Applied Computational Science for providing research space and computing facilities.

The help received from the SEAS academic computing support staff and the time on the Harvard SEAS ``Resonance'' GPU cluster are greatly acknowledged. 

This work used the Extreme Science and Engineering Discovery Environment (XSEDE), which is supported by National Science Foundation grant number OCI-1053575.

The EROS-2 project was funded by the CEA and the CNRS through the IN2P3 and INSU institutes. JBM acknowledges financial support from "Programme National de Physique Stellaire" (PNPS) of CNRS/INSU, France.

\bibliographystyle{apj}
%\addcontentsline{toc}{chapter}{Bibliography}
\bibliography{references}

\begin{thebibliography}{41}
\expandafter\ifx\csname natexlab\endcsname\relax\def\natexlab#1{#1}\fi

\bibitem[{{Alcock} {et~al.}(2000){Alcock}, {Allsman}, {Alves}, {Axelrod},
  {Becker}, {Bennett}, {Cook}, {Dalal}, {Drake}, {Freeman}, {Geha}, {Griest},
  {Lehner}, {Marshall}, {Minniti}, {Nelson}, {Peterson}, {Popowski}, {Pratt},
  {Quinn}, {Stubbs}, {Sutherland}, {Tomaney}, {Vandehei}, \&
  {Welch}}]{Alcock2000}
{Alcock}, C., {Allsman}, R.~A., {Alves}, D.~R., {Axelrod}, T.~S., {Becker},
  A.~C., {Bennett}, D.~P., {Cook}, K.~H., {Dalal}, N., {Drake}, A.~J.,
  {Freeman}, K.~C., {Geha}, M., {Griest}, K., {Lehner}, M.~J., {Marshall},
  S.~L., {Minniti}, D., {Nelson}, C.~A., {Peterson}, B.~A., {Popowski}, P.,
  {Pratt}, M.~R., {Quinn}, P.~J., {Stubbs}, C.~W., {Sutherland}, W., {Tomaney},
  A.~B., {Vandehei}, T., \& {Welch}, D. 2000, The Astrophysical Journal, 542,
  281

\bibitem[{Buhlmann(1999)}]{Buhlmann99}
Buhlmann, P. 1999, Statistical Sciense, 17, 52

\bibitem[{{Cook} {et~al.}(1995){Cook}, {Alcock}, {Allsman}, {Axelrod},
  {Freeman}, {Peterson}, {Quinn}, {Rodgers}, {Bennett}, {Reimann}, {Griest},
  {Marshall}, {Pratt}, {Stubbs}, {Sutherland}, \& {Welch}}]{Cook1995}
{Cook}, K.~H., {Alcock}, C., {Allsman}, H.~A., {Axelrod}, T.~S., {Freeman},
  K.~C., {Peterson}, B.~A., {Quinn}, P.~J., {Rodgers}, A.~W., {Bennett}, D.~P.,
  {Reimann}, J., {Griest}, K., {Marshall}, S.~L., {Pratt}, M.~R., {Stubbs},
  C.~W., {Sutherland}, W., \& {Welch}, D.~L. 1995, in Astronomical Society of
  the Pacific Conference Series, Vol.~83, IAU Colloq. 155: Astrophysical
  Applications of Stellar Pulsation, ed. R.~S. {Stobie} \& P.~A. {Whitelock},
  221

\bibitem[{Edelson \& Krolik(1988)}]{Edelson1988}
Edelson, R.~A., \& Krolik, J. 1988, The Astrophysical Journal, 333, 646

\bibitem[{{Eyer}(1999)}]{Eyer1999}
{Eyer}, L. 1999, Baltic Astronomy, 8, 321

\bibitem[{{Hodapp} {et~al.}(2004){Hodapp}, {Kaiser}, {Aussel}, {Burgett},
  {Chambers}, {Chun}, {Dombeck}, {Douglas}, {Hafner}, {Heasley}, {Hoblitt},
  {Hude}, {Isani}, {Jedicke}, {Jewitt}, {Laux}, {Luppino}, {Lupton}, {Maberry},
  {Magnier}, {Mannery}, {Monet}, {Morgan}, {Onaka}, {Price}, {Ryan},
  {Siegmund}, {Szapudi}, {Tonry}, {Wainscoat}, \& {Waterson}}]{Hodapp2004}
{Hodapp}, K.~W., {Kaiser}, N., {Aussel}, H., {Burgett}, W., {Chambers}, K.~C.,
  {Chun}, M., {Dombeck}, T., {Douglas}, A., {Hafner}, D., {Heasley}, J.,
  {Hoblitt}, J., {Hude}, C., {Isani}, S., {Jedicke}, R., {Jewitt}, D., {Laux},
  U., {Luppino}, G.~A., {Lupton}, R., {Maberry}, M., {Magnier}, E., {Mannery},
  E., {Monet}, D., {Morgan}, J., {Onaka}, P., {Price}, P., {Ryan}, A.,
  {Siegmund}, W., {Szapudi}, I., {Tonry}, J., {Wainscoat}, R., \& {Waterson},
  M. 2004, Astronomische Nachrichten, 325, 636

\bibitem[{Huijse {et~al.}(2012)Huijse, Estevez, Protopapas, Zegers, \&
  Principe}]{Huijse2012}
Huijse, P., Estevez, P.~A., Protopapas, P., Zegers, P., \& Principe, J.~C.
  2012, IEEE Transactions on Signal Processing, 60, 5135

\bibitem[{{Ivezic} {et~al.}(2011){Ivezic}, {Tyson}, {Acosta}, {Allsman},
  {Anderson}, {Andrew}, {Angel}, {Axelrod}, {Barr}, {Becker}, {Becla},
  {Beldica}, {Blandford}, {Bloom}, {Borne}, {Brandt}, {Brown}, {Bullock},
  {Burke}, {Chandrasekharan}, {Chesley}, {Claver}, {Connolly}, {Cook},
  {Cooray}, {Covey}, {Cribbs}, {Cutri}, {Daues}, {Delgado}, {Ferguson},
  {Gawiser}, {Geary}, {Gee}, {Geha}, {Gibson}, {Gilmore}, {Gressler}, {Hogan},
  {Huffer}, {Jacoby}, {Jain}, {Jernigan}, {Jones}, {Juric}, {Kahn}, {Kalirai},
  {Kantor}, {Kessler}, {Kirkby}, {Knox}, {Krabbendam}, {Krughoff}, {Kulkarni},
  {Lambert}, {Levine}, {Liang}, {Lim}, {Lupton}, {Marshall}, {Marshall}, {May},
  {Miller}, {Mills}, {Monet}, {Neill}, {Nordby}, {O'Connor}, {Oliver},
  {Olivier}, {Olsen}, {Owen}, {Peterson}, {Petry}, {Pierfederici},
  {Pietrowicz}, {Pike}, {Pinto}, {Plante}, {Radeka}, {Rasmussen}, {Ridgway},
  {Rosing}, {Saha}, {Schalk}, {Schindler}, {Schneider}, {Schumacher}, {Sebag},
  {Seppala}, {Shipsey}, {Silvestri}, {Smith}, {Smith}, {Strauss}, {Stubbs},
  {Sweeney}, {Szalay}, {Thaler}, {Vanden Berk}, {Walkowicz}, {Warner},
  {Willman}, {Wittman}, {Wolff}, {Wood-Vasey}, {Yoachim}, {Zhan}, \& {for the
  LSST Collaboration}}]{LSST2012}
{Ivezic}, Z., {Tyson}, J.~A., {Acosta}, E., {Allsman}, R., {Anderson}, S.~F.,
  {Andrew}, J., {Angel}, R., {Axelrod}, T., {Barr}, J.~D., {Becker}, A.~C.,
  {Becla}, J., {Beldica}, C., {Blandford}, R.~D., {Bloom}, J.~S., {Borne}, K.,
  {Brandt}, W.~N., {Brown}, M.~E., {Bullock}, J.~S., {Burke}, D.~L.,
  {Chandrasekharan}, S., {Chesley}, S., {Claver}, C.~F., {Connolly}, A.,
  {Cook}, K.~H., {Cooray}, A., {Covey}, K.~R., {Cribbs}, C., {Cutri}, R.,
  {Daues}, G., {Delgado}, F., {Ferguson}, H., {Gawiser}, E., {Geary}, J.~C.,
  {Gee}, P., {Geha}, M., {Gibson}, R.~R., {Gilmore}, D.~K., {Gressler}, W.~J.,
  {Hogan}, C., {Huffer}, M.~E., {Jacoby}, S.~H., {Jain}, B., {Jernigan}, J.~G.,
  {Jones}, R.~L., {Juric}, M., {Kahn}, S.~M., {Kalirai}, J.~S., {Kantor},
  J.~P., {Kessler}, R., {Kirkby}, D., {Knox}, L., {Krabbendam}, V.~L.,
  {Krughoff}, S., {Kulkarni}, S., {Lambert}, R., {Levine}, D., {Liang}, M.,
  {Lim}, K., {Lupton}, R.~H., {Marshall}, P., {Marshall}, S., {May}, M.,
  {Miller}, M., {Mills}, D.~J., {Monet}, D.~G., {Neill}, D.~R., {Nordby}, M.,
  {O'Connor}, P., {Oliver}, J., {Olivier}, S.~S., {Olsen}, K., {Owen}, R.~E.,
  {Peterson}, J.~R., {Petry}, C.~E., {Pierfederici}, F., {Pietrowicz}, S.,
  {Pike}, R., {Pinto}, P.~A., {Plante}, R., {Radeka}, V., {Rasmussen}, A.,
  {Ridgway}, S.~T., {Rosing}, W., {Saha}, A., {Schalk}, T.~L., {Schindler},
  R.~H., {Schneider}, D.~P., {Schumacher}, G., {Sebag}, J., {Seppala}, L.~G.,
  {Shipsey}, I., {Silvestri}, N., {Smith}, J.~A., {Smith}, R.~C., {Strauss},
  M.~A., {Stubbs}, C.~W., {Sweeney}, D., {Szalay}, A., {Thaler}, J.~J., {Vanden
  Berk}, D., {Walkowicz}, L., {Warner}, M., {Willman}, B., {Wittman}, D.,
  {Wolff}, S.~C., {Wood-Vasey}, W.~M., {Yoachim}, P., {Zhan}, H., \& {for the
  LSST Collaboration}. 2011, ArXiv e-prints, living document found at:
  http://www.lsst.org/lsst/overview/

\bibitem[{Jenkins \& Watts(1968)}]{Jenkins1968}
Jenkins, G.~M., \& Watts, D.~G. 1968, Spectral analysis and its applications
  (Holden-day)

\bibitem[{{Larson} {et~al.}(2003){Larson}, {Beshore}, {Hill}, {Christensen},
  {McLean}, {Kolar}, {McNaught}, \& {Garradd}}]{Catalina2003}
{Larson}, S., {Beshore}, E., {Hill}, R., {Christensen}, E., {McLean}, D.,
  {Kolar}, S., {McNaught}, R., \& {Garradd}, G. 2003, in Bulletin of the
  American Astronomical Society, Vol.~35, AAS/Division for Planetary Sciences
  Meeting Abstracts \#35, 982

\bibitem[{{Law} {et~al.}(2009){Law}, {Kulkarni}, {Dekany}, {Ofek}, {Quimby},
  {Nugent}, {Surace}, {Grillmair}, {Bloom}, {Kasliwal}, {Bildsten}, {Brown},
  {Cenko}, {Ciardi}, {Croner}, {Djorgovski}, {van Eyken}, {Filippenko}, {Fox},
  {Gal-Yam}, {Hale}, {Hamam}, {Helou}, {Henning}, {Howell}, {Jacobsen},
  {Laher}, {Mattingly}, {McKenna}, {Pickles}, {Poznanski}, {Rahmer}, {Rau},
  {Rosing}, {Shara}, {Smith}, {Starr}, {Sullivan}, {Velur}, {Walters}, \&
  {Zolkower}}]{TF2009}
{Law}, N.~M., {Kulkarni}, S.~R., {Dekany}, R.~G., {Ofek}, E.~O., {Quimby},
  R.~M., {Nugent}, P.~E., {Surace}, J., {Grillmair}, C.~C., {Bloom}, J.~S.,
  {Kasliwal}, M.~M., {Bildsten}, L., {Brown}, T., {Cenko}, S.~B., {Ciardi}, D.,
  {Croner}, E., {Djorgovski}, S.~G., {van Eyken}, J., {Filippenko}, A.~V.,
  {Fox}, D.~B., {Gal-Yam}, A., {Hale}, D., {Hamam}, N., {Helou}, G., {Henning},
  J., {Howell}, D.~A., {Jacobsen}, J., {Laher}, R., {Mattingly}, S., {McKenna},
  D., {Pickles}, A., {Poznanski}, D., {Rahmer}, G., {Rau}, A., {Rosing}, W.,
  {Shara}, M., {Smith}, R., {Starr}, D., {Sullivan}, M., {Velur}, V.,
  {Walters}, R., \& {Zolkower}, J. 2009, \pasp, 121, 1395

\bibitem[{Mackay(1998)}]{Mackay1998}
Mackay, D. 1998, Introduction to Gaussian Processes, Vol. 168 (Springer,
  Berlin), 133--165

\bibitem[{Marquardt \& Acuff(1984)}]{Marquardt1984}
Marquardt, D., \& Acuff, S. 1984, Direct Quadratic Spectrum Estimation with
  Irregularly Spaced Data (Springer-Verlag), 211--223

\bibitem[{{Marquette} {et~al.}(2009){Marquette}, {Beaulieu}, {Buchler},
  {Szab{\'o}}, {Tisserand}, {Belghith}, {Fouqu{\'e}}, {Lesquoy}, {Milsztajn},
  {Schwarzenberg-Czerny}, {Afonso}, {Albert}, {Andersen}, {Ansari}, {Aubourg},
  {Bareyre}, {Charlot}, {Coutures}, {Ferlet}, {Glicenstein}, {Goldman},
  {Gould}, {Graff}, {Gros}, {Ha{\"\i}ssinski}, {Hamadache}, {de Kat}, {Le
  Guillou}, {Loup}, {Magneville}, {Maurice}, {Maury}, {Moniez},
  {Palanque-Delabrouille}, {Perdereau}, {Rahal}, {Rich}, {Spiro}, \&
  {Vidal-Madjar}}]{Marquette2009}
{Marquette}, J.~B., {Beaulieu}, J.~P., {Buchler}, J.~R., {Szab{\'o}}, R.,
  {Tisserand}, P., {Belghith}, S., {Fouqu{\'e}}, P., {Lesquoy}, {\'E}.,
  {Milsztajn}, A., {Schwarzenberg-Czerny}, A., {Afonso}, C., {Albert}, J.~N.,
  {Andersen}, J., {Ansari}, R., {Aubourg}, {\'E}., {Bareyre}, P., {Charlot},
  X., {Coutures}, C., {Ferlet}, R., {Glicenstein}, J.~F., {Goldman}, B.,
  {Gould}, A., {Graff}, D., {Gros}, M., {Ha{\"\i}ssinski}, J., {Hamadache}, C.,
  {de Kat}, J., {Le Guillou}, L., {Loup}, C., {Magneville}, C., {Maurice},
  {\'E}., {Maury}, A., {Moniez}, M., {Palanque-Delabrouille}, N., {Perdereau},
  O., {Rahal}, Y.~R., {Rich}, J., {Spiro}, M., \& {Vidal-Madjar}, A. 2009,
  \aap, 495, 249

\bibitem[{Michalak(2010)}]{Michalak2010}
Michalak, M. 2010, in Computer Recognition Systems 4 (Berlin: Springer Verlag),
  136--146

\bibitem[{{Moskalik}(2013)}]{Moskalik2012}
{Moskalik}, P. 2013, in Advances in Solid State Physics, Vol.~31, Advances in
  Solid State Physics, ed. J.~C. {Su{\'a}rez}, R.~{Garrido}, L.~A. {Balona}, \&
  J.~{Christensen-Dalsgaard}, 103

\bibitem[{NVIDIA(2012)}]{CUDA2012}
NVIDIA. 2012, CUDA C Programming Guide version 4.2 (NVIDIA)

\bibitem[{Principe(2010)}]{Principe2010}
Principe, J. 2010, Information Theoretic Learning: Renyi's Entropy and Kernel
  Perspectives (New York: Springer Verlag)

\bibitem[{{Rahal} {et~al.}(2009){Rahal}, {Afonso}, {Albert}, {Andersen},
  {Ansari}, {Aubourg}, {Bareyre}, {Beaulieu}, {Charlot}, {Couchot}, {Coutures},
  {Derue}, {Ferlet}, {Fouqu{\'e}}, {Glicenstein}, {Goldman}, {Gould}, {Graff},
  {Gros}, {Ha{\"\i}ssinski}, {Hamadache}, {de Kat}, {Lesquoy}, {Loup}, {Le
  Guillou}, {Magneville}, {Mansoux}, {Marquette}, {Maurice}, {Maury},
  {Milsztajn}, {Moniez}, {Palanque-Delabrouille}, {Perdereau}, {Rahvar},
  {Rich}, {Spiro}, {Tisserand}, {Vidal-Madjar}, \& {EROS-2
  Collaboration}}]{EROS2009}
{Rahal}, Y.~R., {Afonso}, C., {Albert}, J.-N., {Andersen}, J., {Ansari}, R.,
  {Aubourg}, {\'E}., {Bareyre}, P., {Beaulieu}, J.-P., {Charlot}, X.,
  {Couchot}, F., {Coutures}, C., {Derue}, F., {Ferlet}, R., {Fouqu{\'e}}, P.,
  {Glicenstein}, J.-F., {Goldman}, B., {Gould}, A., {Graff}, D., {Gros}, M.,
  {Ha{\"\i}ssinski}, J., {Hamadache}, C., {de Kat}, J., {Lesquoy}, {\'E}.,
  {Loup}, C., {Le Guillou}, L., {Magneville}, C., {Mansoux}, B., {Marquette},
  J.-B., {Maurice}, {\'E}., {Maury}, A., {Milsztajn}, A., {Moniez}, M.,
  {Palanque-Delabrouille}, N., {Perdereau}, O., {Rahvar}, S., {Rich}, J.,
  {Spiro}, M., {Tisserand}, P., {Vidal-Madjar}, A., \& {EROS-2 Collaboration}.
  2009, Astronomy \& Astrophysics, 500, 1027

\bibitem[{Rasmussen \& Williams(2006)}]{Rasmussen2006}
Rasmussen, C.~E., \& Williams, C. K.~I. 2006, Gaussian processes for machine
  learning (MIT Press)

\bibitem[{Reimann(1994)}]{Reimann1994}
Reimann, J.~D. 1994, Frequency Estimation Using Unequally-Spaced Astronomical
  Data (University of California, Berkeley)

\bibitem[{Scargle(1982)}]{Scargle1982}
Scargle, J. 1982, The Astrophysical Journal, 263, 835

\bibitem[{Schmitz \& Schreiber(1999)}]{Schmitz1999}
Schmitz, A., \& Schreiber, T. 1999, Phys. Rev. E, 59, 4044

\bibitem[{Schreiber \& Schmitz(1999)}]{Schreiber1999}
Schreiber, T., \& Schmitz, A. 1999, Physica D: Nonlinear Phenomena, 142, 346

\bibitem[{Sch\text{\"{o}}lkopf \& Smola(2002)}]{Scholkopf2002}
Sch\text{\"{o}}lkopf, B., \& Smola, A. 2002, Learning with Kernels (Cambridge,
  MA: Cambridge, MA: MIT Press)

\bibitem[{{Smolec} {et~al.}(2012){Smolec}, {Soszy{\'n}ski}, {Moskalik},
  {Udalski}, {Szyma{\'n}ski}, {Kubiak}, {Pietrzy{\'n}ski}, {Wyrzykowski},
  {Ulaczyk}, {Poleski}, {Koz{\l}owski}, \& {Pietrukowicz}}]{Smolec2012}
{Smolec}, R., {Soszy{\'n}ski}, I., {Moskalik}, P., {Udalski}, A.,
  {Szyma{\'n}ski}, M.~K., {Kubiak}, M., {Pietrzy{\'n}ski}, G., {Wyrzykowski},
  {\L}., {Ulaczyk}, K., {Poleski}, R., {Koz{\l}owski}, S., \& {Pietrukowicz},
  P. 2012, \mnras, 419, 2407

\bibitem[{{Soszy{\~n}ski} {et~al.}(2010{\natexlab{a}}){Soszy{\~n}ski},
  {Poleski}, {Udalski}, {Szyma{\~n}ski}, {Kubiak}, {Pietrzy{\~n}ski},
  {Wyrzykowski}, {Szewczyk}, \& {Ulaczyk}}]{Soszynski2010-1}
{Soszy{\~n}ski}, I., {Poleski}, R., {Udalski}, A., {Szyma{\~n}ski}, M.~K.,
  {Kubiak}, M., {Pietrzy{\~n}ski}, G., {Wyrzykowski}, {\L}., {Szewczyk}, O., \&
  {Ulaczyk}, K. 2010{\natexlab{a}}, Acta Astron, 60, 17

\bibitem[{{Soszy{\~n}ski} {et~al.}(2010{\natexlab{b}}){Soszy{\~n}ski},
  {Udalski}, {Szyma{\~n}ski}, {Kubiak}, {Pietrzy{\~n}ski}, {Wyrzykowski},
  {Ulaczyk}, \& {Poleski}}]{Soszynski2010-3}
{Soszy{\~n}ski}, I., {Udalski}, A., {Szyma{\~n}ski}, M.~K., {Kubiak}, J.,
  {Pietrzy{\~n}ski}, G., {Wyrzykowski}, {\L}., {Ulaczyk}, K., \& {Poleski}, R.
  2010{\natexlab{b}}, Acta Astron, 60, 165

\bibitem[{{Soszy{\~n}ski} {et~al.}(2009){Soszy{\~n}ski}, {Udalski},
  {Szyma{\~n}ski}, {Kubiak}, {Pietrzy{\~n}ski}, {Wyrzykowski}, {Szewczyk},
  {Ulaczyk}, \& {Poleski}}]{Soszynski2009-2}
{Soszy{\~n}ski}, I., {Udalski}, A., {Szyma{\~n}ski}, M.~K., {Kubiak}, M.,
  {Pietrzy{\~n}ski}, G., {Wyrzykowski}, {\L}., {Szewczyk}, O., {Ulaczyk}, K.,
  \& {Poleski}, R. 2009, Acta Astron, 59, 239

\bibitem[{{Soszy{\~n}ski} {et~al.}(2010{\natexlab{c}}){Soszy{\~n}ski},
  {Udalski}, {Szyma{\~n}ski}, {Kubiak}, {Pietrzy{\~n}ski}, {Wyrzykowski},
  {Ulaczyk}, \& {Poleski}}]{Soszynski2010-2}
{Soszy{\~n}ski}, I., {Udalski}, A., {Szyma{\~n}ski}, M.~K., {Kubiak}, M.,
  {Pietrzy{\~n}ski}, G., {Wyrzykowski}, {\L}., {Ulaczyk}, K., \& {Poleski}, R.
  2010{\natexlab{c}}, Acta Astron, 60, 91

\bibitem[{{Soszynski} {et~al.}(2008){Soszynski}, {Poleski}, {Udalski},
  {Szymanski}, {Kubiak}, {Pietrzynski}, {Wyrzykowski}, {Szewczyk}, \&
  {Ulaczyk}}]{Soszynski2008-1}
{Soszynski}, I., {Poleski}, R., {Udalski}, A., {Szymanski}, M.~K., {Kubiak},
  M., {Pietrzynski}, G., {Wyrzykowski}, L., {Szewczyk}, O., \& {Ulaczyk}, K.
  2008, Acta Astron, 58, 163

\bibitem[{{Soszy{\'n}ski} {et~al.}(2008){Soszy{\'n}ski}, {Udalski},
  {Szyma{\'n}ski}, {Kubiak}, {Pietrzy{\'n}ski}, {Wyrzykowski}, {Szewczyk},
  {Ulaczyk}, \& {Poleski}}]{Soszynski2008-2}
{Soszy{\'n}ski}, I., {Udalski}, A., {Szyma{\'n}ski}, M.~K., {Kubiak}, M.,
  {Pietrzy{\'n}ski}, G., {Wyrzykowski}, {\L}., {Szewczyk}, O., {Ulaczyk}, K.,
  \& {Poleski}, R. 2008, Acta Astron, 58, 293

\bibitem[{{Soszy{\'n}ski} {et~al.}(2009){Soszy{\'n}ski}, {Udalski},
  {Szyma{\'n}ski}, {Kubiak}, {Pietrzy{\'n}ski}, {Wyrzykowski}, {Szewczyk},
  {Ulaczyk}, \& {Poleski}}]{Soszynski2009-1}
---. 2009, Acta Astron, 59, 1

\bibitem[{{Soszy{\'n}ski} {et~al.}(2011){Soszy{\'n}ski}, {Udalski},
  {Szyma{\'n}ski}, {Kubiak}, {Pietrzy{\'n}ski}, {Wyrzykowski}, {Ulaczyk},
  {Poleski}, {Koz{\l}owski}, \& {Pietrukowicz}}]{Soszynski2011}
{Soszy{\'n}ski}, I., {Udalski}, A., {Szyma{\'n}ski}, M.~K., {Kubiak}, M.,
  {Pietrzy{\'n}ski}, G., {Wyrzykowski}, {\L}., {Ulaczyk}, K., {Poleski}, R.,
  {Koz{\l}owski}, S., \& {Pietrukowicz}, P. 2011, Acta Astron, 61, 217

\bibitem[{{Spano} {et~al.}(2009){Spano}, {Mowlavi}, {Eyer}, \&
  {Burki}}]{Spano2009}
{Spano}, M., {Mowlavi}, N., {Eyer}, L., \& {Burki}, G. 2009, in American
  Institute of Physics Conference Series, Vol. 1170, American Institute of
  Physics Conference Series, ed. J.~A. {Guzik} \& P.~A. {Bradley}, 324--326

\bibitem[{{Stellingwerf}(1978)}]{Stellingwerf1978}
{Stellingwerf}, R. 1978, The Astrophysical Journal, 224, 953

\bibitem[{Taylor \& Cristianini(2004)}]{Cristianini2004}
Taylor, J.~S., \& Cristianini, N. 2004, Kernel Methods for Pattern Analysis
  (Cambridge University Press)

\bibitem[{{Tisserand} {et~al.}(2007){Tisserand}, {Le Guillou}, {Afonso},
  {Albert}, {Andersen}, {Ansari}, {Aubourg}, {Bareyre}, {Beaulieu}, {Charlot},
  {Coutures}, {Ferlet}, {Fouqu{\'e}}, {Glicenstein}, {Goldman}, {Gould},
  {Graff}, {Gros}, {Haissinski}, {Hamadache}, {de Kat}, {Lasserre}, {Lesquoy},
  {Loup}, {Magneville}, {Marquette}, {Maurice}, {Maury}, {Milsztajn}, {Moniez},
  {Palanque-Delabrouille}, {Perdereau}, {Rahal}, {Rich}, {Spiro},
  {Vidal-Madjar}, {Vigroux}, {Zylberajch}, \& {EROS-2
  Collaboration}}]{EROS2007}
{Tisserand}, P., {Le Guillou}, L., {Afonso}, C., {Albert}, J.~N., {Andersen},
  J., {Ansari}, R., {Aubourg}, {\'E}., {Bareyre}, P., {Beaulieu}, J.~P.,
  {Charlot}, X., {Coutures}, C., {Ferlet}, R., {Fouqu{\'e}}, P., {Glicenstein},
  J.~F., {Goldman}, B., {Gould}, A., {Graff}, D., {Gros}, M., {Haissinski}, J.,
  {Hamadache}, C., {de Kat}, J., {Lasserre}, T., {Lesquoy}, {\'E}., {Loup}, C.,
  {Magneville}, C., {Marquette}, J.~B., {Maurice}, {\'E}., {Maury}, A.,
  {Milsztajn}, A., {Moniez}, M., {Palanque-Delabrouille}, N., {Perdereau}, O.,
  {Rahal}, Y.~R., {Rich}, J., {Spiro}, M., {Vidal-Madjar}, A., {Vigroux}, L.,
  {Zylberajch}, S., \& {EROS-2 Collaboration}. 2007, Astronomy \& Astrophysics,
  496, 387

\bibitem[{Udalski {et~al.}(1997)Udalski, Kubiak, \& Szymanski}]{Udalski1997}
Udalski, A., Kubiak, M., \& Szymanski, M. 1997, Acta Astronomica, 47, 319

\bibitem[{{Wang} {et~al.}(2012){Wang}, {Khardon}, \& {Protopapas}}]{Wang2012}
{Wang}, Y., {Khardon}, R., \& {Protopapas}, P. 2012, \apj, 756, 67

\bibitem[{{York} {et~al.}(2000){York}, {Adelman}, {Anderson}, {Anderson},
  {Annis}, {Bahcall}, {Bakken}, {Barkhouser}, {Bastian}, {Berman}, {Boroski},
  {Bracker}, {Briegel}, {Briggs}, {Brinkmann}, {Brunner}, {Burles}, {Carey},
  {Carr}, {Castander}, {Chen}, {Colestock}, {Connolly}, {Crocker}, {Csabai},
  {Czarapata}, {Davis}, {Doi}, {Dombeck}, {Eisenstein}, {Ellman}, {Elms},
  {Evans}, {Fan}, {Federwitz}, {Fiscelli}, {Friedman}, {Frieman}, {Fukugita},
  {Gillespie}, {Gunn}, {Gurbani}, {de Haas}, {Haldeman}, {Harris}, {Hayes},
  {Heckman}, {Hennessy}, {Hindsley}, {Holm}, {Holmgren}, {Huang}, {Hull},
  {Husby}, {Ichikawa}, {Ichikawa}, {Ivezi{\'c}}, {Kent}, {Kim}, {Kinney},
  {Klaene}, {Kleinman}, {Kleinman}, {Knapp}, {Korienek}, {Kron}, {Kunszt},
  {Lamb}, {Lee}, {Leger}, {Limmongkol}, {Lindenmeyer}, {Long}, {Loomis},
  {Loveday}, {Lucinio}, {Lupton}, {MacKinnon}, {Mannery}, {Mantsch}, {Margon},
  {McGehee}, {McKay}, {Meiksin}, {Merelli}, {Monet}, {Munn}, {Narayanan},
  {Nash}, {Neilsen}, {Neswold}, {Newberg}, {Nichol}, {Nicinski}, {Nonino},
  {Okada}, {Okamura}, {Ostriker}, {Owen}, {Pauls}, {Peoples}, {Peterson},
  {Petravick}, {Pier}, {Pope}, {Pordes}, {Prosapio}, {Rechenmacher}, {Quinn},
  {Richards}, {Richmond}, {Rivetta}, {Rockosi}, {Ruthmansdorfer}, {Sandford},
  {Schlegel}, {Schneider}, {Sekiguchi}, {Sergey}, {Shimasaku}, {Siegmund},
  {Smee}, {Smith}, {Snedden}, {Stone}, {Stoughton}, {Strauss}, {Stubbs},
  {SubbaRao}, {Szalay}, {Szapudi}, {Szokoly}, {Thakar}, {Tremonti}, {Tucker},
  {Uomoto}, {Vanden Berk}, {Vogeley}, {Waddell}, {Wang}, {Watanabe},
  {Weinberg}, {Yanny}, {Yasuda}, \& {SDSS Collaboration}}]{York2000}
{York}, D.~G., {Adelman}, J., {Anderson}, Jr., J.~E., {Anderson}, S.~F.,
  {Annis}, J., {Bahcall}, N.~A., {Bakken}, J.~A., {Barkhouser}, R., {Bastian},
  S., {Berman}, E., {Boroski}, W.~N., {Bracker}, S., {Briegel}, C., {Briggs},
  J.~W., {Brinkmann}, J., {Brunner}, R., {Burles}, S., {Carey}, L., {Carr},
  M.~A., {Castander}, F.~J., {Chen}, B., {Colestock}, P.~L., {Connolly}, A.~J.,
  {Crocker}, J.~H., {Csabai}, I., {Czarapata}, P.~C., {Davis}, J.~E., {Doi},
  M., {Dombeck}, T., {Eisenstein}, D., {Ellman}, N., {Elms}, B.~R., {Evans},
  M.~L., {Fan}, X., {Federwitz}, G.~R., {Fiscelli}, L., {Friedman}, S.,
  {Frieman}, J.~A., {Fukugita}, M., {Gillespie}, B., {Gunn}, J.~E., {Gurbani},
  V.~K., {de Haas}, E., {Haldeman}, M., {Harris}, F.~H., {Hayes}, J.,
  {Heckman}, T.~M., {Hennessy}, G.~S., {Hindsley}, R.~B., {Holm}, S.,
  {Holmgren}, D.~J., {Huang}, C.-h., {Hull}, C., {Husby}, D., {Ichikawa},
  S.-I., {Ichikawa}, T., {Ivezi{\'c}}, {\v Z}., {Kent}, S., {Kim}, R.~S.~J.,
  {Kinney}, E., {Klaene}, M., {Kleinman}, A.~N., {Kleinman}, S., {Knapp},
  G.~R., {Korienek}, J., {Kron}, R.~G., {Kunszt}, P.~Z., {Lamb}, D.~Q., {Lee},
  B., {Leger}, R.~F., {Limmongkol}, S., {Lindenmeyer}, C., {Long}, D.~C.,
  {Loomis}, C., {Loveday}, J., {Lucinio}, R., {Lupton}, R.~H., {MacKinnon}, B.,
  {Mannery}, E.~J., {Mantsch}, P.~M., {Margon}, B., {McGehee}, P., {McKay},
  T.~A., {Meiksin}, A., {Merelli}, A., {Monet}, D.~G., {Munn}, J.~A.,
  {Narayanan}, V.~K., {Nash}, T., {Neilsen}, E., {Neswold}, R., {Newberg},
  H.~J., {Nichol}, R.~C., {Nicinski}, T., {Nonino}, M., {Okada}, N., {Okamura},
  S., {Ostriker}, J.~P., {Owen}, R., {Pauls}, A.~G., {Peoples}, J., {Peterson},
  R.~L., {Petravick}, D., {Pier}, J.~R., {Pope}, A., {Pordes}, R., {Prosapio},
  A., {Rechenmacher}, R., {Quinn}, T.~R., {Richards}, G.~T., {Richmond}, M.~W.,
  {Rivetta}, C.~H., {Rockosi}, C.~M., {Ruthmansdorfer}, K., {Sandford}, D.,
  {Schlegel}, D.~J., {Schneider}, D.~P., {Sekiguchi}, M., {Sergey}, G.,
  {Shimasaku}, K., {Siegmund}, W.~A., {Smee}, S., {Smith}, J.~A., {Snedden},
  S., {Stone}, R., {Stoughton}, C., {Strauss}, M.~A., {Stubbs}, C., {SubbaRao},
  M., {Szalay}, A.~S., {Szapudi}, I., {Szokoly}, G.~P., {Thakar}, A.~R.,
  {Tremonti}, C., {Tucker}, D.~L., {Uomoto}, A., {Vanden Berk}, D., {Vogeley},
  M.~S., {Waddell}, P., {Wang}, S.-i., {Watanabe}, M., {Weinberg}, D.~H.,
  {Yanny}, B., {Yasuda}, N., \& {SDSS Collaboration}. 2000, The Astronomical
  Journal, 120, 1579

\end{thebibliography}

\end{document}